\documentclass[aps,reprint,twocolumn,footinbib,longbibliography,superscriptaddress]{revtex4-2}
\usepackage{amsmath}
\usepackage{amsfonts}
\usepackage{amssymb}
\usepackage{times}
\usepackage[dvipdfmx]{graphicx}
\usepackage[usenames,dvipsnames]{xcolor}
\usepackage{bm}
\usepackage{amsthm}
\usepackage[english]{babel}
\usepackage{blindtext}
\usepackage{tikz}
\usetikzlibrary{intersections, calc, arrows.meta}

\usepackage{physics}
\usepackage{mathrsfs}

\usepackage[hidelinks]{hyperref}
\hypersetup{
  colorlinks   = true, %Colours links instead of ugly boxes
  urlcolor     = blue, %Colour for external hyperlinks
  linkcolor    = blue, %Colour of internal links
  citecolor   = blue %Colour of citations, could be ``red''
}

\usepackage{color}% for colors
\RequirePackage[dvipsnames,usenames]{xcolor}

\newcommand{\im}{\mathrm{im}\,}

\DeclareMathOperator*{\argmin}{arg\,min}

\newcommand{\transpose}{\mathsf{T}}

\newcommand{\addd}[1]{#1}

\begin{document}

\title{Housekeeping and excess entropy production for general nonlinear dynamics}
\date{\today}

\author{Kohei Yoshimura}
\email{kyoshimura@ubi.s.u-tokyo.ac.jp}
\affiliation{Department of Physics, The University of Tokyo, 7-3-1 Hongo, Bunkyo-ku, Tokyo 113-0033, Japan}
\author{Artemy Kolchinsky}
\affiliation{Universal Biology Institute, The University of Tokyo, 7-3-1 Hongo, Bunkyo-ku, Tokyo 113-0033, Japan}
\author{Andreas Dechant}
\affiliation{Department of Physics No. 1, Graduate School of Science, Kyoto University, Kyoto 606-8502, Japan}
\author{Sosuke Ito}
\affiliation{Department of Physics, The University of Tokyo, 7-3-1 Hongo, Bunkyo-ku, Tokyo 113-0033, Japan}
\affiliation{Universal Biology Institute, The University of Tokyo, 7-3-1 Hongo, Bunkyo-ku, Tokyo 113-0033, Japan}

\begin{abstract}
    We propose a housekeeping/excess decomposition of entropy production for general nonlinear dynamics in a discrete space, including chemical reaction networks and discrete stochastic systems. We exploit the geometric structure of thermodynamic forces to define the decomposition; this does not rely on the notion of a steady state, and even applies to systems that exhibit multistability, limit cycles, and chaos. In the decomposition, distinct aspects of the dynamics contribute separately to entropy production: the housekeeping part stems from a cyclic mode that arises from external driving, generalizing Schnakenberg's cyclic decomposition to non-steady states, while the excess part stems from an instantaneous relaxation mode that arises from conservative forces. Our decomposition refines previously known thermodynamic uncertainty relations and speed limits. In particular, it not only improves an optimal-transport-theoretic speed limit, but also extends the optimal transport theory of discrete systems to nonlinear and nonconservative settings.
\end{abstract}

\maketitle
\section{Introduction}

The notion of ``nonequilibrium'' has two aspects: breaking of detailed balance and nonstationarity. 
The housekeeping/excess decomposition of entropy production rate (EPR) is a way to deal with these two aspects of nonequilibrium systems~\cite{oono1998steady,hatano2001steady,esposito2010three,maes2014nonequilibrium}. 
Such a decomposition has previously been formulated based on steady states: the housekeeping EPR characterizes how detailed balance is broken in a steady state, while the excess part quantifies the additional dissipation that is needed to transition between steady states. 
The best known decomposition of this kind was proposed by Hatano and Sasa (HS)~\cite{hatano2001steady} (also known as the  adiabatic/nonadiabatic decomposition~\cite{esposito2010three}), while an important alternative decomposition was proposed by Maes and Neto\v{c}n\'{y} (MN)~\cite{maes2014nonequilibrium}.  
The practical utility of these decompositions is that they provide meaningful thermodynamic bounds for nonequilibrium systems where the integrated entropy production (EP) diverges, by discounting the housekeeping EP, which also diverges, to give a finite excess EP~\cite{oono1998steady}.

However, previously known decompositions have a crucial limitation. Namely, they are not defined for general nonlinear dynamics, such as general chemical reaction networks. Such systems are not always globally stable and can exhibit various nontrivial phenomena such as limit cycles, bifurcations and multistability, in contrast to existing EPR decompositions which assume global stability. 
More precisely, the HS decomposition has been generalized to chemical reaction networks only if there is a special steady state, called complex balanced steady state~\cite{rao2016nonequilibrium,ge2016nonequilibrium}, which is always globally stable~\footnote{More precisely, systems with complex balanced steady states are globally stable within each stoichiometric compatibility class~\cite{horn1972general,craciun2015toric,anderson2011proof}.}. 
However, even if a chemical reaction network has a stable steady state, it may not be complex balanced. In that case, it is possible for the HS decomposition to give a negative value of excess EPR, which makes it difficult to interpret the decomposition physically. 
Because a complex balanced steady state is ``almost detailed balanced'' in the sense that it always shows global stability and cannot exhibit genuine nonlinear phenomena such as multistability or chaos~\cite{feinberg2019foundations}, this particular generalization does not help us understand the various physically interesting situations which play a key role in biology and other complex chemical systems~\cite{beard2008chemical}. 

\begin{figure}
    \centering
    \includegraphics[width=\linewidth]{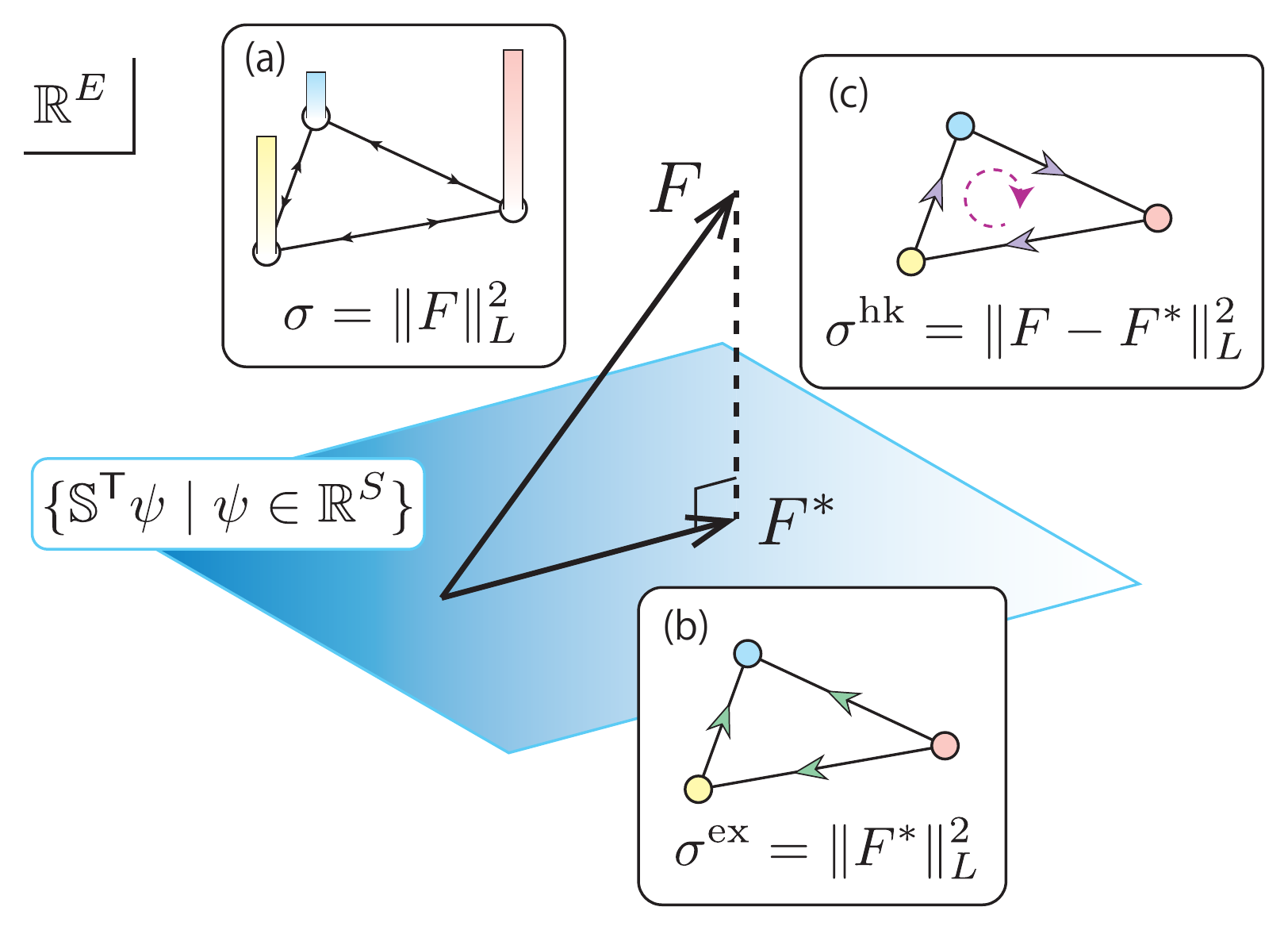}
    \caption{
    We propose a decomposition of (a) the total entropy production rate (EPR) in general nonlinear dynamics in a discrete space (e.g., a chemical reaction network, and a Markov jump process) into (b) the excess part that reflects a relaxation mode and (c) the housekeeping part that reflects a cyclic mode. 
    The total EPR $\sigma$ is expressed as the square norm of the thermodynamic force $F\in\mathbb{R}^E$ with the metric $L$. The projection of the force onto the space of conservative forces $\{\mathbb{S}^\transpose \psi\mid\psi\in\mathbb{R}^S\}\subset\mathbb{R}^E$ defined in Eq.~\eqref{eq:projection} determines the conservative force $F^*$ that provides the Onsager excess EPR $\sigma^\mathrm{ex}$. The difference $F-F^*$ gives the Onsager housekeeping EPR $\sigma^\mathrm{hk}$, which can be expressed as the sum of cyclic contributions as in Eq.~\eqref{eq:hkcyclic}. 
    Because $F^*$ and $F-F^*$ are orthogonal with respect to the inner product given by $L$, $\sigma^\mathrm{ex}$ and $\sigma^\mathrm{hk}$ decompose the total EPR, $\sigma=\sigma^\mathrm{ex}+\sigma^\mathrm{hk}$.}
    \label{fig:concept}
\end{figure}

In this paper, we propose an EPR decomposition that applies to general nonlinear dynamics in a discrete space, which include Markov jump processes and chemical reaction networks as special cases (Fig.~\ref{fig:concept}). 
Because the decomposition is defined based on the geometric structure of thermodynamic forces, it does not require any information about the steady state or even its existence.
Since it is not related to steady states, the decomposition is not only applicable to a wider range of dynamics, but it also provides crucial insights into the physical meaning of the individual terms: it allows us to interpret each term of the decomposition based on physical modes of dynamics, beyond the notion of stationarity. 
The housekeeping part reflects a cyclic mode on the graph of states that does not affect the apparent dynamics, and it quantifies how external driving breaks detailed balance at a given instant of time. As a result, we generalize Schnakenberg's cyclic decomposition of steady-state dissipation into contributions stemming from cycles in the graph of states~\cite{schnakenberg1976network}. 
On the other hand, the excess term expresses a ``pseudo-relaxation'' mode even if the system does not relax to a steady state, by extracting a conservative component of the thermodynamic force. Because a conservative force is given by a potential difference of the internal states, it represents an internal free-energy-like contribution to dissipation. Using a corresponding instantaneous ``pseudo-equilibrium'', we obtain a gradient-flow form of an equation of motion, which was previously used to describe the relaxation to an equilibrium state~\cite{mielke2011gradient,mielke2013geodesic}. 
In summary, our result not only allows us to decompose the EPR in general nonlinear dynamics in a discrete space, but also to separate the effects on dissipation stemming from the external cyclic mode and the internal relaxation mode. 

While this physical interpretation of our decomposition is different from previous results, it is also consistent with existing theory of nonequilibrium thermodynamics and effectively refines several results. 
It gives much tighter thermodynamic uncertainty relations (TURs) and speed limits, which is concordant with the original philosophy of EPR decomposition~\cite{oono1998steady} and strengthens its physical validity. 
Concretely, we derive refined versions of short-time and finite-time TURs for Markov jump processes~\cite{gingrich2016dissipation,pietzonka2016universal,horowitz2017proof,proesmans2017discrete,pietzonka2017finite, dechant2018current,dechant2018multidimensional, hasegawa2019fluctuation,falasco2020unifying,wolpert2020uncertainty,otsubo2020estimating,manikandan2020inferring,liu2020thermodynamic,dechant2020fluctuation,dechant2021continuous,dechant2022geometric1,dechant2022geometric2}, and a short-time TUR for chemical reaction networks~\cite{yoshimura2021thermodynamic}. 
We also generalize an optimal-transport-theoretic speed limit derived in~\cite{van2021geometrical} to general nonequilibrium systems. 

More generally, we discuss connections between our decomposition and optimal transport theory. 
Optimal transport theory interprets the minimum transportation cost from one probability distribution to another as a distance between them~\cite{villani2009optimal}, providing a natural geometry for probability distributions. 
Recent studies have revealed that one can obtain thermodynamic bounds in continuous systems through this geometrical approach, associating thermodynamic costs with the optimal-transport-theoretic distance~\cite{aurell2011optimal,aurell2012refined,bo2013optimal,dechant2019thermodynamic,chen2019stochastic,fu2021maximal,nakazato2021geometrical,dechant2022geometric1,dechant2022geometric2,miangolarra2022geometry}. 
In particular, a distance called $L^2$-Wasserstein distance provides a very intuitive type of speed limit $\tau\geq \mathcal{W}^2/\Sigma$, with $\tau$ being the time the system takes to change, $\mathcal{W}$ the $L^2$-Wasserstein distance between the initial and final state, and $\Sigma$ the integrated EP. 
However, such a simple bound has not been know for general discrete systems, except for a result restricted to detailed balanced stochastic systems~\cite{van2020unified}, because the discretized $L^2$-Wasserstein distance has been only defined for such systems~\cite{maas2011gradient,erbar2012ricci,erbar2019geometry}. 
In this paper, we show that the local geometrical structure which defines our EPR decomposition reveals the physical meaning of the discrete $L^2$-Wasserstein distance as minimum excess EP, and allows us to generalize it to general nonlinear dynamics in a discrete space. As a result, we obtain a simple speed limit for general nonlinear dynamics in a discrete space, in which the total integrated EP is replaced by the integrated excess EP, tightening the bound.

The paper is organized as follows. 
In Sec.~\ref{sec:preliminary}, we introduce basic concepts of nonlinear dynamics in a discrete space, which we characterize as a graph, and propose a general equation of motion that contains master equations and chemical rate equations as special cases. We also define the edgewise Onsager coefficients and the induced inner product, which determine the aforementioned geometrical structure
of thermodynamic forces. In addition, we review some relevant previous studies. 
In Sec.~\ref{sec:decomp}, we derive the decomposition of the EPR and examine the properties of the excess and housekeeping EPRs, in particular the cyclic decomposition of the housekeeping EPR and a connection between gradient flow and the excess EPR.
In Sec.~\ref{sec:tur}, we apply this decomposition to derive refined versions of several TURs. 
In Sec.~\ref{sec:wasserstein}, we introduce the generalized $L^2$-Wasserstein distance and examine its mathematical and physical features. We relate it to the excess EPR in Sec.~\ref{sec:wasserstein:excess}, and obtain the speed limit in Sec.~\ref{sec:wasserstein:sl}. 
Section~\ref{sec:example} is devoted to the study of two examples that demonstrate how the decomposition of the EPR is obtained and characterize 
the
distinct dynamical modes separately, both analytically and numerically.
The first example is a two level system attached to two heat baths at different temperatures, where the relaxation mode is described by a ``mean'' temperature of the two. 
The second example is the Brusselator model of chemical oscillation, which exhibits a limit cycle. 
The appendixes contain mathematical details of the derivations and results. 

\section{General formalism of dynamics and thermodynamics}
\label{sec:preliminary}
\subsection{Dynamics}
\label{sec:preliminary:dynamics}
\begin{figure}
    \centering
    \includegraphics[width=\linewidth]{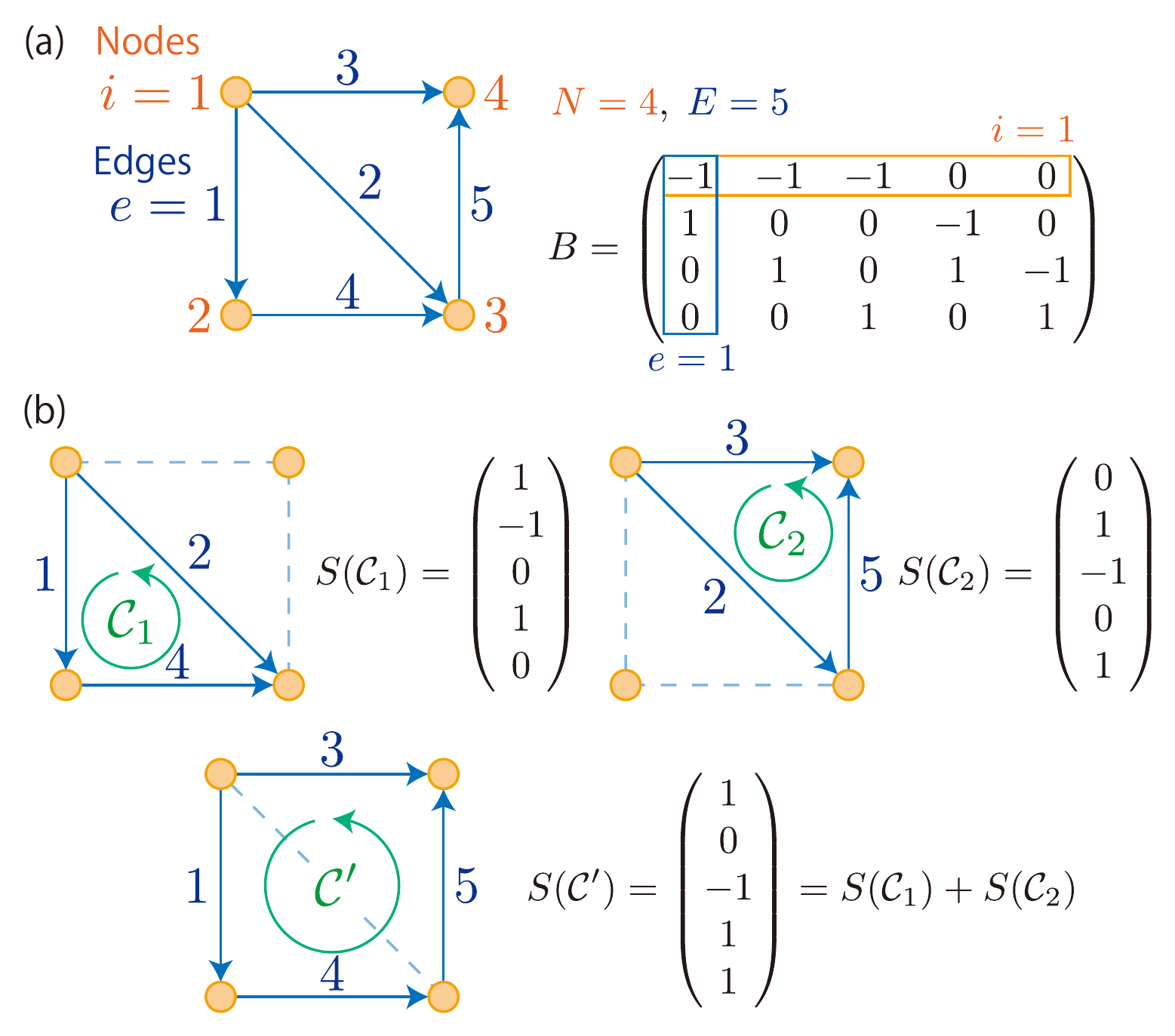}
    \caption{
    \addd{Graph theoretical concepts are shown. (a) An example of a graph. There are $N=4$ nodes and $E=5$ directed edges. Each edge indicates a pair of reversible transitions or reactions. The incidence matrix $B$ encodes the structural information of the graph. (b) In a Markov jump process, where each node is identified with a physical state, we define a cycle as a chain of transitions that leaves the system unchanged. Now there are two independent cycles, $\mathcal{C}_1$ and $\mathcal{C}_2$, whose corresponding vectors $S(\mathcal{C}_1)$ and $S(\mathcal{C}_2)$ span linear space $\ker B$. The other cycles, such as $\mathcal{C}'$, are given by superposing the two cycles. The notion of cycle differs a bit in chemical reaction network as we explain in the next figure. }}
    \label{fig:graph}
\end{figure}

\begin{figure}
    \centering
    \includegraphics[width=\linewidth]{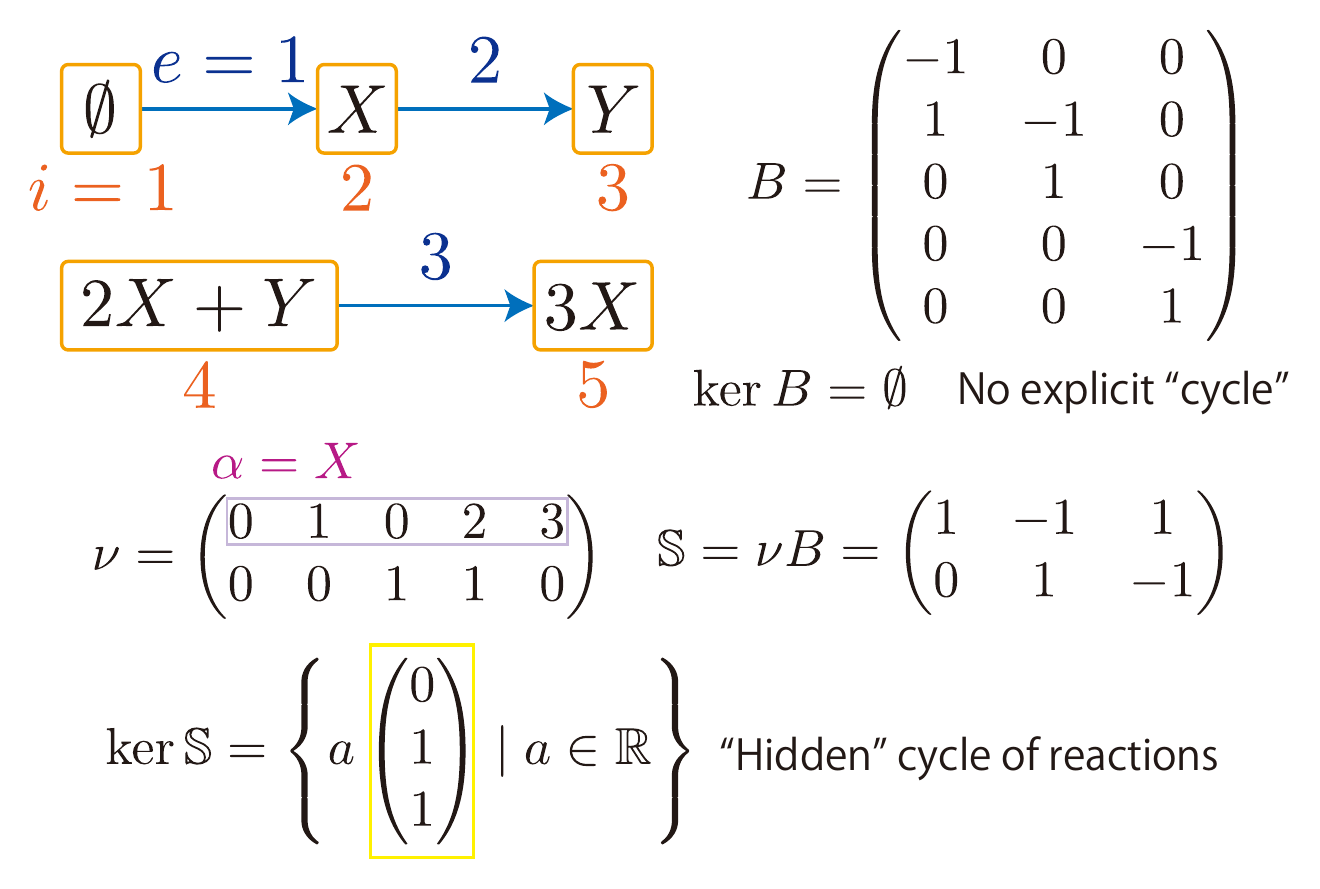}
    \caption{
    \addd{An example of graph theoretic description of a chemical reaction network. Note that although each edge has a specific direction, it represents a set of reversible reactions. The reaction network is the Brusselator, a paradigmatic model of chemical oscillation, which we consider as an example in Section~\ref{sec:example:2}. Unlike Markov jump processes, a node in a chemical reaction network represents a so-called \textit{complex} that contains some chemical species as represented by $\nu$. Complex $\emptyset$ corresponds to external species whose concentrations are kept constant. Here we cannot find a ``cycle'' by just looking at the diagram, or equivalently, the incidence matrix $B$. However, the stoichiometric coefficient matrix $\mathbb{S}$, which governs the dynamics, has a nonempty kernel, which proves that there exists a cycle of reactions that does not change the system. }}
    \label{fig:chemgraph}
\end{figure}

Let us consider general nonlinear Markovian dynamics on a graph whose nodes may include several ``states'' or ``species''. 
It will be shown that the dynamics can lead to master equations and rate equations.
Graph theoretical notions introduced below are summarized in Figs.~\ref{fig:graph} and \ref{fig:chemgraph} through examples of Markov jump process and chemical reaction network respectively. 

We consider a directed graph with nodes $\{1,2,\dots, N\}$ and edges $\{1,2,\dots,E\}$. 
The stoichiometry of state or species $\alpha\in\{1,2,\dots, S\}$ at node $i$ is indicated by $\nu_{\alpha i}\in\mathbb{Z}_{\geq 0}$.
The distribution (concentration vector or statistical state) of the system at time $t$ is indicated by a vector $x(t)=(x_\alpha(t))\in\mathbb{R}_{\geq 0}^S$.

An edge represents a pair of reversible transitions or reactions and connects two nodes. We denote the starting node of edge $e\in\{1,\dots,E\}$ as $\iota(e)$ and its ending node by $\iota'(e)$. The reversed direction of edge $e$ is indicated by $-e$. 
The structure of the graph is expressed by the incidence matrix $B=(B_{ie})$ defined by 
\begin{align}
    B_{ie}=
    \begin{cases}
        1&\text{if $\iota'(e)=i$}\\
        -1&\text{if $\iota(e)=i$}\\
        0&\text{otherwise}. 
    \end{cases}
\end{align}
We write the possibly nonlinear and time-dependent flux of $\iota(e)\to\iota'(e)$ as $\mathcal{K}_e(x)$ and $\iota'(e)\to\iota(e)$ as $\mathcal{K}_{-e}(x)$. 
\addd{We assume the fluxes are always positive $\mathcal{K}_{\pm e}(x)>0$. }
The current $J_e(x)$ on edge $e$ is given by $J_e(x)=\mathcal{K}_e(x)-\mathcal{K}_{-e}(x)$. 
Therefore, the net inflow to node $i$ amounts to $\sum_e B_{ie}J_e(x)$. 
Defining $\mathbb{S}=\nu B$, where $\nu=(\nu_{\alpha i})$, 
we obtain a continuity equation
\begin{align}
    \frac{dx}{dt}=\mathbb{S}J(x(t)). \label{eq:conteq}
\end{align}
Throughout this paper, we consider general Markovian dynamics of a distribution $x$ governed by this equation. It can represent Markov jump processes and deterministic chemical reaction networks, as we show below. 
We sometimes assume that the flux is given by the law of mass action,
\begin{align}
    \mathcal{K}_e(x)=k_e\prod_\alpha x_\alpha^{\nu_{\alpha\iota(e)}},  \label{eq:massactionlaw}
\end{align}
which however is not essential to our general results. 
Below, we explicitly note whenever we make the assumption of mass actions kinetics. 
In addition, when mass action is used, the coefficients $k_e(t)>0$ may depend on time.

We sometimes abbreviate the dependence of some quantity $f(x(t),t)$ on the state $x(t)$ at time $t$ and the explicit dependence on time $t$ as $f(t)$ if there is no ambiguity; we make the dependence implicit in the majority of the paper.

\subsubsection{Example: Markov jump process}

Consider the case where each node $i$ is identified with a state $\alpha$, so $\nu=I$, the identity. 
Then, by writing $x$ as $p$ and assuming the mass action law, Eq.~\eqref{eq:conteq} becomes the master equation
\begin{align}
    \frac{dp_i}{dt} = \sum_e B_{ie}(k_ep_{\iota(e)}-k_{-e}p_{\iota'(e)}), \label{eq:mastereq}
\end{align}
where $p_i$ can be interpreted as the occupation probability of state $i$ if normalized $\sum_i p_i=1$, and $k_{\pm e}$ the transition rate along edge $\pm e$. 
It can be written as
\begin{align}
    \frac{dp_i}{dt} = \sum_{j(\neq i)}(k_{j\to i}p_j-k_{i\to j}p_i)
\end{align}
because if $e=i\to j$, $B_{ie}=-1$ and if $e=j\to i$, $B_{ie}=1$, and only one of the two directions is contained in the summation in Eq.~\eqref{eq:mastereq}. 
We note that the linear form of flux is an example of mass action kinetics since
\begin{align}
    \mathcal{K}_e(p)=k_e\prod_{\alpha}p_\alpha^{\delta_{\alpha\iota(e)}}
    =k_ep_{\iota(e)},
\end{align}
where $\delta$ is Kronecker's delta. 

\subsubsection{Example: Chemical reaction network}
The general equation~\eqref{eq:conteq} can represent an open chemical reaction network by interpreting $x$ as a concentration distribution $c$~\cite{rao2016nonequilibrium}. 
A node is interpreted as an aggregate of chemical species, technically called \textit{complex}, which contains $\nu_{\alpha i}$ molecules of chemical species $\alpha$ and appears on the left- or right-hand side of a chemical equation. 
Because the $(\alpha,e)$-element of $\mathbb{S}=\nu B$ reads 
\begin{align}
    \sum_i \nu_{\alpha i}B_{ie}=\nu_{\alpha\iota'(e)}-\nu_{\alpha\iota(e)},
\end{align}
$\mathbb{S}$ is the conventional stoichiometric matrix. 

When we consider chemical reaction networks, we do not assume the mass action kinetics in general so that we can treat a broad class of non-ideal chemical reaction networks~\cite{avanzini2021nonequilibrium}. 
We put the assumption only when 1) we review previous studies which require it, e.g., gradient flow equations~\cite{mielke2011gradient} and the HS decomposition~\cite{rao2016nonequilibrium,ge2016nonequilibrium}, 2) we illustrate our results numerically where we need to provide a concrete form of $\mathcal{K}$. 

\bigskip
When we discuss a result particular to a Markov jump process, we use the variable as $p$ instead of $x$ to indicate that the variable is a probability distribution; on the other hand, when we focus on a chemical reaction network, we will use $c$ to indicate a concentration distribution. 

\subsection{Thermodynamics}
\label{sec:preliminary:thermodynamics}
Let us define the thermodynamic force on edge $e$ by
\begin{align}
    F_e(x):=\ln\frac{\mathcal{K}_e(x)}{\mathcal{K}_{-e}(x)}. \label{eq:localdb}
\end{align}
We assume the local detailed balance condition, namely that $F_e$ gives the total change of the reduced entropy on the process that edge $e$ represents~\cite{schnakenberg1976network,rao2016nonequilibrium,avanzini2021nonequilibrium}, where ``reduced'' means that it is divided by a physical constant, the Boltzmann constant $k_\mathrm{B}$ for Markov jump processes, or the gas constant $R$ for chemical reaction networks. 
We do not distinguish the entropy and the reduced one by setting the constant we consider to one. 
Then, we obtain the following form of the entropy production rate (EPR):
\begin{align}
    \sigma(x)=\sum_e J_e(x)F_e(x). 
    \label{eq:eprintro}
\end{align}
For simplicity, hereafter we do not write the argument $x$ explicitly. 

In the seminal paper \cite{schnakenberg1976network}, Schnakenberg obtained an expression of the EPR by decomposing the network into cycles. 
He introduced the fundamental set of cycles $\{\mathcal{C}_\mu\}_{\mu=1,\dots,M}$ of the graph defined by states and transitions in a Markov jump process. 
Each cycle corresponds to a vector $S(\mathcal{C}_\mu)=(S_e(\mathcal{C}_\mu))_{e=1,\dots,E}\in\ker B$ and the set $\{S(\mathcal{C}_\mu)\}_{\mu=1,\dots,M}$ forms a basis of $\ker B$, where $\ker$ indicates the kernel of the matrix. 
Such a set of vectors can also be introduced for a chemical reaction network, where it spans $\ker\mathbb{S}$~\cite{polettini2014irreversible}. 
In general, we refer to a basis of $\ker\mathbb{S}$ rooted in the cycles of a graph as a cycle basis. 
A cycle basis is not unique in general; hereafter, we fix some choice of the basis. 
See also Figs.~\ref{fig:graph} and \ref{fig:chemgraph}, where we provide concrete examples of cycles and clarify the slight difference between Markov jump processes and chemical reaction networks. 

When the system is in a steady state, the current $J^\mathrm{st}$ is in the kernel of $\mathbb{S}$, so it can be expanded in the cycle basis as
\begin{align}
    J_e^\mathrm{st}=\sum_\mu S_e(\mathcal{C}_\mu)\mathcal{J}_\mu.
    \label{eq:cyclicflux}
\end{align}
Note that the superscript $^\mathrm{st}$ is used to indicate the value of a quantity in the steady state.
Each expansion coefficient $\mathcal{J}_\mu$ quantifies the current flowing in the cycle $\mathcal{C}_\mu$. 
The conjugated cyclic forces are defined by 
\begin{align}
    \mathcal{F}_\mu:=\sum_e S_e(\mathcal{C}_\mu)F_e^\mathrm{st}.
    \label{eq:cyclicforce}
\end{align}
In terms of the currents and forces, we obtain the expression of the EPR as in Refs.~\cite{schnakenberg1976network,polettini2014irreversible},
\begin{align}
    \sigma^\mathrm{st}=\sum_\mu \mathcal{J}_\mu \mathcal{F}_\mu,
    \label{eq:epcycles}
\end{align}
which shows that the steady-state EPR can be divided into contributions from cycles, whose dissipation can be analyzed separately~\cite{andrieux2007fluctuation,horowitz2014thermodynamics,van2022thermodynamic}. We will generalize this expression out of steady states in Sec.~\ref{sec:decomp:cyclic}. 

\subsection{Edgewise Onsager coefficient, gradient flow, and Wasserstein distance}
\label{sec:preliminary:onsager}
We define the edgewise Onsager coefficient by 
\begin{align}
    \ell_e(x):=
    \frac{\mathcal{K}_e(x)-\mathcal{K}_{-e}(x)}{\ln(\mathcal{K}_e(x)/\mathcal{K}_{-e}(x))}
    =\frac{J_e(x)}{F_e(x)} \label{eq:onscoeff}
\end{align}
and the diagonal matrix $L(x):=\mathrm{diag}(\ell_e(x))$. 
When $\mathcal{K}_e(x)=\mathcal{K}_{-e}(x)$, by continuity we see $\ell_e(x)$ is given by $\mathcal{K}_e(x)$. 
It is also easy to see that $\ell_e(x)>0$ and $L(x)$ is positive definite. 
For simplicity, we make $L$ and $\ell$'s dependence on $x$ implicit, but we stress that we always choose the current distribution $x$ as the argument.
We write $L^{1/2}=\mathrm{diag}(\sqrt{\ell_e})$. 
For notational convenience, we define an inner product $\langle u,v\rangle_L:=u^\transpose Lv$ and the corresponding norm $\|u\|_L:=\sqrt{\langle u,u\rangle_L}$. 
We can use this inner product to express the EPR as
\begin{align}
    \sigma=\|F\|_L^2, \label{eq:epassqnorm}
\end{align}
which will be the starting point of our results~\footnote{
We note that in a withdrawn preprint~\cite{ge2011thermodynamics}, Ge and coauthors also found this characterization of the EPR~\eqref{eq:epassqnorm} as the squared norm of the thermodynamic force with the Onsager coefficient being the metric. They tried to express the HS decomposition as an orthogonal decomposition, but discovered an error in their proof and withdrew their preprint. In Sec.~\ref{sec:decomp:hs}, we discuss that, in general, the HS decomposition is not given geometrically.}.
Note that if we choose any other state like a steady state as the argument of $L(\cdot)$, then we can no longer give the EPR as a squared norm with $L$ being the metric. 

We remark that as a general property of the logarithmic mean, $\ell_e(x)$ is sandwiched between arithmetic mean $(\mathcal{K}_e(x)+\mathcal{K}_{-e}(x))/2$ and geometric mean $\sqrt{\mathcal{K}_e(x)\mathcal{K}_{-e}(x)}$ as
\begin{align}
    \sqrt{\mathcal{K}_e(x)\mathcal{K}_{-e}(x)}
    \leq\ell_e(x)
    \leq \frac{\mathcal{K}_e(x)+\mathcal{K}_{-e}(x)}{2},
    \label{eq:logmeanineq}
\end{align}
where the equalities hold if and only if $\mathcal{K}_e(x)=\mathcal{K}_{-e}(x)$. 

In Ref.~\cite{schnakenberg1976network}, Schnakenberg introduced a symmetric matrix $\mathcal{L}$ and the reciprocal relation 
\begin{align}
    \mathcal{J}_\mu=\sum_{\mu'}\mathcal{L}_{\mu\mu'}\mathcal{F}_{\mu'} \label{eq:schnons}
\end{align}
for the steady-state cycle current and force defined in Eqs.~\eqref{eq:cyclicflux} and~\eqref{eq:cyclicforce}. 
By the definitions of $\mathcal{J}$ and $\mathcal{F}$, we find that
\begin{align}
    \tilde{L}_{ee'}:=\sum_{\mu,\mu'} S_e(\mathcal{C}_\mu)\mathcal{L}_{\mu\mu'}S_{e'}(\mathcal{C}_{\mu'})
\end{align}
connects $J^\mathrm{st}$ and $F^\mathrm{st}$ as $J^\mathrm{st}=\tilde{L}F^\mathrm{st}$. However, $\tilde{L}$ is not diagonal in general and does not coincide with $L^\mathrm{st}$.

Nevertheless, the edgewise Onsager coefficient is useful to rewrite the equation of motion~\eqref{eq:conteq} into a gradient flow~\cite{mielke2011gradient,mielke2013geodesic}, which we review here. 
This is possible when there exists an equilibrium steady state $x^\mathrm{eq}$ that satisfies the detailed balance condition
\begin{align}
    \mathcal{K}_e(x^\mathrm{eq})=\mathcal{K}_{-e}(x^\mathrm{eq}), 
\end{align}
and the flux $\mathcal{K}$ is given by the mass action law~\eqref{eq:massactionlaw}. 
Then the thermodynamic force is given by
\begin{align}
    F=-\mathbb{S}^\transpose \nabla D(x\|x^\mathrm{eq}), \label{eq:gradientforce}
\end{align}
where $\nabla=(\partial/\partial x_\alpha)$ is a column vector of the partial differential operators, and $D(\cdot\|\cdot)$ is the generalized Kullback--Leibler (KL) divergence,
\begin{align}
    D(x\|x'):=\sum_\alpha\pqty{x_\alpha\ln\frac{x_\alpha}{x'_\alpha}-x_\alpha+x'_\alpha},
\end{align}
which reduces to the usual KL divergence in the Markov jump case.
Let us prove Eq.~\eqref{eq:gradientforce}. 
The detailed balance condition leads to
\begin{align}
    \frac{k_e}{k_{-e}}=\frac{\prod_\alpha(x_\alpha^\mathrm{eq})^{\nu_{\alpha\iota'(e)}}}{\prod_\alpha(x_\alpha^\mathrm{eq})^{\nu_{\alpha\iota(e)}}}. 
\end{align}
Substituting this into the definition of the thermodynamic force~\eqref{eq:localdb}, we have
\begin{align}
    F_e
    &=\sum_\alpha (\nu_{\alpha\iota(e)}-\nu_{\alpha\iota'(e)})\ln\frac{x_{\alpha}}{x^\mathrm{eq}_{\alpha}} \notag\\
    &=-[\mathbb{S}^\transpose\Psi]_e
\end{align}
with $\Psi =\ln(x/x^\mathrm{eq})$, where we define the log of a vector as the vector of the logs of the elements. 
The proof ends by calculating the derivative of the KL divergence to show
\begin{align}
    \frac{\partial}{\partial x_\alpha}D(x\|x^\mathrm{eq})
    =\ln\frac{x_\alpha}{x_\alpha^\mathrm{eq}}
    =\Psi_\alpha. 
\end{align}
Because $LF=J$, by substituting Eq.~\eqref{eq:gradientforce} into Eq.~\eqref{eq:conteq}, we obtain the gradient flow expression~\cite{maas2011gradient,mielke2011gradient}
\begin{align}
    \frac{dx}{dt}=-\mathbb{S}L\mathbb{S}^\transpose \nabla D(x\|x^\mathrm{eq}). \label{eq:gradientflow}
\end{align}
Note that the matrix $\mathbb{S}L\mathbb{S}^\transpose $ is positive semidefinite. 
When we consider a Markov jump process, $\mathbb{S}L\mathbb{S}^\transpose $ reads $BLB^\transpose $, which is called the Laplacian matrix of a weighted graph~\cite{godsil2001algebraic}. 

A consequence of the above is that the KL divergence is a Lyapunov function when the system is autonomous, thus $x^\mathrm{eq}$ is time invariant. The divergence monotonically decreases as 
\begin{align}
    \frac{d}{dt}D(x\| x^\mathrm{eq})
    &=\pqty{\frac{dx}{dt}}^\transpose \nabla D(x\|x^\mathrm{eq})\notag\\
    &=-\left\|\mathbb{S}^\transpose \nabla D(x\|x^\mathrm{eq})\right\|_L^2\leq 0 \label{eq:lyapunov}. 
\end{align}
Using Eqs.~\eqref{eq:epassqnorm} and \eqref{eq:gradientforce}, we can confirm that the time derivative of the KL divergence is equal to the negative of the EPR. 

For autonomous Markovian dynamics with a detailed balanced steady state, the gradient flow structure has also been utilized to measure the distance between two states. 
Maas proposed in Ref.~\cite{maas2011gradient} an $L^2$-Wasserstein distance measure 
\begin{align}
    \mathcal{W}(p^{(0)},p^{(1)}):=
    \inf_{p,\psi}\pqty{
  \int_0^1 dt\;\|B^\transpose \psi(t)\|_{L(p(t))}^2
    }^{1/2} \label{eq:wassersteinmarkov}
\end{align}
where the infimum is taken under the conditions
\begin{align}
    \frac{dp}{dt}=BL(p)B^\transpose \psi,\;
    p(t=0)=p^{(0)},\; 
    p(t=1)=p^{(1)}. \label{eq:ourcondmarkov}
\end{align} 
We can change the time interval from the unit one into an arbitrary one $[0,\mathcal{T}]$ as
\begin{align}
    \mathcal{W}(p^{(0)},p^{(1)})
    =\inf_{p,\psi}\pqty{
    \mathcal{T}\int_0^\mathcal{T}
    dt\,\|B^\transpose \psi(t)\|_{L(p(t))}^2
    }^{1/2}, \label{eq:wassersteinmarkov2}
\end{align}
with the last condition of Eq.~\eqref{eq:ourcondmarkov} replaced with $p(\mathcal{T})=p^{(1)}$
~\footnote{
Let $\{p^*(t)\}_{t\in[0,1]}$ and $\{\psi^*(t)\}_{t\in[0,1]}$ give the Wasserstein distance $\mathcal{W}(p^{(0)},p^{(1)})$. Then for any $\mathcal{T}>0$, $p'(t'):=p^*(t'/\mathcal{T})$ and $\psi'(t'):=\psi^*(t'/\mathcal{T})/\mathcal{T}$ satisfy $d_{t'}p'(t')=BL(p'(t'))B^\transpose\psi'(t')$ and provides $\mathcal{W}(p^{(0)},p^{(1)})/\mathcal{T}$ by integrating $\int_0^\mathcal{T}dt'\|B^\transpose\psi'(t')\|_{L(p'(t'))}^2$. This discussion is valid for more general settings like Eqs.~\eqref{eq:generalwasserstein0} and~\eqref{eq:generalwasserstein2}
}. 
Note that the distance is a dynamics-wise measure because $L$ depends not only on the current state but also on the given kinetics as in Eq.~\eqref{eq:onscoeff}, or more explicitly as 
\begin{align*}
    \ell_e(p)=\frac{k_ep_{\iota(e)}-k_{-e}p_{\iota'(e)}}{\ln(k_ep_{\iota(e)}/ k_{-e}p_{\iota'(e)})}.
\end{align*}
Our definition looks unlike the original one~\cite{maas2011gradient}, but the equivalence is easily proved (see Appendix~\ref{app:equivalence} for the proof). 

Vu and Hasegawa~\cite{van2021geometrical} found a speed limit by using the Wasserstein distance as
\begin{align}
    \tau\geq \frac{\mathcal{W}(p(0),p(\tau))^2}{\int_0^\tau dt\;\sigma}, \label{eq:speedlimitdbmarkov}
\end{align}
where $\sigma$ is the EPR of the solution of a master equation $\{p(t)\}_{t\in[0,\tau]}$. 
The speed limit is obtained because the solution of a master equation satisfies the first condition in Eq.~\eqref{eq:ourcondmarkov} as
\begin{align}
    \frac{dp}{dt}=BL(p)B^\transpose \Psi
\end{align}
with $\Psi_i = -(\partial/\partial p_i)D(p\|p^\mathrm{eq})$. 
Since Eq.~\eqref{eq:wassersteinmarkov2} involves minimization with respect to $p$ and $\psi$, the Wasserstein distance is smaller than the corresponding expression evaluated for the original dynamics. By choosing the time interval of the solution $\tau$ as the time interval in Eq.~\eqref{eq:wassersteinmarkov2}, we obtain the lower bound.

\section{Onsager-projective decomposition of EPR}
\label{sec:decomp}
\begin{figure}
    \centering
    \includegraphics[width=\linewidth]{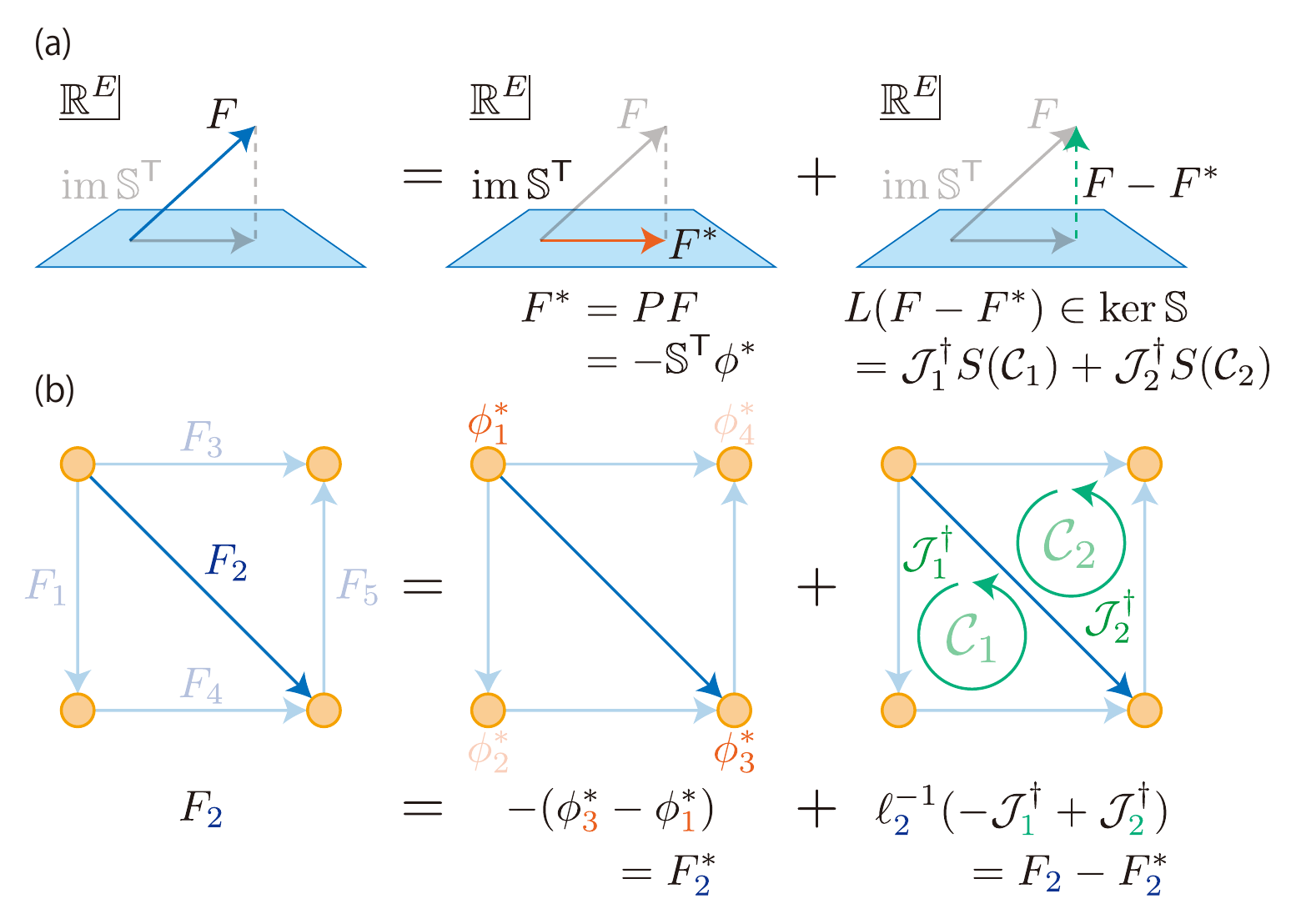}
    \caption{\addd{Schematic diagrams of the decompostion of force, which is central to the decomposition of entropy production rate~\eqref{eq:decompdef}. 
    (a) Geometrical illustration of decomposing force $F$. Projector $P$ onto the space of conservative forces $\im\mathbb{S}$ is given in Eq.~\eqref{eq:projection}. The projected force $F^*$ is orthogonal to the remainder $F-F^*$ under the inner product $\langle \cdot,\cdot\rangle_L$ introduced in Sec.~\ref{sec:preliminary:onsager}. 
    The remaining current $J^\mathrm{cyc}=L(F-F^*)$ belongs to $\ker\mathbb{S}^\transpose$, therefore it is cyclic and can be expanded in a cyclic basis as in Eq.~\eqref{eq:cycliccurrent}. 
    (b) The decomposition of force is exemplified on edge $e=2$ in the system considered in Fig.~\ref{fig:graph}. 
    The conservative force $F^*$ is given by the difference of $\phi^*$ which is assigned a value on each node. 
    On the other hand, the remaining force $F-F^*$ is associated with the coefficients of the linear combination of cycles.}}
    \label{fig:forcedecomp}
\end{figure}

We state our main result, a decomposition of the EPR in Markov jump processes and chemical reaction networks, which can be seen as the generalization of the Maes--Neto\v{c}n\'{y} decomposition to such systems~\cite{maes2014nonequilibrium}. Because we define the decomposition in a geometrical way by using a projection operator and the edgewise Onsager coefficient~\eqref{eq:onscoeff}, we term it \textit{the Onsager-projective decomposition}. 
\addd{The basic idea of the technical details discussed in the following two paragraphs is summarized in Fig.~\ref{fig:forcedecomp}. }

Let us consider a matrix $P$ that is the projection onto the image of $\mathbb{S}^\transpose $ and orthogonal with respect to the matrix $L$. Formally, this means that for any $f,g\in\mathbb{R}^E$, $P$ satisfies $\langle Pf,(I-P)g\rangle_L=0$, where $I$ is the identity. 
This projection matrix is given by~\cite{puntanen2011matrix}
\begin{align}
    P=\mathbb{S}^\transpose (\mathbb{S}L\mathbb{S}^\transpose)^-\mathbb{S}L, \label{eq:projection}
\end{align}
with $A^-$ denoting the generalized inverse of a matrix $A$.
\addd{
Since $\mathbb{S}$ does not have full rank in general, we need to consider a generalized inverse. 
Nevertheless, we may set it to a theoretically tractable one, like the Moore--Penrose inverse, because $P$ does not depend on the choice of the generalized inverse~\cite{puntanen2011matrix}. 
To be more concrete, let the eigendecomposition of $\mathbb{S}L\mathbb{S}^\transpose$ be $U\Lambda U^{-1}$ with orthogonal matrix $U$ and diagonal matrix $\Lambda=\mathrm{diag}(\lambda_1,\dots,\lambda_r,0\dots,0)$ where $\lambda_\alpha\neq 0$ and $r=\rank\mathbb{S}$. Then, the Moore--Penrose inverse of $\mathbb{S}L\mathbb{S}^\transpose$ can be computed as $U\Lambda^+U^{-1}$ with $\Lambda^+=\mathrm{diag}(\lambda_1^{-1},\dots,\lambda_r^{-1},0\dots,0)$, which is the Moore--Penrose inverse of $\Lambda$. } 

For any $f\in\mathbb{R}^E$, $Pf$ is also characterized by 
\begin{align}
    Pf=\argmin_{g\in \im\mathbb{S}^\transpose }\|f-g\|_L^2. \label{eq:pminimize}
\end{align}
Because $Pf\in\im\mathbb{S}^\transpose $, there exists $\phi(f)\in\mathbb{R}^S$ that satisfies $-\mathbb{S}^\transpose \phi(f)=Pf$. 
The solution $\phi(f)$ is not unique in general because the transformation $\phi(f)\to \phi(f)+c$, where $c$ is a left null vector of $\mathbb{S}$, does not change $-\mathbb{S}^\transpose \phi(f)$. 
If $\mathbb{S}$ is the incidence matrix $B$ of a connected graph, $c$ is given by a vector of ones times any scalar. When we consider chemical reaction networks, such a null vector is called a conservation law~\cite{rao2016nonequilibrium}. 
Nonetheless, $Pf$ is unique, so we can consider $\phi(f)$ as fixed. 

Now, we can define the conservative force that plays a central role in the Onsager-projective decomposition. 
Define $F^*:=PF$ and $\phi^*=\phi(F)$ such that $F^*=-\mathbb{S}^\transpose \phi^*$. 
The force $F^*$ can be seen as a conservative force that reproduces the original (possibly nonconservative) dynamics because $\mathbb{S}LF^*=\mathbb{S}LF=\mathbb{S}J=d_tx$. 
This can be shown as follows. First, we have a decomposition
\begin{align}
    \mathbb{S}LF
    &=\mathbb{S}L(P+I-P)F\notag\\
    &=\mathbb{S}LF^*+\mathbb{S}L(I-P)F.
\end{align}
The second term $\mathbb{S}L(I-P)$ is zero because $P$ is the orthogonal projection onto $\im \mathbb{S}^\transpose $ with respect to the metric $L$, i.e., for any $\varphi\in\mathbb{R}^S$ and $f\in\mathbb{R}^E$, $P$ satisfies $\langle\mathbb{S}^\transpose \varphi,(I-P)f\rangle_L=\varphi^\transpose \mathbb{S}L(I-P)f=0$. 
Because $\varphi$ is now arbitrary, by setting $f=F$, we obtain $\mathbb{S}L(I-P)F=0$. Thus $d_tx =\mathbb{S}LF^*$ holds.

Using the conservative force $F^*$, we find a decomposition of the EPR that can be regarded as a generalization of the Maes--Neto\v{c}n\'{y} decomposition in Langevin systems to Markov jump processes and chemical reaction networks~\cite{maes2014nonequilibrium}. 
We define the \textit{Onsager excess} and \textit{housekeeping EPR} by
\begin{align}
    \sigma^\mathrm{ex}:=\|F^*\|_L^2,\quad
    \sigma^\mathrm{hk}:=\|F-F^*\|_L^2
    \label{eq:decompdef}
\end{align}
respectively.
These terms give a nonnegative decomposition of the EPR,
\begin{gather}
    \sigma=\sigma^\mathrm{ex}+\sigma^\mathrm{hk}, \label{eq:mndecomp}\\
    \sigma^\mathrm{ex},\sigma^\mathrm{hk}\geq 0
\end{gather}
for both Markov jump processes and CRNs. 
They are always nonnegative because $L$ is positive definite.
From the orthogonality relation
\begin{align}
    \langle F^*,F-F^*\rangle_L
    =\langle PF,(I-P)F\rangle_L=0,
\end{align}
the decomposition is proved as follows
\begin{align}
    \sigma
    &=\langle F^*+(F-F^*),F^*+(F-F^*)\rangle_L\notag\\
    &=\langle F^*,F^*\rangle_L
    +\langle F-F^*,F-F^*\rangle_L.
\end{align}
The two orthogonal components $F^*$ and $F-F^*$ of the thermodynamic force allow us to interpret the decomposition in a geometrical way, see Figs.~\ref{fig:concept} and \ref{fig:forcedecomp}. 
This is similar to the interpretation in the Langevin case~\cite{dechant2022geometric1,dechant2022geometric2}.

In a steady state, where $\mathbb{S}J^\mathrm{st}=0$ holds, the excess EPR vanishes and $\sigma=\sigma^\mathrm{hk}$ because 
\begin{align}
    \sigma^\mathrm{ex}
    =\langle F^*,F^*\rangle_{L^\mathrm{st}}
    =-(\phi^*)^\transpose \mathbb{S}L^\mathrm{st}F^*
    =-(\phi^*)^\transpose \mathbb{S}J^\mathrm{st}=0. 
\end{align}
\addd{On the other hand, the housekeeping EPR is identically zero if and only if the steady state is detailed balanced~\footnote{
\addd{
It is shown under the assumption that the dynamics obeys the law of mass action by seeing the equivalence between the following two conditions. One is that the force is conservative, which is equivalent to $\sigma^\mathrm{hk}=0$. The other is the so-called generalized Wegscheider condition $\ln(k^+/k^-)\in\im\mathbb{S}^\transpose$,
which is a necessary and sufficient condition for the steady state to be detailed balanced~\cite{schuster1989generalization}.}
}.}
Therefore, the decomposition~\eqref{eq:mndecomp} separates the EPR into a transient contribution $\sigma^\mathrm{ex}$ and a contribution $\sigma^\mathrm{hk}$ required to sustain a nonequilibrium steady state if a steady state exists. 

\addd{
Given the eigendecomposition $\mathbb{S}L\mathbb{S}^\transpose=U\Lambda U^{-1}$ discussed earlier, we can write down the excess and housekeeping EPR as 
\begin{equation}
    \begin{aligned}
        \sigma^\mathrm{ex}&=\sum_{\alpha=1}^r\lambda_\alpha^{-1}|[U^{-1}\mathbb{S}LF]_\alpha|^2,\\
        \sigma^\mathrm{hk}&=\sigma-\sum_{\alpha=1}^r\lambda_\alpha^{-1}|[U^{-1}\mathbb{S}LF]_\alpha|^2.
    \end{aligned}
\end{equation}}

\addd{
In what follows in this section, we discuss several properties and consequences of the decomposition, which we summarize in Table~\ref{tab:decomp} with corresponding equation numbers. 
Except for the operational implications discussed in Sec.~\ref{sec:decomp:minimumep} using the variational formulas from  Sec.~\ref{sec:decomp:others}, these results are basically independent from each other. 
}

\begin{table}[h]
    \centering
    \begin{tabular}{|c|c|}
        \hline
        \begin{tabular}{c}
            Generalization of \\[-3pt]
            Schnakenberg decomposition
        \end{tabular} & Eq.~\eqref{eq:hkcyclic} \\\hline
        \begin{tabular}{c}
            Generalization of\\[-3pt]
            MN decomposition 
        \end{tabular}& Eq.~\eqref{eq:mepp} \\\hline
        Variational formulas & Eqs.~\eqref{eq:mnexsup}--\eqref{eq:mnhksup} \\\hline
        Operational implications & Eq.~\eqref{eq:minent}\\\hline
        Gradient flow & Eq.~\eqref{eq:relaxation}\\\hline
    \end{tabular}
    \caption{
    \addd{Summary of consequences of the Onsager-projective decomposition discussed in Sec.~\ref{sec:decomp}. }}
    \label{tab:decomp}
\end{table}

\subsection{Generalization of Schnakenberg's cyclic decomposition}
\label{sec:decomp:cyclic}

Our decomposition leads to a generalization of Schnakenberg's cyclic decomposition of the steady-state EPR~\eqref{eq:epcycles} to non-steady states. 
Consider the current corresponding to $F-F^*$, that is $J^\mathrm{cyc}:=L(F-F^*)$. 
This current is in the kernel of $\mathbb{S}$ because $\mathbb{S}LF=\mathbb{S}LF^*$, therefore it is a cyclic current and can be expanded in the cycle basis as
\begin{align}
    J_e^\mathrm{cyc}=\sum_\mu S_e(\mathcal{C}_\mu)\mathcal{J}_\mu^\dagger. \label{eq:cycliccurrent}
\end{align}
This expansion is available whether the system is in a steady state or not. 
The conjugated force is also given by
\begin{align}
    \mathcal{F}_\mu^\dagger
    :=\sum_e S_e(\mathcal{C}_\mu)(F_e-F_e^*)
    =\sum_e S_e(\mathcal{C}_\mu)F_e,
\end{align}
where we used $\sum_e S_e(\mathcal{C}_\mu)F_e^*=-(\phi^*)^\transpose \mathbb{S}S(\mathcal{C}_\mu)=0$. 
Thus, we have generalized the decomposition in Eq.~\eqref{eq:cyclicforce}, which applies only in a steady state, to arbitrary states.
Importantly, these cycle quantities allow us to write the Onsager housekeeping EPR as
\begin{align}
    \sigma^\mathrm{hk}=\sum_\mu\mathcal{J}_\mu^\dagger\mathcal{F}_\mu^\dagger \label{eq:hkcyclic}
\end{align}
because
\begin{align}
    \sum_\mu\mathcal{J}_\mu^\dagger\mathcal{F}_\mu^\dagger
    &=\sum_\mu\mathcal{J}_\mu^\dagger\sum_{e}S_e(\mathcal{C}_\mu)(F_e-F_e^*) \notag\\
    &=\sum_eJ_e^\mathrm{cyc}(F_e-F_e^*)
    =\|F-F^*\|_L^2. 
\end{align}
Therefore, our decomposition generalizes Schnakenberg's cyclic decomposing of the steady-state EPR~\cite{schnakenberg1976network} to a cyclic decomposition of the housekeeping EPR in non-steady-state systems.

\subsection{Generalization of Maes--Neto\v{c}n\'{y} decomposition}
\label{sec:decomp:mndecomp}

Next, we show that the decomposition can also be understood as a generalization of the MN decomposition~\cite{maes2014nonequilibrium}. 
In their paper, Maes and Neto\v{c}n\'{y} regarded the EPR as a functional of the potential landscape, and minimized it by changing the landscape. 
The result is the following minimum EP principle: the potential landscape minimizing the EPR is the one for which the instantaneous distribution is a steady state. 
In our case, using the minimizing property of the projection operator~\eqref{eq:pminimize}, we have the expression
\begin{align}
    \sigma^\mathrm{hk}=\inf_{\psi}\|F-(-\mathbb{S}^\transpose \psi)\|_L^2. \label{eq:mepp}
\end{align}
In a Markov jump process, $F$ is written as $F=-B^\transpose (u+\ln p)+F^\mathrm{nc}$, where $u$ is some potential landscape and $F^\mathrm{nc}$ is a nonconservative driving force. 
We define $J(v):=-LB^\transpose (v+\ln p)+LF^\mathrm{nc}$ as a function of the potential landscape. 
Using these expressions, we find a formulation resembling Ref.~\cite{maes2014nonequilibrium},
\begin{align}
    \sigma^\mathrm{hk}=\inf_{v}J(v)^\transpose L^{-1}J(v). 
\end{align}
Minimization leads to the condition 
\begin{align}
    BJ(v)=0.
    \label{eq:mnlikecond}
\end{align}
Thus, among all potential landscapes with Onsager matrix $L$, the entropy is minimized for the one in which the instantaneous distribution $p$ is a steady state. 
This is the analogue of Maes and Neto\v{c}n\'{y}'s minimum EP principle~\cite{maes2014nonequilibrium}. 
We remark that fixing $L$ is connected to fixing the diffusion matrix in the Langevin case (see Appendix~\ref{app:onsdiff}). 

\subsection{Variational formulas}
\label{sec:decomp:others}
We can also extend other results derived in the Langevin case~\cite{dechant2022geometric1,dechant2022geometric2} to discrete dynamics.
First, the following variational expressions hold: 
\begin{align}
    \sigma^\mathrm{ex}
    &=\sup_{\psi}\frac{\pqty{\ev{\mathbb{S}^\transpose \psi,F}_L}^2}{\ev{\mathbb{S}^\transpose \psi,\mathbb{S}^\transpose \psi}_L}, \label{eq:mnexsup}\\
    &=\inf_{F'\in\ker \mathbb{S}L}\ev{F-F',F-F'}_L, \label{eq:mnexinf}\\
    \sigma^\mathrm{hk}
    &=\sup_{F'\in\ker \mathbb{S}L}\frac{\pqty{\ev{F',F}_L}^2}{\ev{F',F'}_L}. \label{eq:mnhksup}
\end{align}
Note that $\mathbb{S}L$ is the adjoint of $\mathbb{S}^\transpose$ with respect to the inner product $\ev{\cdot,\cdot}_L$.
As an important example, $F-F^*$ is in its kernel because $\mathbb{S}LF=\mathbb{S}LF^*$. 
This space can be interpreted as the space of forces that do not contribute to the dynamics if we regard $LF'$ as a probability current for any force $F'$.

The first expression~\eqref{eq:mnexsup} is obtained by combining the Cauchy--Schwarz inequality
\begin{align}
    \pqty{\ev{\mathbb{S}^\transpose \psi,F^*}_L}^2\leq\ev{\mathbb{S}^\transpose \psi,\mathbb{S}^\transpose \psi}_L\ev{F^*,F^*}_L,
    \label{eq:cs1}
\end{align}
and the following equality, 
\begin{align}
    \ev{\mathbb{S}^\transpose \psi,F^*}_L
    =\psi^\transpose \mathbb{S}LF^*
    =\psi^\transpose \mathbb{S}LF
    =\ev{\mathbb{S}^\transpose \psi,F}_L.
\end{align}
Equality is obtained in~\eqref{eq:cs1} when $\mathbb{S}^\transpose \psi$ is parallel to $F^*$. 
The infimum representation~\eqref{eq:mnexinf} for the excess EPR is proven by the following calculation: When $F'\in\ker\mathbb{S}L$, 
\begin{align}
    &\ev{F-F',F-F'}_L \notag\\
    &=\ev{F^*+F-F^*-F',F^*+F-F^*-F'}_L\notag\\
    &=\ev{F^*,F^*}_L+\ev{F-F^*-F',F-F^*-F'}_L,
    \label{eq:vareqpf}
\end{align}
because 
\begin{align}
    \ev{F^*,F-F^*-F'}_L
    =(\phi^*)^\transpose\mathbb{S}L(F-F^*-F')=0. 
\end{align}
The second term in the third line of Eq.~\eqref{eq:vareqpf} is nonnegative and vanishes when $F'=F-F^*\in\ker \mathbb{S}L$; thus the minimum value gives the excess EPR. 

The supremum formula~\eqref{eq:mnhksup} of the housekeeping EPR can be shown in a parallel way to the excess case. 
For $F'\in\ker\mathbb{S}L$, we have $(\ev{F',F-F^*}_L)^2\leq \ev{F',F'}_L\ev{F-F^*,F-F^*}_L$ and $\ev{F',F-F^*}_L=\ev{F',F}_L$.
The latter equality is because 
\begin{align}
    \ev{F',F^*}_L=-(\phi^*)^\transpose\mathbb{S}LF'=0.
\end{align}
Thus, the inequality
\begin{align}
    \sigma^\mathrm{hk}
    \geq\frac{\pqty{\ev{F',F}_L}^2}{\ev{F',F'}_L}
\end{align}
is obtained for any $F'\in\ker\mathbb{S}L$. 
The equality is attained when $F'$ is proportional to $F-F^*$, which also belongs to $\ker\mathbb{S}L$.

\subsection{Minimum entropy production}
\label{sec:decomp:minimumep}

The variational formula~\eqref{eq:mnexinf} allows us to interpret the excess EPR as a minimum EPR. In this section, we focus on Markov jump processes.
In general, when we control rate constants to minimize the EPR with the dynamics $d_tp$ unchanged, the EPR can be made arbitrarily small unless we put further assumptions~\cite{dechant2022minimum}. 
To consider minimum EPR under additional constraints, recent studies have proposed to fix the geometrical mean of the transition rates $\sqrt{k_ek_{-e}}$ on each edge~\cite{remlein2021optimality}, or to fix the backward rates $k_{-e}$~\cite{ilker2022shortcuts}. 
They found that such constraints lead to nonconservative optimal forces. 
On the other hand, the excess EPR can be interpreted as the minimum EPR when we change transition rates while keeping the value of $L(p)$ fixed, since Eq.~\eqref{eq:mnexinf} can be rewritten as
\begin{align}
    \sigma^\mathrm{ex}
    &=\inf_{F'|d_tp=BLF'}
    \langle F',F'\rangle_L\notag\\
    &=\inf_{k'|d_tp=BLF_{k'},L_{k'}=L}
    \langle F_{k'},F_{k'}\rangle_L\notag\\
    &=\inf_{k'|d_tp=BLF_{k'},L_{k'}=L}
    \sigma_{k'}, \label{eq:minent}
\end{align}
where $F_{k'}$, $L_{k'}$ and $\sigma_{k'}$ are the force, the Onsager matrix, and the EPR given by transition rates $k'$ instead of the actual dynamics $k$, which provides $L$.
The definition of the excess EPR~\eqref{eq:decompdef} also shows that we can obtain a conservative optimal force in this minimization. 
Quite recently, a similar study~\cite{vanvu2022thermodynamic} discussed minimization of integrated EP by fixing the time average of the sum of the edgewise coefficients, not the instantaneous coefficients.
We summarize these minimizations in Table~\ref{tab:minepr}.

Although the minimization is mathematically well defined, fixing the Onsager coefficients is not always interpretable from a physical point of view. Nonetheless, we can find a concrete meaning of this constraint in the continuum limit. In this limit, the edgewise Onsager coefficient becomes $\ell_e(p)\approx D_{e}(\iota(e))p_{\iota(e)}/(\Delta x)^2$ with $\Delta x$ being the distance between nearest sites going to zero, and $D_e(i)$ a kind of ``diffusion coefficient’’~\footnote{This relation was proved for one-dimensional systems in \cite{vanvu2022thermodynamic}. In Appendix~\ref{app:onsdiff}, we generalized it to multidimensional cases.}. Therefore, fixing $L(p)$ is equivalent to assuming that the diffusion coefficients are constant, which is a natural assumption when we consider Langevin dynamics. For more detailed discussion on the asymptotic relation, see Appendix~\ref{app:onsdiff}.

\begin{table}[h]
    \centering
    \begin{tabular}{c|c|c|c|c|}
        & Ours 
        & Remlein~\cite{remlein2021optimality} 
        & Ilker~\cite{ilker2022shortcuts} 
        & Vu~\cite{vanvu2022thermodynamic} 
        \\\hline
        Constraint 
        & $\ell_e(p)$
        & $\sqrt{k_ek_{-e}}$ 
        & $k_{-e}$ 
        & $\tau^{-1}\int_0^\tau \ell_e(p)dt$ 
        \\\hline
        Force & C & NC & NC & $?$ \\\hline
    \end{tabular}
    \caption{Comparison of dynamical constraints and optimal thermodynamic forces that minimize the EPR (or integrated EP). Here, NC and C stand for nonconservative and conservative. Vu~\cite{vanvu2022thermodynamic} did not mention whether the optimal force is conservative or not. }
    \label{tab:minepr}
\end{table}

\subsection{Gradient flow}
\label{sec:decomp:gradientflow}

We now generalize Eqs.~\eqref{eq:gradientflow} and \eqref{eq:lyapunov} using the Onsager-projective decomposition.
Let $\phi^*(t)$ be the potential that gives $F^*(t)=-\mathbb{S}^\transpose \phi^*(t)$ at time $t$, and define a pseudo-canonical distribution $x^\mathrm{pcan}(t)$ by 
\begin{align}
    x_\alpha^\mathrm{pcan}(t)
    =\frac{1}{Z_\alpha(t)}x_\alpha(t)e^{-\phi^*_\alpha(t)}. \label{eq:pcan}
\end{align}
In a Markov jump process, $Z_\alpha(t)=\sum_i p_i(t)e^{-\phi_i^*(t)}$ is a normalization constant that does not depend on $\alpha$. In a chemical reaction network, $Z(t)$ is any vector whose log becomes a conservation law, i.e., $\ln Z(t)=(\ln Z_\alpha(t))\in\ker\mathbb{S}^\transpose $ holds. If there is no conservation law, $Z(t)$ will be the vector of ones. 
Then, we have the relation
\begin{align}
    \sigma^\mathrm{ex}=\eval{-\frac{d}{dt}D(x(t)\|x^\mathrm{pcan}(s))}_{s=t}. \label{eq:relaxation}
\end{align}
This result follows because 
\begin{align}
    \eval{\frac{d}{dt}D(x(t)\|x^\mathrm{pcan}(s))}_{s=t}
    =\sum_\alpha\frac{dx_\alpha}{dt}(\phi_\alpha^*+\ln Z_\alpha)\notag\\
    =J^\transpose \mathbb{S}^\transpose \phi^*=-\langle F,F^*\rangle_L=-\langle F^*,F^*\rangle_L,
\end{align}
where the second equality follows from $\sum_iB_{ie}=0$ for Markov jump processes and from $\mathbb{S}^\transpose \ln Z=0$ for chemical reaction networks. 
In particular, in a Markov jump process, by the decomposition $\phi^*=\epsilon^*+\ln p$, the pseudo-canonical distribution reads
\begin{align}
    p_i^\mathrm{pcan}=\frac{1}{Z}e^{-\epsilon_i^*}. \label{eq:canonicalform}
\end{align}
If $F^*=F$, the pseudo-energy level $\epsilon^*$ becomes the actual energy level of each state. 
The expression~\eqref{eq:canonicalform} is the reason why we refer to $x^\mathrm{pcan}$ as the pseudo-canonical distribution. 
Although this pseudo-canonical distribution depends on the instantaneous state, 
it indicates the momentary ``goal'' of the dynamics, even if the system is attracted to a limit cycle. 

Along with the pseudo-canonical distribution $x^\mathrm{pcan}$, we can formally obtain a gradient flow representation of the continuity equation~\eqref{eq:conteq} even without detailed balance because 
\begin{align}
    \mathbb{S}^\transpose
    \nabla D(x\|x^\mathrm{pcan})
    &=\mathbb{S}^\transpose\ln \frac{x}{x^\mathrm{pcan}}\notag\\
    &=\mathbb{S}^\transpose(\phi^*+\ln Z)\notag\\
    &=-F^*, 
\end{align}
so that 
\begin{align}
    -\mathbb{S}L\mathbb{S}^\transpose
    \nabla D(x\|x^\mathrm{pcan})
    =\mathbb{S}LF^*=\frac{dx}{dt}. 
\end{align}
This is a generalization of the gradient flow~\eqref{eq:gradientflow}. We remark that, even if the rates entering the dynamics via \eqref{eq:massactionlaw} are time-independent, $x^\mathrm{pcan}$ generally depends on time, see Eq.~\eqref{eq:pcan}.

\subsection{Brief comparison with Hatano--Sasa decomposition}
\label{sec:decomp:hs}
There are other ways to decompose the EPR into excess and housekeeping contributions. 
One well-known approach is the Hatano--Sasa (HS) decomposition~\cite{hatano2001steady}, also called the adiabatic--nonadiabatic decomposition~\cite{esposito2010three}, which is given by
\begin{align}
    \sigma&=\sigma_\mathrm{HS}^\mathrm{ex}+\sigma_\mathrm{HS}^\mathrm{hk},\\
    \sigma_\mathrm{HS}^\mathrm{ex}
    &=J^\transpose(F-F^\mathrm{st}),\\
    \sigma_\mathrm{HS}^\mathrm{hk}
    &=J^\transpose F^\mathrm{st}.
\end{align}

If the flux is written in mass action form, the difference between the forces becomes
\begin{align}
    F_e-F_e^\mathrm{st}
    &=\sum_\alpha (\nu_{\alpha\iota(e)}-\nu_{\alpha\iota'(e)})\ln\frac{x_{\alpha}}{x^\mathrm{st}_{\alpha}}\notag\\
    &=-[\mathbb{S}^\transpose \Psi]_e \label{eq:ffpsi}
\end{align}
with $\Psi_\alpha = \ln(x_\alpha/x_\alpha^\mathrm{st})$. Then, we obtain another expression for the HS excess EPR
\begin{align}
    \sigma_\mathrm{HS}^\mathrm{ex}
    &=-J^\transpose\mathbb{S}^\transpose \Psi,\\
    \Psi&=\ln\frac{x}{x^\mathrm{st}}, \label{eq:HS-potential}
\end{align}
which is similar to the expression $\sigma^\mathrm{ex}=-J^\transpose\mathbb{S}^\transpose\psi^*$ that follows from the definition~\eqref{eq:decompdef} and $-\mathbb{S}L\mathbb{S}^\transpose\phi^*=\mathbb{S}J$. 
The nonnegativity of the HS excess and housekeeping EPR can be proven only when the steady state satisfies $BJ^\mathrm{st}=0$. 
For a Markov-jump process, this condition is precisely the steady-state condition.
For a chemical reaction network, however, this condition is called \textit{complex balance} condition and is stricter than the steady state condition $\mathbb{S} J^\mathrm{st} =\nu BJ^\mathrm{st}=0$~\cite{ge2016nonequilibrium,rao2016nonequilibrium}.
On the other hand, the Onsager excess EPR is defined even in the absence of stable steady state, which we will show in Sec.~\ref{sec:example:1} through the Brusselator model that exhibits a limit cycle. For more detailed discussion about the relation between our approach and the HS decomposition, see Appendix~\ref{app:HS}.

\section{Thermodynamic uncertainty relations}
\label{sec:tur}

\begin{table}[h]
    \centering
    \begin{tabular}{c|c|c|}
         & Short time & Finite time \\\hline
         Excess & \begin{tabular}{c}
              Eq.~\eqref{eq:MNTUR2} \\
              For CRN, Eq.~\eqref{eq:chemTUR}
         \end{tabular} & Eq.~\eqref{eq:finiteturex}\\\hline
         Housekeeping & Eq.~\eqref{eq:hkshorttur} & Eq.~\eqref{eq:hktur}\\\hline
    \end{tabular}
    \caption{\addd{Summary of TURs we derive in section~\ref{sec:tur}. Unless noted, the result is obtained only for Markov jump processes. Here, CRN stands for chemical reaction network. }}
    \label{tab:tur}
\end{table}

\subsection{Short-time TUR: Markov jump process}
\label{sec:tur:shortmarkov}
The variational expression~\eqref{eq:mnexsup} allows us to obtain a set of refined short-time TURs.
In this section, we focus on Markov jump processes. 
We define the edgewise dynamical activity (traffic) $\chi_e$ by~\cite{maes2008canonical}
\begin{align}
    \chi_e:=k_ep_{\iota(e)}+k_{-e}p_{\iota'(e)}. 
\end{align}
From the inequality in Eq.~\eqref{eq:logmeanineq}, we have 
\begin{align}
    \ell_e=\frac{J_e}{F_e}\leq \frac{1}{2}\chi_e. \label{eq:logmeanineq2}
\end{align}
The equality holds if and only if $p$ satisfies $k_ep_{\iota(e)}=k_{-e}p_{\iota'(e)}$, i.e., in the equilibrium state.

The numerator of the right-hand side in Eq.~\eqref{eq:mnexsup}, $\left\langle\mathbb{S}^\transpose \psi,F\right\rangle_L$, which reads $\langle B^\transpose \psi,F\rangle_L$ now, is the time derivative of the expectation value $\langle\psi\rangle=\sum_i p_i \psi_i$ because
\begin{align}
    \frac{d}{dt}\langle{\psi}\rangle
    =\psi^\transpose BJ
    =\psi^\transpose BLF
    =\langle B^\transpose \psi,F\rangle_L. 
\end{align}
Note that here we assume that $\psi$ does not have any time-dependence.
Thus, from the variational formula~\eqref{eq:mnexsup}, we have a TUR
\begin{align}
    \frac{(d_t\langle \psi\rangle)^2}{\langle B^\transpose \psi,B^\transpose \psi\rangle_L}\leq \sigma^\mathrm{ex}
    \label{eq:MNTUR1}
\end{align}
for any observable $\psi$. 
Therefore, as long as we know the edgewise Onsager matrix $L$, we can estimate the excess EPR by measuring the rate of change of the average $\langle\psi\rangle$.

Because of the inequality in Eq.~\eqref{eq:logmeanineq2}, the denominator is bounded from above by the short-time limit of the variance as
\begin{align}
    \langle B^\transpose \psi,B^\transpose \psi\rangle_L\leq \lim_{\tau\to 0}\frac{\mathrm{Var}(\Delta_\tau \psi)}{2\tau}=:\frac{1}{2}D_\psi, 
\end{align}
where $\Delta_\tau \psi=\psi_{i(t+\tau)}-\psi_{i(t)}$ and the variance is asymptotically given by~\cite{otsubo2020estimating}
\begin{align}
    \mathrm{Var}\,(\Delta_\tau \psi)
    =\tau\sum_{e}\chi_e ([B^\transpose \psi]_e)^2+o(\tau). 
\end{align}
This inequality shows a weaker bound
\begin{align}
    \frac{2(d_t\langle\psi\rangle)^2}{D_\psi}\leq \sigma^\mathrm{ex}. \label{eq:MNTUR2}
\end{align}
Although equality can only be achieved in equilibrium, where we trivially have $\sigma^\mathrm{ex}=0$, the inequality can be tight near equilibrium, where the difference between the arithmetic and logarithmic mean in Eq.~\eqref{eq:logmeanineq2} is small.
Moreover, Eq.~\eqref{eq:MNTUR2} does not require information about the Onsager matrix $L$. Therefore, is much more tractable than Eq.~\eqref{eq:MNTUR1} and gives an estimate of the excess EPR that depends only on the rate of change of the average and the short-time fluctuations of some observable.
This inequality partly generalizes the short-time TUR derived in Ref.~\cite{otsubo2020estimating} for the case where currents are expressed as changes of observables as above. 

\addd{
While the interpretation of the inequalities derived from Eq.~\eqref{eq:mnexsup} as TURs was straightforward, 
we need to develop a trajectory level description to understand the meaning of the inequality suggested by  Eq.~\eqref{eq:mnhksup}. 
This is done later when we discuss finite-time TURs. 
}

\subsection{Short-time TUR: Chemical Reaction Network}
\label{sec:tur:shortcrn}
Next, we consider chemical reaction networks. 
Because of the inequality between logarithmic mean and arithmetic mean~\eqref{eq:logmeanineq}, we have
\begin{align}
    \langle \mathbb{S}^\transpose \psi,\mathbb{S}^\transpose \psi\rangle_L
    &=\sum_e ([\mathbb{S}^\transpose \psi]_e)^2\frac{J_e}{F_e}\notag\\
    &\leq \frac{1}{2}\sum_e ([\mathbb{S}^\transpose \psi]_e)^2(\mathcal{K}_e(c)+\mathcal{K}_{-e}(c)). \label{eq:pdp}
\end{align}
The scaled diffusion coefficient of a chemical reaction network is defined by~\cite{ge2016nonequilibrium,yoshimura2021thermodynamic}
\begin{align}
    \tilde{D}_{\alpha\beta}
    :=\frac{1}{2}\sum_e \mathbb{S}_{\alpha e}\mathbb{S}_{\beta e}(\mathcal{K}_e(c)+\mathcal{K}_{-e}(c)). 
\end{align}
Using this quantity, the right-hand side of Eq.~\eqref{eq:pdp} becomes
\begin{align}
    \frac{1}{2}\sum_e ([\mathbb{S}^\transpose \psi]_e)^2(\mathcal{K}_e(c)+\mathcal{K}_{-e}(c))
    =\psi^\transpose \tilde{D}\psi.
\end{align}
Moreover, because
\begin{align}
    \langle \mathbb{S}^\transpose \psi,F\rangle_L
    =\psi^\transpose \mathbb{S}J
    =\frac{d}{dt}(\psi^\transpose c),
\end{align}
we obtain a TUR for a chemical reaction network
\begin{align}
    \frac{1}{\psi^\transpose \tilde{D}\psi}
    \bqty{\frac{d}{dt}(\psi^\transpose c)}^2
    \leq \sigma^\mathrm{ex}. \label{eq:chemTUR}
\end{align}
This inequality generalizes the previous result in Ref.~\cite{yoshimura2021thermodynamic}: because $\sigma^\mathrm{ex}\leq \sigma$,
\begin{align}
    \max_\alpha\frac{1}{\tilde{D}_{\alpha\alpha}}\pqty{\frac{dc_\alpha}{dt}}^2\leq \sigma,
\end{align}
follows by setting $\psi_{\alpha'}=\delta_{\alpha'\alpha}$.

\subsection{Finite-time TURs}
\label{sec:tur:finitemarkov}
The above results relate the excess EPR to the short-time fluctuations of observables. 
In the following, we obtain finite-time TURs for Markov jump process. 
Proofs of these relations are found in Appendix~\ref{app:TUR}. Note that these proofs rely on the expression for the path probability of the system, so their generalization to chemical reaction networks is challenging and we leave it for future work.

In general, we consider a stochastic current observable $\mathcal{J}_w$ with weight $w_e(t)$,
\begin{align}
    \mathcal{J}_w
    =\sum_{k}w_{e_k}(t_k)
\end{align}
for a stochastic trajectory with jumps $e_k$ at time $t_k$ ($k=1,2,\dots$). We assume the weight satisfies $w_e=-w_{-e}$. 
The average during a time interval $[0,\tau]$ is given by 
\begin{align}
    \langle \mathcal{J}_{w}\rangle_\tau=\int_0^\tau dt\;\sum_e w_e(t) J_e(t) \label{eq:currentintegral}
\end{align}
We introduce a decomposition of the current observable into the excess and the housekeeping contribution~\cite{dechant2022geometric2} as
\begin{align}
    \langle \mathcal{J}_{w}^\mathrm{ex}\rangle_\tau
    &:=\int_0^\tau dt\;\sum_e w_e(t) J_e^*(t)\\
    \langle \mathcal{J}_{w}^\mathrm{hk}\rangle_\tau
    &:=\int_0^\tau dt\;\sum_e w_e(t) (J_e(t)-J_e^*(t)),
\end{align}
where we defined $J_e^*(t):=\ell_e(t)F_e^*(t)$. As shown above, these currents induce the same time evolution as $J$, since $BJ^*=BLF^*=BJ$. 

By using the housekeeping component of a current observable, we can show the TUR for the Onsager housekeeping EP
\begin{align}
    \frac{(\langle \mathcal{J}_{w}^\mathrm{hk}\rangle_\tau)^2}{\mathrm{Var}(\mathcal{J}_{w})}\leq\frac{1}{2}\int_0^\tau dt\;\sigma^\mathrm{hk}, \label{eq:hktur}
\end{align}
with variance $\mathrm{Var}(\mathcal{J}_{w}):=\langle (\mathcal{J}_{w}-\langle\mathcal{J}_{w}\rangle_\tau)^2\rangle_\tau$.
Unlike the short-time TURs discussed above, this relation is valid for any finite time duration. 
It states that the conventional form of the steady-state TUR remains valid if we consider the respective housekeeping contributions for both the current and the entropy production.

\addd{
Now, we can derive a short-time TUR for the housekeeping EPR.
Since the time interval $[0,\tau]$ can be replaced with $[t,t+\Delta t]$, by taking limit $\Delta t\to 0$, we obtain the short-time TUR
\begin{align}
    \frac{2(d_t\langle\mathcal{J}_w^\mathrm{hk}\rangle_t)^2}{\mathcal{D}_w}\leq\sigma^\mathrm{hk} \label{eq:hkshorttur}
\end{align}
with
\begin{align}
    \mathcal{D}_w=\sum_e \chi_ew_e^2=\lim_{\Delta t\to 0}\frac{\mathrm{Var}(\mathcal{J}_w)}{\Delta t}
\end{align}
because 
\begin{align}
    d_t&\langle\mathcal{J}_w^\mathrm{hk}\rangle_t
    =\sum_e w_e(t)(J_e(t)-J_e^*(t))\notag\\
    &=\lim_{\Delta t\to 0}\frac{1}{\Delta t}\int_t^{t+\Delta t}dt'
    \sum_e w_e(t')(J_e(t')-J_e^*(t')). 
\end{align}
When we assume the weight satisfies $w=(w_e)_{e=1,\dots,E}\in\ker{\mathbb{S}L}$, we see that $w^\transpose J^*= w^\mathsf{T}LF^*=0$ because $F^*\in\im\mathbb{S}^\transpose$. Thus, we find another TUR
\begin{align}
    \frac{2(d_t\langle\mathcal{J}_w\rangle_t)^2}{\mathcal{D}_w}\leq\sigma^\mathrm{hk}
\end{align}
for $w\in\ker\mathbb{S}L$. Because $d_t\langle\mathcal{J}_w\rangle_t=w^\transpose J=\langle w,F\rangle_L$, this is derived from Eq.~\eqref{eq:mnhksup} in the same way as we did in Sec.~\ref{sec:tur:shortmarkov}. 
What is important here is that, by introducing a current observable, we can interpret the inequality we get as a TUR.
}

If we change the time interval as in Refs.~\cite{koyuk2020thermodynamic,dechant2022geometric2}, we can also derive a finite-time TUR for the Onsager excess EP. We assume that the transition rates $k_e(t)$ depend on $t$ and $\tau$ in the form of $t/\tau$ only.
For example, with $k_0$ constant, $k_e(t)=k_0\exp(t/\tau)$ is acceptable but $k_e(t)=k_0\exp(t)$ and $k_e(t)=\tau^{-1}\exp(t/\tau)$ are not. 
Physically, this assumption means that any external operation on the system is accelerated when $\tau$ becomes smaller, and vice versa. 
We put the same assumption on the time dependence of the weight $w$. 
Under these assumptions, we can prove the finite-time TUR
\begin{align}
    \frac{(\tau\partial_\tau \langle \mathcal{J}_{w}\rangle_\tau-\langle\mathcal{J}_{w}^\mathrm{hk}\rangle_\tau)^2}{\mathrm{Var}(\mathcal{J}_{w})}\leq \frac{1}{2}\int_0^\tau dt\;\sigma^\mathrm{ex}, \label{eq:finiteturex}
\end{align}
where $\partial_\tau$ is taken while holding fixed the normalized time $t/\tau$ that defines $k_e$ (and enters into the definition of $\langle \mathcal{J}_w\rangle_\tau$, Eq.~\eqref{eq:currentintegral}, via $J_e$). 

When the weight is given by a scalar time-dependent observable $\psi(t/\tau)$ as $w=B^\transpose \psi$, $\mathcal{J}_{w}$ becomes the change $\Delta_\tau\psi=\psi(1)-\psi(0)$ and we have the relations
\begin{align}
    \langle\mathcal{J}_{w}\rangle_\tau
    &=\int_0^\tau dt\;\psi^\transpose BJ
    =\int_0^\tau dt\;\frac{d}{dt}\langle\psi\rangle\notag\\
    &=\langle\psi(1)\rangle_{t=\tau}-\langle\psi(0)\rangle_{t=0}\\
    \langle\mathcal{J}_{w}^\mathrm{hk}\rangle_\tau
    &=\int_0^\tau dt\;\psi^\transpose B(J-J^*)=0. 
\end{align}
Then, equation~\eqref{eq:finiteturex} reads
\begin{align}
    \frac{(\tau d_\tau\langle\psi(1)\rangle_{t=\tau})^2}{\mathrm{Var}(\Delta_\tau\psi)}\leq \frac{1}{2}\int_0^\tau dt\;\sigma^\mathrm{ex}. 
\end{align}
This can be understood as a refinement of the TUR for time-dependent driving derived in Ref.~\cite{koyuk2020thermodynamic}. In particular, when one considers the change of scalar observables, rather than currents, the resulting TUR can only estimate the excess part of the entropy production.
By taking the limit $\tau\to 0$, we recover the short-time TUR~\eqref{eq:MNTUR2}.

Since the proofs of the finite-time TURs are lengthy and technically similar to existing results, we give them in Appendix~\ref{app:TUR}, only explaining their underlying idea briefly at this point. 
The structures of the proofs are essentially the same. 
The first step is to parameterize the rate constant and compare the two path probabilities for two close parameters. 
The difference between the two path probabilities is measured by a KL divergence which bounds from above the ratio of two quantities: (1) the difference between the averages of the current observable given by the different parameters and (2) the variance of the current observable. This inequality is obtained from the fluctuation-response inequality~\cite{dechant2020fluctuation}. 
The difference between the observables can be shown to be equal to $\langle\mathcal{J}_{w}^\mathrm{hk}\rangle$ or $\tau\partial_\tau\langle\mathcal{J}_{w}\rangle_\tau-\langle\mathcal{J}_{w}^\mathrm{hk}\rangle$ depending on the way the rates are parameterized. 
At the same time, the KL divergence bounds from below the Onsager excess or housekeeping EP. Combining the two inequalities, we obtain the uncertainty relations. 

\section{Wasserstein geometry}
\label{sec:wasserstein}
\begin{table}[h]
    \centering
    \begin{tabular}{|c|c|}
        \hline
        Generalization of distance & Eq.~\eqref{eq:generalwasserstein} \\\hline
        Constant speed property & Eq.~\eqref{eq:constantspeed} \\\hline
        Duality formula & Eq.~\eqref{eq:dual} \\\hline
        Connection between distance and excess EPR & Eq.~\eqref{eq:wandex} \\\hline
        Speed limit with excess EP & Eq.~\eqref{eq:speedlimit4} \\\hline
    \end{tabular}
    \caption{
    \addd{Summary of main results we obtain in section~\ref{sec:wasserstein}. Here ``distance'' means Wasserstein distance. The first three are mathematically novel, while the last two reveals physical relations between the Wasserstein geometry and the excess EP(R).}}
    \label{tab:wasserstein}
\end{table}
\subsection{Generalization of the Wasserstein distance}
\label{sec:wasserstein:generalization}
In our final set of main theoretical results, we provide a generalization of the Wasserstein distance measure~\eqref{eq:wassersteinmarkov} and the speed limit~\eqref{eq:speedlimitdbmarkov}. In this section, we assume that the kinetics $\mathcal{K}$ do not depend on time explicitly. 
% If it does, we cannot prove the triangle inequality and constant speed property of a geodesic in a conventional way. 

As we reviewed in Sec.~\ref{sec:preliminary:onsager}, the Wasserstein distance has been defined for Markov jump processes with detailed balanced steady states. 
Now, we find a general definition applicable to both Markov jump processes and chemical reaction networks:
\begin{align}
    \mathcal{W}(x^{(0)},x^{(1)})
    :=\inf_{x,f}\pqty{\int_0^1dt\;\|f(t)\|_{L(x(t))}^2}^{1/2}
    \label{eq:generalwasserstein0}
\end{align}
under the following conditions
\begin{align}
    \frac{dx}{dt}=\mathbb{S}
    L(x)f,\quad
    x(t=0)=x^{(0)},\;\; x(t=1)=x^{(1)}. 
    \label{eq:genwasscondition0}
\end{align}
As in Eq.~\eqref{eq:wassersteinmarkov2}, the unit time interval can be replaced with an arbitrary one $[0,\mathcal{T}]$ and we have
\begin{align}
    \mathcal{W}(x^{(0)},x^{(1)})
    =\inf_{x,f}\pqty{\mathcal{T}\int_0^\mathcal{T} dt\;\|f(t)\|_{L(x(t))}^2}^{1/2}
    \label{eq:generalwasserstein2}
\end{align}
for any $\mathcal{T}>0$, with the last condition of Eq.~\eqref{eq:genwasscondition0} replaced with $x(\mathcal{T})=x^{(1)}$. 
Because $L(x)=\mathrm{diag}(\ell_e(x))$ is defined by 
\begin{align}
    \ell_e(x)=\frac{\mathcal{K}_e(x)-\mathcal{K}_{-e}(x)}{\ln(\mathcal{K}_e(x)/\mathcal{K}_{-e}(x))},
\end{align}
$\mathcal{W}$ is defined in reference to some kinetics $\mathcal{K}$.
Here, $x(t)$ and $f(t)$ are taken as independent variables in the minimization which are related by $d_tx = \mathbb{S}L(x)f$; in particular, $f(t)$ is not necessarily equal to the original thermodynamic force $F(x(t))$. 
If there is no pair of $x$ and $f$ that satisfies the continuity equation in Eq.~\eqref{eq:genwasscondition0} for a given $(x^{(0)},x^{(1)})$, we may let $\mathcal{W}=\infty$. However, we do not consider such a situation, because if $(x^{(0)},x^{(1)})$ are realized as the initial and final state of the dynamics \eqref{eq:conteq}, then $x(t)$ and $F(x(t))$ are a solution.

We can show that the optimal force that provides the value of $\mathcal{W}$ is conservative. That is, we have the formula
\begin{align}
    \mathcal{W}(x^{(0)},x^{(1)})
    =\inf_{x,\psi}\pqty{\mathcal{T}\int_0^\mathcal{T} dt\;\|\mathbb{S}^\transpose \psi(t)\|_{L(x(t))}^2}^{1/2} \label{eq:generalwasserstein}
\end{align}
with the conditions
\begin{align}
    \frac{dx}{dt}=\mathbb{S}
    L(x)
    \mathbb{S}^{\transpose }
    \psi,\quad
    x(t=0)=x^{(0)},\quad x(t=\mathcal{T})=x^{(1)}. \label{eq:genwasscondition}
\end{align}
We give the proof of this formula in Appendix~\ref{app:wasserstein}. 

The mathematical features of the generalized Wasserstein distance are expected to be similar to those of Maas's distance~\cite{maas2011gradient}. 
In this paper, we demonstrate two important properties of the Wasserstein distance, namely the constant speed property of a geodesic~\cite{erbar2012ricci} and the Kantorovich duality~\cite{erbar2019geometry}. 

We first explain that a geodesic with respect to the Wasserstein distance has constant speed. 
In Ref.~\cite{erbar2012ricci}, the authors proved that there exists a geodesic $p^*$ accompanied by potential $\psi^*$ with respect to their original definition of the Wasserstein distance~\eqref{eq:wassersteinmarkov}. 
This geodesic has constant speed, meaning that $\|B^\transpose\psi^*(t)\|_{L(p^*(t))}=\mathrm{const}$. As we show, the proof is also valid if we assume there is a geodesic $x^*$ with potential $\psi^*$ for the generalized Wasserstein distance~\eqref{eq:generalwasserstein}. Because of the Cauchy--Schwarz inequality, for an arbitrary $\{x,\psi\}$, we have
\begin{align}
    \mathcal{T}\int_0^\mathcal{T} dt\;\|\mathbb{S}^\transpose\psi(t)\|_{L(x(t))}^2
    \geq
    \left(\int_0^\mathcal{T} dt\;\|\mathbb{S}^\transpose\psi(t)\|_{L(x(t))}\right)^2. \label{eq:wasscsineq}
\end{align}
Moreover, using a reparametrization discussed in Ref.~\cite{dolbeault2009new},
we can also prove the opposite inequality 
\begin{align}
    \mathcal{T}\int_0^\mathcal{T} dt\;\|\mathbb{S}^\transpose\psi^*(t)\|_{L(x^*(t))}^2
    \leq
    \left(\int_0^\mathcal{T}dt
    \;\|\mathbb{S}^\transpose\psi^*(t)\|_{L(x^*(t))}\right)^2. \label{eq:wasscsineqopposite}
\end{align}
for the geodesic pair $\{x^*,\psi^*\}$. 
We provide the detail of this technique in Appendix~\ref{app:reparametrization}. 
Because the equality of the Cauchy--Schwarz inequality~\eqref{eq:wasscsineq} is attained only if $\|\mathbb{S}^\transpose\psi(t)\|_{L(x(t))}$ is constant, the geodesic has the constant speed
\begin{align}
    \|\mathbb{S}^\transpose\psi^*(t)\|_{L(x^*(t))}=\frac{\mathcal{W}(x^{(0)},x^{(1)})}{\mathcal{T}}. \label{eq:constantspeed}
\end{align}

Next, for a Markov jump process, we prove the Kantorovich duality representation~\cite{villani2009optimal} of the Wasserstein distance as in Ref.~\cite{erbar2019geometry}: 
\begin{align}
    \frac{1}{2\mathcal{T}}\mathcal{W}(p^{(0)},p^{(1)})^2=\sup_{\psi}\pqty{\langle\psi(\mathcal{T})\rangle_1-\langle\psi(0)\rangle_0}, \label{eq:dual}
\end{align}
where $\psi(t)\in\mathbb{R}^S$ $(t\in[0,\mathcal{T}])$ has to satisfy
\begin{align}
    q^\transpose d_t\psi(t)+\frac{1}{2}\|B^\transpose\psi(t)\|_{L(q)}^2\leq 0 \label{eq:hjsubeq}
\end{align}
for any probability distribution $q$, and $\langle\cdot\rangle_{i}$ ($i=0,1$) is the expectation value under $p^{(i)}$. 
The maximizer $\psi^{**}$ of Eq.~\eqref{eq:dual} and the minimizer $\{p^*,\psi^{*}\}$ of Eq.~\eqref{eq:generalwasserstein} provide the same conservative force, $B^\transpose\psi^*=B^\transpose\psi^{**}$. 
Furthermore, it can be shown that $\psi^{**}$ and $p^{*}$ attain the equality in Eq.~\eqref{eq:hjsubeq} as
\begin{align}
    (p^*(t))^\transpose d_t\psi^{**}(t)+\frac{1}{2}\|B^\transpose\psi^{**}(t)\|_{L(p^*(t))}^2=0, \label{eq:hjeq}
\end{align}
which can be seen as a generalization of the Hamilton--Jacobi equation. 
While the duality representation characterizes the Wasserstein distance as the maximum potential difference without considering the time evolution of the distribution, 
we can use the Hamilton--Jacobi equation with the differential equation in Eq.~\eqref{eq:genwasscondition} to obtain the optimal solution. 
We provide a proof of the duality formula~\eqref{eq:dual} in Appendix~\ref{app:dual}; there, we assume the existence of an optimal time evolution of $p$ and $\psi$, that is, a geodesic. A proof that is more rigorous but restricted to systems with detailed balanced steady states can be found in Ref.~\cite{erbar2019geometry}. 

\subsection{Wasserstein distance and Onsager excess EPR}
\label{sec:wasserstein:excess}

Here we show that, analogously to the continuous case~\cite{aurell2012refined,dechant2019thermodynamic,nakazato2021geometrical,dechant2022geometric1,dechant2022geometric2}, the Wasserstein distance is connected with the excess entropy production. 
To this end, we employ the definition given in Eqs.~\eqref{eq:generalwasserstein2} and \eqref{eq:genwasscondition0}. 

Let $\{\hat{x}(t)\}_{t\in[0,\tau]}$ be a solution of the continuity equation~\eqref{eq:conteq}. 
Note that $\tau$ indicates the physical time interval we focus on while $\mathcal{T}$ is an arbitrary time parameter. 
If we set $x^{(0)}=\hat{x}(t)$, $x^{(1)}=\hat{x}(t+dt)$, and $\mathcal{T} =dt$ in Eq.~\eqref{eq:generalwasserstein2}, condition~\eqref{eq:genwasscondition0} reads
\begin{equation}
    \begin{gathered}
        x(dt)-x(0)=\mathbb{S}L(\hat{x}(t))f(0) dt+o(dt),\\
        x(0)=\hat{x}(t),\quad x(dt)=\hat{x}(t+dt).
    \end{gathered}
\end{equation}
Hence for an infinitesimal time interval $dt$, the Wasserstein distance can be expanded as
\begin{align}
    &\mathcal{W}(\hat{x}(t),\hat{x}(t+dt))^2\notag\\
    &=dt^2\inf_{f|d_t\hat{x}(t)
    =\mathbb{S}L(\hat{x}(t))f}\|f\|_{L(\hat{x}(t))}^2+o(dt^2).
    \label{eq:wexpansion}
\end{align}
The condition of minimization can be translated into $\{f\mid F(\hat{x}(t))-f\in\ker\mathbb{S}L\}$, hence the leading term is identified with the excess EPR $\sigma^\mathrm{ex}$ from Eq.~\eqref{eq:mnexinf}. 
We finally obtain the relation
\begin{align}
    \pqty{\lim_{dt\to0}\frac{\mathcal{W}(\hat{x}(t),\hat{x}(t+dt))}{dt}}^2=\sigma^\mathrm{ex}. \label{eq:wandex}
\end{align}
Intuitively, in the definition of the Wasserstein distance Eq.~\eqref{eq:generalwasserstein2}, we optimize over the thermodynamic forces and the corresponding evolution of the state connecting the initial and final state.
By contrast, in computing the excess EPR, we restrict the evolution of the state to the one of the original dynamics and only optimize over the forces.
In the short-time limit of a single time-step, there is no evolution of the state, so both problems are equivalent and we can identify the Wasserstein distance and the excess EPR with each other.

\subsection{Speed limit}
\label{sec:wasserstein:sl}
We define $\dot{l}$ and $l$ by
\begin{align}
    \dot{l}(t):=\lim_{dt\to0}\frac{\mathcal{W}(\hat{x}(t),\hat{x}(t+dt))}{dt},\quad l:=\int_0^\tau dt\;\dot{l}(t).
\end{align}
While $\mathcal{W}(\hat{x}(0),\hat{x}(\tau))$ gives the shortest length between the initial state $\hat{x}(0)$ and the final state $\hat{x}(t)$, $l$ quantifies the length of the path $\{\hat{x}(t)\}_{t\in[0,\tau]}$ in terms of the Wasserstein metric. 
Using $l$, we have the speed limit, as in Ref.~\cite{nakazato2021geometrical},
\begin{align}
    l^2\leq \tau \int_0^\tau dt\;\sigma^\mathrm{ex}. \label{eq:speedlimit1}
\end{align}
This is because of the Cauchy--Schwarz inequality $l^2=(\int_0^\tau dt\;\dot{l})^2\leq (\int_0^\tau dt)(\int_0^\tau dt\;\dot{l}^2)$ and the equality~\eqref{eq:wandex}.

By the constant speed property, the following triangle inequality is proven for any $t_1, t_2, t_3\in[0,\tau]$: 
\begin{align}
    \mathcal{W}(\hat{x}(t_1),\hat{x}(t_3))
    \leq \mathcal{W}(\hat{x}(t_1),\hat{x}(t_2))
    +\mathcal{W}(\hat{x}(t_2),\hat{x}(t_3)). \label{eq:trineq}
\end{align}
This relation leads to an inequality between the two distance measures as
\begin{align}
    \mathcal{W}(\hat{x}(0),\hat{x}(\tau))
    \leq \int_0^\tau dt\;\dot{l}(t)=l. \label{eq:wandl}
\end{align}
The equality holds only when the distribution changes at the constant speed
\begin{align}
    \dot{l}(t)=\frac{\mathcal{W}(\hat{x}(0),\hat{x}(\tau))}{\tau}.
\end{align}
By combining Eqs.~\eqref{eq:speedlimit1} and~\eqref{eq:wandl}, we obtain the speed limit
\begin{align}
    \mathcal{W}(\hat{x}(0),\hat{x}(\tau))^2
    \leq \tau \int_0^\tau dt\;\sigma^\mathrm{ex}.
    \label{eq:speedlimit4}
\end{align}
This generalizes the Vu and Hasegawa speed limit~\eqref{eq:speedlimitdbmarkov}, which was introduced in Ref.~\cite{van2021geometrical} for systems with detailed balanced steady states, to general Markovian dynamics on a graph which may have no stable steady state. 

\subsection{Relation to $L^1$-Wasserstein distance}
\label{sec:wasserstein:l1l2}
Another distance that has been considered in the theory of optimal transport is the $L^1$-Wasserstein distance~\cite{villani2009optimal,dechant2022minimum}. 
This distance is defined by
\begin{align}
    \mathcal{W}_1(p^{(0)},p^{(1)}):=\inf_\Pi \sum_{i,j}d_{ij}\Pi_{ij},
\end{align}
where the infimum is taken over the couplings between the probability distributions $p^{(0)}$ and $p^{(1)}$, that is, 
\begin{align}
    \Pi_{ij}\geq 0, \quad \sum_{j}\Pi_{ij}=p_i^{(0)}, \quad \sum_{i}\Pi_{ij}=p_j^{(1)},
\end{align}
and $d_{ij}$ is a weight that satisfies the axioms of distance: $d_{ij}=d_{ji}\geq 0$, $d_{ij}=0$ if and only if $i=j$, and $d_{ij}\leq d_{ik}+d_{kj}$. 
The $(i,j)$-element of a coupling represents how much of the mass on $i$ is moved to $j$ and $d$ evaluates the cost of the transportation. 
Here we set $d_{ij}=1$ if $i$ and $j$ are directly connected, and assume that the graph of states is connected. 

In Ref.~\cite{dechant2022minimum}, one of us derived the Kantorovich duality for this $L^1$-Wasserstein distance 
\begin{align}
    \mathcal{W}_1(p^{(0)},p^{(1)})
    =\sum_i \xi_i(p_i^{(1)}-p_i^{(0)}), \label{eq:l1kantrovich}
\end{align}
where $\xi$ is a potential that satisfies $\xi_j-\xi_i=d_{ij}=1$ if and only if the optimal coupling $\Pi^*$ is positive on $(i,j)$, $\Pi_{ij}^*>0$. Using this formula, we can prove the following relation between the $L^2$- and $L^1$-Wasserstein distance,
\begin{align}
    \frac{\bar{\mathcal{A}}[p^*]}{2}\mathcal{W}_2(p^{(0)},p^{(1)})^2\geq \mathcal{W}_1(p^{(0)},p^{(1)})^2, \label{eq:l1l2ineq}
\end{align}
where we write the $L^2$-Wasserstein distance as $\mathcal{W}_2$ and $\bar{\mathcal{A}}[p^*]$ is the time-averaged dynamical activity of the minimizer $p^*$ of the $L^2$-Wasserstein distance
\begin{align}
    \bar{\mathcal{A}}[p^*]
    =\frac{1}{\tau}\int_0^\tau dt\;\sum_e(k_ep_{\iota(e)}^*+k_{-e}p_{\iota'(e)}^*). 
\end{align}
This is similar to the inequality $W_2(p_0,p_\tau)\geq W_1(p_0,p_\tau)$ for the $L^2$- and $L^1$- Wasserstein distances of continuous probability distributions~\cite{villani2009optimal}, apart from the prefactor that represents how frequently transitions occur in the optimal path on average. 
We provide the proof in Appendix~\ref{app:l1l2ineq}. 
We remark that in the discrete case, the $L^2$-distance defined above is a dynamics-wise measure and depends on the concrete values of the rates.
By contrast, the $L^1$-distance depends only on the connectivity of the state network.
This differs from the continuous case, where both distances are defined in terms of the Euclidean distance.

\section{Examples}
\label{sec:example}
In this section, we discuss two examples. 
One clearly shows how the Onsager-projective decomposition works and separates the two aspects of dynamics into excess and housekeeping terms. It is illustrates that the short-time TUR~\eqref{eq:MNTUR2} improves the bound given in Ref~\cite{otsubo2020estimating}.
The other demonstrates that the decomposition can also be obtained for a nonlinear chemical reaction network that exhibits a limit cycle. We numerically show that in such a system, the conventional HS decomposition fails. 
On the other hand, we do not discuss other results such as the generalization of finite-time TURs and the Wasserstein speed limit because they have been well studied in~\cite{liu2020thermodynamic,koyuk2020thermodynamic} and~\cite{vanvu2022thermodynamic}, respectively. 

\subsection{Two-level system attached to two reservoirs}
\label{sec:example:1}
\begin{figure}
    \centering
    \includegraphics[width=\linewidth]{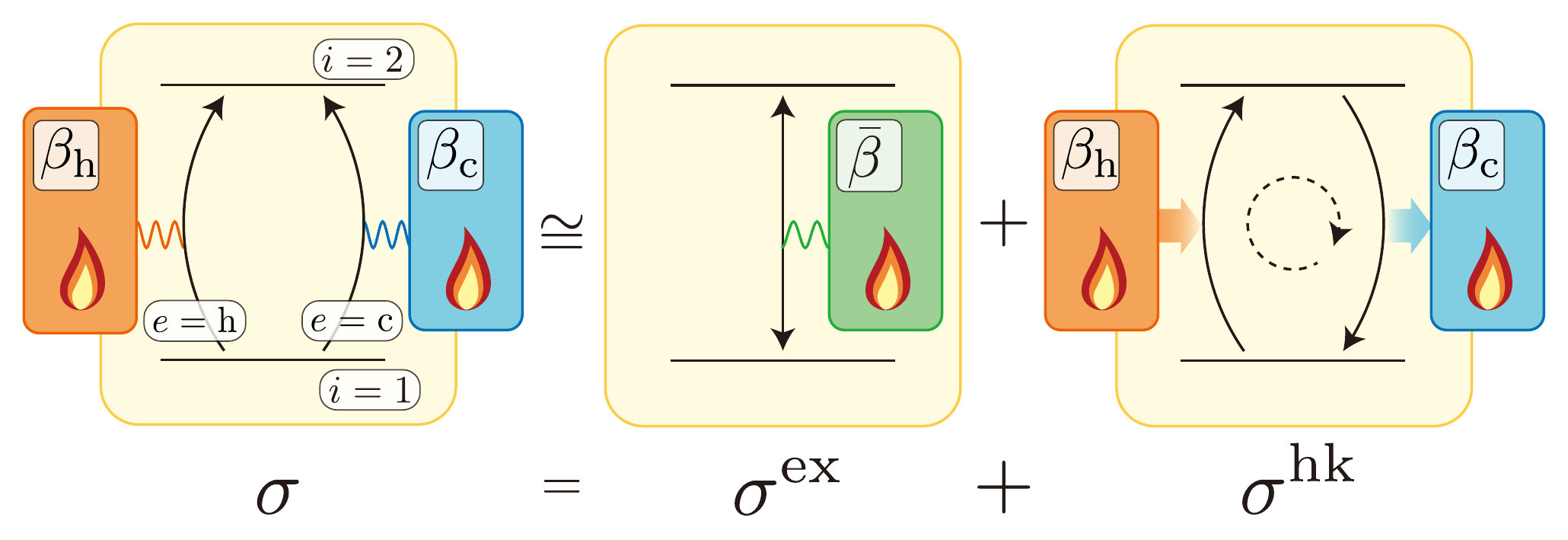}
    \caption{Conceptional diagram of example~\ref{sec:example:1}. A two-level system is attached to two heat reservoirs at different temperatures. The total dissipation (EPR) $\sigma$ can be divided into excess EPR $\sigma^\mathrm{ex}$ and housekeeping EPR $\sigma^\mathrm{hk}$. Both are clearly interpreted as follows:
    $\sigma^\mathrm{ex}$ corresponds to a relaxation caused by a single reservoir at the ``mean'' temperature $\bar{\beta}$, while $\sigma^\mathrm{hk}$ is attributed to heat transfer that does not change the state.}
    \label{fig:ex1_image}
\end{figure}
\begin{figure}
    \centering
    \includegraphics[width=\linewidth]{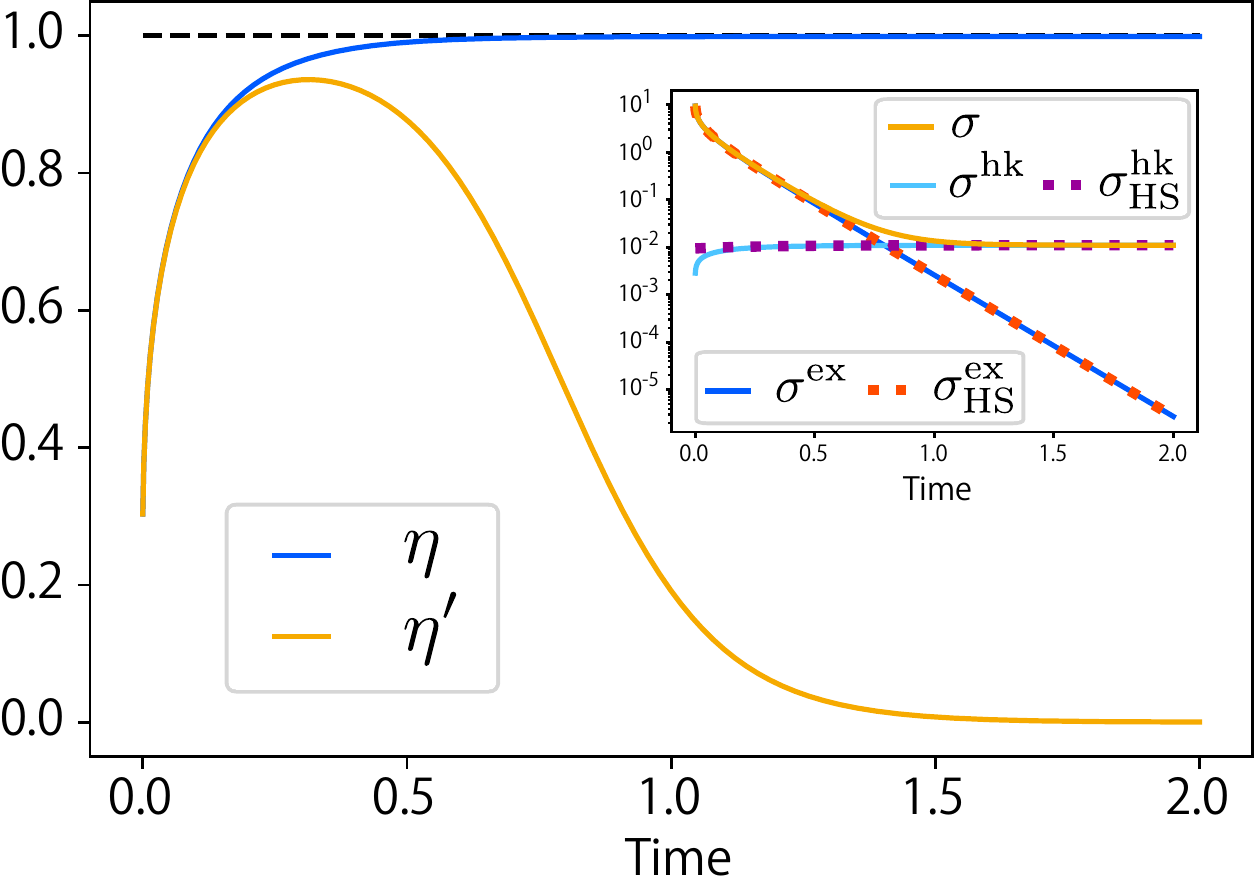}
    \caption{Verification of the short-time TUR of Eq.~\eqref{eq:MNTUR2}. The ratios between the lower bound $2(d_t\langle\psi\rangle)^2/D_\psi$ vs. EPR $\sigma$ and excess EPR $\sigma^\mathrm{ex}$ are shown. $\eta$ corresponds to $\sigma^\mathrm{ex}$ and $\eta'$ to $\sigma$. We choose the observable $\psi=(1,0)^\transpose $ but, in this system, the lower bound does not depend on the choice of  $\psi \ne 0$. $\eta'$ decreases and tends to zero after its peak because $\sigma$ does not vanish when $t\to\infty$.
    On the other hand, $\eta$ approaches the upper bound of $1$, although $\sigma^\mathrm{ex}$ diminishes at the same time. 
    In the inset, we compare our decomposition with the HS decomposition. In this example, where a stable steady state exists, they behave in almost the same way. }
    \label{fig:ex1_tur}
\end{figure}
We illustrate our results in a simple two-level stochastic system. 
Consider a two-level system attached to two heat reservoirs at inverse temperature $\beta_\mathrm{h}$ and $\beta_\mathrm{c}$ respectively. 
We can calculate the analytical forms of the excess and housekeeping EPR. 
Let the energy of state $i$ be $\epsilon_i$ for $i=1,2$ and $\epsilon_2>\epsilon_1$. 
There are two kinds of transition associated with the distinct reservoirs. 
We label each transition with the reservoir at inverse temperature $\beta_e$ by $e=\mathrm{h}$ or $\mathrm{c}$, instead of $1$ or $2$, to avoid confusion. The incidence matrix is then given by
\begin{align}
    B=\begin{pmatrix}-1&-1\\1&1
    \end{pmatrix}.
\end{align}
This setup is illustrated in Fig.~\ref{fig:ex1_image}. 

By using the eigendecomposition of $BLB^\transpose$ and formula~\eqref{eq:projection}, 
we can find the projection matrix
\begin{align}
    P=\frac{1}{\ell_\mathrm{h}+\ell_\mathrm{c}}\begin{pmatrix}
    \ell_\mathrm{h}&\ell_\mathrm{c}\\
    \ell_\mathrm{h}&\ell_\mathrm{c}
    \end{pmatrix},
\end{align}
where $\ell_e$ denotes the Onsager coefficient associated with a transition driven by heat bath $e$.
It leads to the conservative force $F^*$ as
\begin{align}
    F^*&=PF
    =\frac{\ell_\mathrm{h}F_\mathrm{h}+\ell_\mathrm{c}F_\mathrm{c}}{\ell_\mathrm{h}+\ell_\mathrm{c}}\begin{pmatrix}1\\1\end{pmatrix}\notag\\
    &=-B^\transpose \phi^*,
\end{align}
where, up to an additive constant, 
\begin{align}
    \phi^*=-\frac{\ell_\mathrm{h}F_\mathrm{h}+\ell_\mathrm{c}F_\mathrm{c}}{\ell_\mathrm{h}+\ell_\mathrm{c}}\begin{pmatrix}0\\1\end{pmatrix}.
\end{align}
The choice of $\phi^*$ is not unique, but $\phi_2^*-\phi_1^*=-F^*_\mathrm{h}=-F_\mathrm{c}^*$ is uniquely determined, so we write $\phi_2^*-\phi_1^*$ as $\Delta\phi^*$. 
If we do not consider any explicit external force, the forces read
\begin{align}
    F_e=-\beta_e\Delta\epsilon -\Delta\ln p
\end{align}
for $e=\mathrm{h},\mathrm{c}$, where $\Delta\epsilon =\epsilon_2-\epsilon_1$ and $\Delta\ln p=\ln p_2-\ln p_1$. 
Then, we obtain the expression
\begin{align}
    \Delta \phi^*
    =\bar{\beta}\Delta\epsilon+\Delta \ln p
\end{align}
with
\begin{align}
    \bar{\beta}=\frac{\ell_\mathrm{h}\beta_\mathrm{h}+\ell_\mathrm{c}\beta_\mathrm{c}}{\ell_\mathrm{h}+\ell_\mathrm{c}}. 
\end{align}
Therefore, $\phi^*$ can be thought of as a  mean potential of the system attached with a single bath at inverse temperature $\bar{\beta}$. 
The excess EPR is then interpreted as the dissipation due to the relaxation to the pseudo-canonical distribution $p_i^\mathrm{pcan}\propto e^{-\bar{\beta}\epsilon_i}$, in view of Eq.~\eqref{eq:relaxation}. 

On the other hand, $F-F^*$ is written as
\begin{align}
    F-F^*=\frac{\Delta\epsilon\Delta\beta}{\ell_\mathrm{h}+\ell_\mathrm{c}}\begin{pmatrix}\ell_\mathrm{c}\\-\ell_\mathrm{h}\end{pmatrix}.
\end{align}
where $\Delta\beta:=\beta_\mathrm{c}-\beta_\mathrm{h}$. 
Then, the corresponding current
\begin{align}
    J^\mathrm{cyc}:=L(F-F^*)
    =\frac{\ell_\mathrm{h}\ell_\mathrm{c}\Delta\epsilon\Delta\beta}{\ell_\mathrm{h}+\ell_\mathrm{c}}\begin{pmatrix}1\\-1\end{pmatrix}.
\end{align}
is cyclic ($BJ^\mathrm{cyc}=0$). 
Moreover, the housekeeping EPR reads
\begin{align}
    \sigma^\mathrm{hk}
    =\frac{\ell_\mathrm{h}\ell_\mathrm{c}}{\ell_\mathrm{h}+\ell_\mathrm{c}}(\Delta \epsilon)^2(\Delta \beta)^2. \label{eq:ex1_hk}
\end{align}
From this expression, we can see that the (inverse) temperature difference leads to a nonequilibrium steady state with a positive rate of dissipation. 

We illustrate the TUR of Eq.~\eqref{eq:MNTUR2} and compare our decomposition with the HS decomposition in Fig.~\ref{fig:ex1_tur}.
We plot the time evolution of the ratios
\begin{align}
    \eta&=\frac{2(d_t\langle\psi\rangle)^2}{D_\psi\sigma^\mathrm{ex}}\leq 1, \quad \eta'=\frac{2(d_t\langle\psi\rangle)^2}{D_\psi\sigma}\leq 1,
\end{align}
where we choose the observable $\psi=(1,0)^\transpose $. Note, however, that in this case $\eta$ and $\eta'$ do not depend on the choice of $\psi$, as long as it is nonzero. 
We find that $\eta'$ has a maximum value of $t\simeq 0.3$ and cannot attain the upper bound $1$, 
while $\eta$ approaches 1 as $t\to \infty$ (even though $\sigma^\mathrm{ex}$ vanishes at the same time). 
This calculation was done with parameters $\beta_\mathrm{h}=1,\beta_\mathrm{c}=2, \epsilon_2-\epsilon_1=1$, $k_{-\mathrm{h}}=k_{-\mathrm{c}}=1$ and $p_0(0)/p_1(0)=10^3$. 

The inset shows the time evolution of the excess and housekeeping EPR for the two decompositions. 
Both the decompositions give almost the same values in this system. 
We note that we can also express the HS housekeeping EPR with another mean temperature $\bar{\bar{\beta}}$ can be defined by
\begin{align}
    -\Delta \epsilon\bar{\bar{\beta}}
    :=\ln\frac{k_\mathrm{h}^++k_\mathrm{c}^+}{k_\mathrm{h}^-+k_\mathrm{c}^-}. \label{eq:hsmeant}
\end{align}
It satisfies $\beta_\mathrm{h}\leq\bar{\bar{\beta}}\leq\beta_\mathrm{c}$ if $\beta_\mathrm{h}\leq\beta_\mathrm{c}$ and let us rewrite the HS housekeeping EPR as
\begin{align}
    \sigma_\mathrm{HS}^\mathrm{hk}
    =\Delta\epsilon[(\bar{\bar{\beta}}-\beta_\mathrm{h})J_\mathrm{h}
    +(\bar{\bar{\beta}}-\beta_\mathrm{c})J_\mathrm{c}]. 
\end{align}
However, the definition of $\bar{\bar{\beta}}$ is not straightforward and this expression is less easy to interpret than Eq.~\eqref{eq:ex1_hk}. 

\subsection{Brusselator model of chemical oscillation}
\label{sec:example:2}
\begin{figure}
    \centering
    \includegraphics[width=\linewidth]{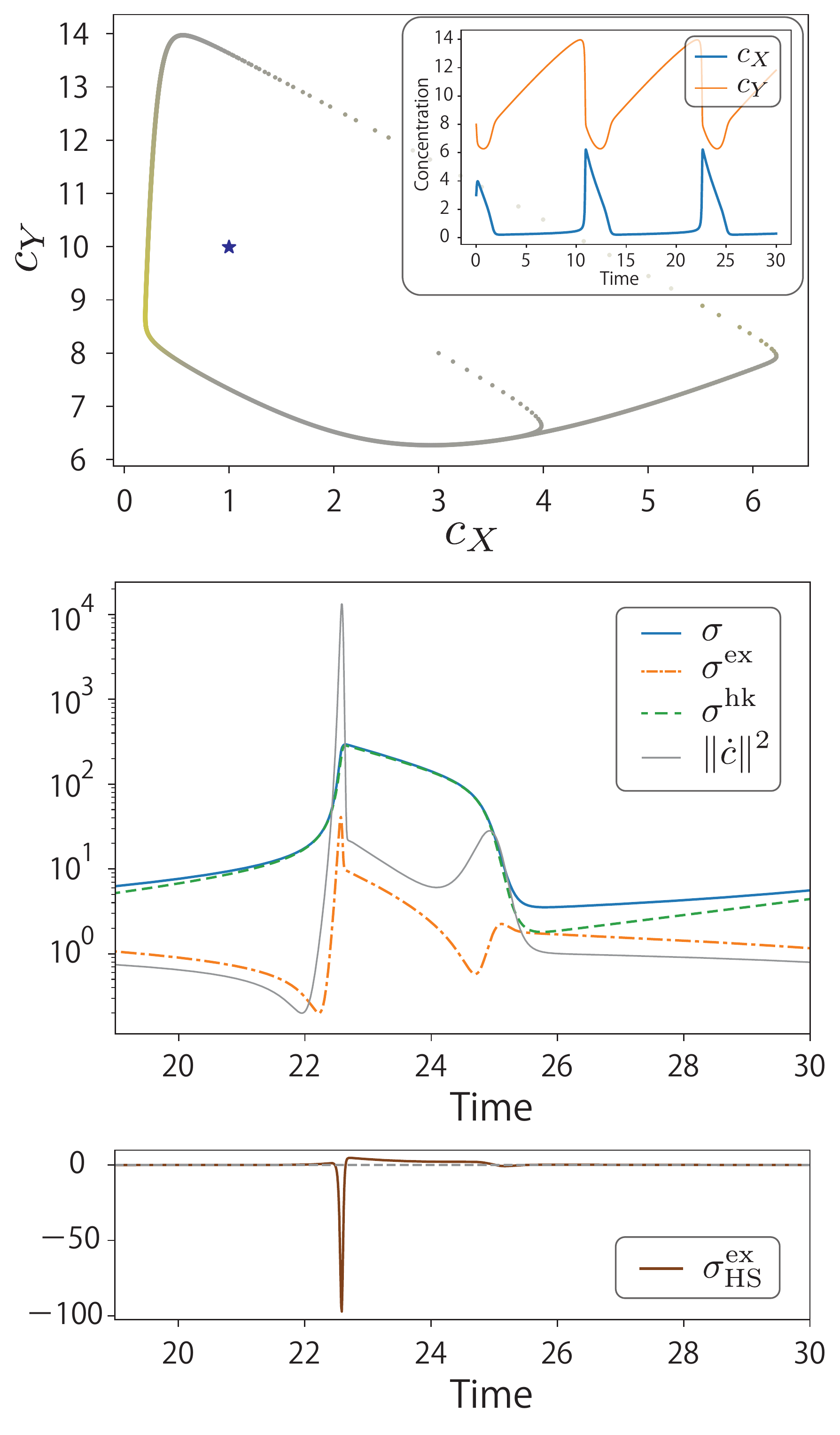}
    \caption{
    Housekeeping/excess EPR decomposition in the Brusselator model of chemical oscillations. We show the time evolution of the concentration $c=(c_X,c_Y)$ from Eq.~\eqref{eq:ex2_rateeq} in the inset of the uppermost panel. The uppermost panel shows the trajectory of $c$, which starts from $c(0)=(3,8)$ and rapidly converges to the limit cycle. The blue star indicates the unstable fixed point $c=(1,10)$. 
    A point in the trajectory is yellower (brighter) when $\sigma^\mathrm{ex}$ is large relative to $\sigma^\mathrm{hk}$. 
    The middle figure shows EPRs and the squared norm $\|\dot{c}\|^2=\dot{c}_X^2+\dot{c}_Y^2$ as a function of time. 
    The norm is normalized so that it fits in the graph with EPRs, thus the value is in arbitrary units.
    Note the correlation between the Onsager excess EPR $\sigma^\mathrm{ex}$ and the speed of the concentration vector $\|\dot{c}\|^2$ in the middle panel. 
    As shown in the lowermost panel, in this system with an unstable steady state, the HS excess EPR can be negative.}
    \label{fig:ex2}
\end{figure}
Let us consider the Brusselator model of a chemical oscillation. This model  consists of  three reactions,
\begin{align}
    \emptyset\rightleftharpoons X,\quad
    X\rightleftharpoons Y,\quad
    2X+Y\rightleftharpoons 3X.
\end{align}
We assume the mass action law and set the rate constants $k_1=k_3=k_{-1}=k_{-3}=1$, $k_2=10$, and $k_{-2}=0.1$. The concentrations of $X$ and $Y$ obey the following rate equation:
\begin{align}
    \frac{dc_X}{dt}&=1-c_X-10c_X+0.1c_Y+c_X^2c_Y-c_X^3,\notag\\
    \frac{dc_Y}{dt}&=10c_X-0.1c_Y-c_X^2c_Y+c_X^3. \label{eq:ex2_rateeq}
\end{align}
Here we denote the concentrations by $c$ rather than $x$. 
This equation admits an unstable steady state $(c_X,c_Y)=(1,10)$ and has a limit cycle. 
The time evolution is shown in the uppermost panel of Fig.~\ref{fig:ex2}. In the smaller inset, we show the oscillation of the concentrations $c_X$ and $c_Y$. The bigger graph shows its cyclic behavior. The blue star indicates the unstable fixed point. 

We show the behavior of the EPRs along with $\|\dot{c}\|^2=\dot{c}_X^2+\dot{c}_Y^2$ in the middle panel. $\|\dot{c}\|^2$ is normalized so that it fits in the graph with the EPRs. 
This figure shows the correlation between $\sigma^\mathrm{ex}$ and $\|\dot{c}\|^2$. 
In this model, the faster the concentration changes, the bigger the Onsager excess EPR becomes.
The color of dots in the uppermost panel indicates the size of $\sigma^\mathrm{ex}$ relative to $\sigma^\mathrm{hk}$. The yellow (bright) color corresponds to parts of the trajectory where $\sigma^\mathrm{ex}$ is large. This situation corresponds to around $t\simeq 25.8$ in the middle panel. 

Below the middle figure, we show that the HS excess EPR can be negative if one defines it using the unstable steady state $x^\mathrm{st}$. 
Let us prove this fact in the general case. 
$\sigma^\mathrm{ex}_\mathrm{HS}$ is given by the time derivative of the KL divergence as
\begin{align}
    \frac{d}{dt}D(x\|x^\mathrm{st})
    =-\sigma_\mathrm{HS}^\mathrm{ex}
\end{align}
when the flux obeys the mass action law~\cite{esposito2010three,ge2016nonequilibrium,rao2016nonequilibrium}.
For a cyclic solution $x$ with a period $T>0$, we have
\begin{align}
    \int_0^T dt\; \sigma_\mathrm{HS}^\mathrm{ex}
    =D(x(0)\|x^\mathrm{st})-D(x(T)\|x^\mathrm{st})=0.
\end{align}
This implies that, unless $\sigma_\mathrm{HS}^\mathrm{ex}$ vanishes identically, it has to take negative values at some times during the cycle.
We stress that this is in contrast to the Onsager projective decomposition, whose excess part is positive by construction.

\section{Discussion}
In this paper, we developed a geometrical perspective on dissipation in general Markovian dynamics on networks. 
In particular, we found a geometrical excess/housekeeping decomposition of the entropy production rate that helps us understand dynamics in terms of dissipation.
% , extending Schnakenberg's cyclic decomposition of dissipation as a result 
Our decomposition leads to refinements of several thermodynamic uncertainty relations. 
We have also generalized the $L^2$-Wasserstein geometry of a Markovian dynamics on a network, which allows us to derive the thermodynamic speed limit. 
Because our results are valid for very general dynamics, we expect that they will offer novel insights in various concrete systems. 

The geometrical decomposition given by the projection in~\eqref{eq:projection} shows how dynamics can be divided into two modes, one being cyclic and one being a relaxation driven by a conservative force, which can be written as a gradient flow. 
Even if a stable steady state does not exist, we can identify an instantaneous target of the relaxation mode as in Eq.~\eqref{eq:pcan}. The Onsager excess entropy production quantifies the dissipation due to this relaxation. 
On the other hand, the cyclic mode keeps the system out of equilibrium. 
The size of the cyclic mode defines the Onsager housekeeping entropy production, which can be further decomposed into contributions from individual cycles, generalizing Schnakenberg's decomposition as in Eq.~\eqref{eq:hkcyclic}.
Because these interpretations apply to both Markov jump processes and chemical reaction networks, our decomposition provides a clear and general thermodynamic separation of motion. 

We note that one may define a decomposition of the EPR using an orthogonal projection onto the image of $B^\transpose$ instead of $\mathbb{S}^\transpose$ as we have done. 
For a Markov jump process, it provides the same results, while for a chemical reaction network, it gives a different decomposition. 
The excess part of this decomposition vanishes in a steady state only if the steady state is complex balanced, but both parts are always nonnegative in contrast to the HS decomposition, thus this decomposition could provide another perspective to chemical reaction networks. 
However, it may be argued that the projection onto the image of $\mathbb{S}^\transpose$, as investigated here, is more physically meaningful. This is because it corresponds to conservative forces defined in terms of the stoichiometric coefficients of the reactions,  as standard in chemical thermodynamics, rather than forces that depend explicitly on the reactant and product chemical complexes.

From the experimental point of view, the edgewise Onsager coefficient, and therefore the Onsager excess and housekeeping EPRs, require information about the kinetics.  
For this reason, it may difficult to access them experimentally. 
However, recent studies have shown that TURs can be utilized to obtain estimates of the entropy production~\cite{manikandan2020inferring,otsubo2020estimating}. 
In a similar manner, we can use the refined TURs to estimate the excess and housekeeping EPR using measurable quantities.
Although we cannot generally attain equality in the TUR for a discrete system out of equilibrium, since part of the estimation error stems from the log mean inequality between the edgewise Onsager coefficient and the edgewise activity~\cite{shiraishi2021optimal}, this error gets smaller as the system gets closer to a steady state. 
In a two-level system, we can approach equality, as shown in Fig.~\ref{fig:ex1_tur} of Sec.~\ref{sec:example:1}, because the steady state is ``detailed balanced'' in the sense that there is no net current between the states, even though there is a steady flux between the reservoirs. 

We have also extended Maas's definition of $L^2$-Wasserstein distance to a general Markovian dynamics on a network. 
A mathematical advantage of Maas's distance is that it can be related to master equation dynamics via the Wasserstein gradient flow~\cite{maas2011gradient}, 
compared with other optimal transport theoretical distances between two discrete distributions, e.g., the $L^1$-Wasserstein distance~\cite{villani2009optimal,dechant2022minimum} and another discretization of the $L^2$-Wasserstein distance~\cite{chow2012fokker}. 
We expect that a deterministic rate equation with a detailed-balanced steady state is also characterized as a Wasserstein gradient flow with the aid of our generalized Wasserstein distance, beyond the usual gradient flow expression as in Eq.~\eqref{eq:gradientflow}~\cite{mielke2011gradient}. 

The relation between the geometrical decomposition and the generalized Wasserstein distance presented in Eq.~\eqref{eq:wandex} is parallel to recent results for Langevin systems~\cite{dechant2022geometric1,dechant2022geometric2}. 
Therefore, our generalization of Maas's $L^2$-Wasserstein distance is not only important from a mathematical point of view, but also has a physical interpretation as the minimum excess entropy of a process connecting the initial and final state.
As discussed above, obtaining a meaningful ``minimum entropy production'' requires a constraint, that is, keeping some quantity fixed during the minimization~\cite{dechant2022minimum,remlein2021optimality,muratore2013heat,ilker2022shortcuts,vanvu2022thermodynamic}.
In the present study, we fix the Onsager coefficients, whereas in previous studies \cite{muratore2013heat,remlein2021optimality,ilker2022shortcuts} the symmetric parts or one of the two directions of the transition rates were fixed.
While the latter may be the intuitive choice from an operational point of view, our approach establishes the connection between minimum entropy production, conservative forces and optimal transport.
Although we have shown how the excess EPR can be understood as a minimum EPR at each time point,
further study is needed to investigate how the finite-time optimization of the Wasserstein distance works practically and how the different approaches are related to each other.

One drawback of the discrete $L^2$-Wasserstein distance is that it needs a kinetics to refer to. 
Since the reference kinetics has to be time-independent, we cannot obtain the Wasserstein speed limit in externally driven systems. 
We remark a point that prevents generalizing the Wasserstein distance to the case where the reference kinetics is time dependent. 
While in the original definition, the time interval is not definite, we should care about how to define the time integral for a time-dependent reference. 
Then, however we choose the way to do so, we cannot use the conventional reparametrization technique, which is well described in Appendix~\ref{app:reparametrization}, because it requires the reference kinetics to be time independent.  
As a result, it becomes difficult to prove several properties, such as the triangle inequality and constant speed property of a geodesic. 

Let us mention a complementary study that we have recently done. 
In this paper, we considered an EPR decomposition by studying the space of thermodynamic forces using Euclidean geometry. However, it is also possible to define such a decomposition using the non-Euclidean information geometry. 
We develop this perspective in Ref.~\cite{Kolchinsky2022information}, where derive several thermodynamic bounds based on an information-geometric excess/housekeeping decomposition. 

Finally, we remark about a missing piece: nonlinear systems in continuous space. Two of the present authors, with another coauthor, presented a geometric decomposition of the EPR for linear stochastic systems in continuous space~\cite{dechant2022geometric1,dechant2022geometric2}. In this work, 
we developed it for Markov jump processes (linear stochastic system) and chemical master equations (nonlinear deterministic systems) in discrete space. The story remains to be written for nonlinear systems in continuous space, such as reaction-diffusion chemical systems and hydrodynamic systems. For such systems, geometric techniques for decomposing EPR may shed light on new kinds of thermodynamic constraints and tradeoffs.

\begin{acknowledgments}
K.~Y.~and S.~I.~thank Ryuna Nagayama and Kiyoshi Kanazawa for fruitful discussions.
K.~Y.~is supported by Grant-in-Aid for JSPS Fellows (Grant No.~22J21619). 
A.~D.~is supported by JSPS KAKENHI Grants No.~19H05795, and No.~22K13974.
S.~I.~is supported by JSPS KAKENHI Grants No.~19H05796, No.~21H01560, and No.~22H01141, and UTEC-UTokyo FSI Research Grant Program.

A.~K.~and A.~D.~contributed equally to this work. 
\end{acknowledgments}

\appendix
\section{Equivalence of definitions of Wasserstein distance}
\label{app:equivalence}
Let us show our definitions of the Wasserstein distance~\eqref{eq:wassersteinmarkov} and~\eqref{eq:generalwasserstein} are consistent with previous studies, e.g., Ref.~\cite{maas2011gradient}.
That is, the approach from \cite{maas2011gradient} is equivalent to ours, if detailed balance holds. 
In Ref.~\cite{maas2011gradient}, the author defines the relative distribution 
\begin{align}
    \rho_i:=\frac{p_i}{p_i^\mathrm{eq}}
\end{align}
and defines the Wasserstein distance by 
%\begin{widetext}
\begin{align}
    \begin{aligned}
    &\mathcal{W}'(p^{(0)},p^{(1)})
    :=\\
    & \inf_{p,\psi}\left({\int_0^1dt\;
    \frac{1}{2}\sum_{i,j}(\psi_i-\psi_j)^2
    k_{i\to j}p^\mathrm{eq}_{i}\mathbb{L}(\rho_i,\rho_j)}\right)^{1/2}
    \end{aligned}
\end{align}
with conditions
\begin{gather}
\frac{d}{dt}\rho_i+\sum_j(\psi_j-\psi_i)k_{i\to j}\mathbb{L}(\rho_i,\rho_j)=0, \label{eq:cond1}\\
p(t=0)=p^{(0)},\quad p(t=1)=p^{(1)},
\end{gather}
%\end{widetext}
where $\mathbb{L}$ is the logarithmic mean
\begin{align}
    \mathbb{L}(a,b):=\frac{a-b}{\ln a-\ln b}.
\end{align}
First, by symmetry, the integrand is rewritten as
\begin{align}
    &\frac{1}{2}\sum_{i,j}(\psi_i-\psi_j)^2
    k_{i\to j}p^\mathrm{eq}_{i}\mathbb{L}(\rho_i,\rho_j)\notag\\
    &=\sum_{e}([B^\transpose \psi]_e)^2k_ep_{\iota(e)}^\mathrm{eq}\mathbb{L}(\rho_{\iota(e)},\rho_{\iota'(e)}),
\end{align}
where we used the detailed balance condition $k_ep_{\iota(e)}^\mathrm{eq}=k_{-e}p_{\iota'(e)}^\mathrm{eq}$. 
Moreover, the numerator of the logarithmic mean times $k_ep_{\iota(e)}^\mathrm{eq}$ gives $J_e$ because
\begin{gather}
    k_ep_{\iota(e)}^\mathrm{eq}\rho_{\iota(e)}
    =k_ep_{\iota(e)}\\
    k_ep_{\iota(e)}^\mathrm{eq}\rho_{\iota'(e)}
    =k_{-e}p_{\iota'(e)}^\mathrm{eq}\rho_{\iota'(e)}=k_{-e}p_{\iota'(e)}. 
\end{gather}
Furthermore, the denominator reads
\begin{align}
    \ln{\rho_{\iota(e)}}-\ln {\rho_{\iota'(e)}}
    =\ln\frac{k_ep_{\iota(e)}}{k_{-e}p_{\iota'(e)}}=F_e 
\end{align}
due to the detailed balance condition. 
Therefore, we have 
\begin{align}
    k_ep_{\iota(e)}^\mathrm{eq}\mathbb{L}(\rho_{\iota(e)},\rho_{\iota'(e)})
    =\frac{J_e}{F_e}=\ell_e, \label{eq:logmeanandonsager}
\end{align}
and the integrand is equal to
\begin{align}
    \sum_e ([B^\transpose \psi]_e)^2\ell_e=\|B^\transpose \psi\|_L^2. 
\end{align}

We next consider the conditions for minimization. 
From Eq.~\eqref{eq:logmeanandonsager}, we find condition~\eqref{eq:cond1} is equivalent to
\begin{align}
    \frac{d}{dt}p_i+\sum_j(\psi_j-\psi_i)\ell_{i\to j}=0,
\end{align}
where $\ell_{i\to j}=\ell_e$ if the edge $e$ corresponding to the transition $i\to j$ is in the set of edges $\{1,\dots,E\}$, and $\ell_{i\to j}=0$ if not (so $k_{i\to j}=k_{j\to i}=0$).
Due to the symmetry $\ell_{i\to j}=\ell_{j\to i}$, we have
\begin{align}
    &\sum_j(\psi_j-\psi_i)\ell_{i\to j}\notag\\
    &=\sum_{e:i\to *} \ell_e \sum_j B_{je}\psi_j
    -\sum_{e:*\to i} \ell_e \sum_j B_{je}\psi_j\notag\\
    &=-\sum_e B_{ie}\ell_e\sum_jB_{je}\psi_j
    =-[BLB^\transpose \psi]_i, 
\end{align}
therefore the conditions are equivalent to ours in Eq.~\eqref{eq:ourcondmarkov}. 

% Let $\{p(s),\psi(s)\}_{s\in[0,1]}$ be the solution of the minimization problem. If we change the variable by $s\to t=\tau s$, $\{p'(t)=p(t/\tau),\psi'(t)=\psi(t/\tau)\}_{t\in[0,\tau]}$ will not satisfy
% \begin{align}
%     \frac{d}{dt}p'_i+\sum_j (\psi'_j-\psi_i')\ell_{i\to j}=0. \label{eq:condb17}
% \end{align}
% because 
% \begin{align}
%     \frac{d}{dt}p'_i
%     =\frac{1}{\tau}\frac{d}{ds}p_i
%     =\frac{1}{\tau}\sum_j(\psi'_j-\psi_i')\ell_{i\to j}. 
% \end{align}
% Instead, $\{p'(t)=p(t/\tau),\psi''(t)=\tau\psi(t/\tau)\}_{t\in[0,\tau]}$ will satisfy condition~\eqref{eq:condb17} and solve the minimization
% \begin{align}
%     \inf_{p,\psi}\frac{1}{2}\int_0^\tau dt\;\sum_{i,j}(\psi_i-\psi_j)^2
%     k_{i\to j}p^\mathrm{eq}_{i}\mathbb{L}(\rho_i,\rho_j). 
% \end{align}
% For this solution, we have
% %\begin{widetext}
% \begin{align*}
%     &\int_0^\tau dt\;(\psi''_i(t)-\psi''_j(t))^2
%     k_{i\to j}p^\mathrm{eq}_{i}\mathbb{L}(\rho'_i(t),\rho'_j(t)) \nonumber \\
%     &=\tau^2\int_0^\tau dt\;(\psi_i({\textstyle \frac{t}{\tau}})-\psi_j({\textstyle \frac{t}{\tau}}))^2k_{i\to j}p_i^\mathrm{eq} \mathbb{L}(\rho_i({\textstyle \frac{t}{\tau}}),\rho_j({\textstyle \frac{t}{\tau}}))\notag\\
%     &=\tau\int_0^1 ds\;(\psi_i(s)-\psi_j(s))^2k_{i\to j}p_i^\mathrm{eq} \mathbb{L}(\rho_i(s),\rho_j(s)). 
% \end{align*}
% As a result, 
Finally, we complete proving the equivalence
\begin{multline}
    \inf_{p,\psi}\left({\frac{1}{2}\int_0^1dt\;\sum_{i,j}(\psi_i-\psi_j)^2
    k_{i\to j}p^\mathrm{eq}_{i}\mathbb{L}(\rho_i,\rho_j)}\right)^{1/2}\\
    =\inf_{p,\psi}\left({\int_0^1 dt\;\|B^\transpose \psi\|_L^2}\right)^{1/2}
\end{multline}
with equivalent conditions 
\begin{gather}
\frac{d}{dt}\rho_i+\sum_j(\psi_j-\psi_i)k_{i\to j}\mathbb{L}(\rho_i,\rho_j)=0, \\
p(t=0)=p^{(0)},\quad p(t=1)=p^{(1)}
\end{gather}
for the left-hand side and
\begin{gather}
    \frac{d}{dt}p=BLB^\transpose \psi,\; 
    p(t=0)=p^{(0)},\;p(t=1)=p^{(1)}
\end{gather}
for the right-hand side. 
%\end{widetext}

\section{Onsager coefficient and diffusion coefficient}
\label{app:onsdiff}

We discuss how the Onsager coefficients are connected to the diffusion coefficient in Langevin systems when we consider a Markov jump process on a square lattice that approximates a continuous space.
The result here was obtained in Ref.~\cite{vanvu2022thermodynamic} for one-dimensional cases and we generalize it to $n$-dimensional systems for an arbitrary dimension $n$. 

Let $i$ denote a point on $n$-dimensional square lattice with the lattice constant $\Delta x\ll1$. The discrete probability distribution $p_i$ can be approximated by a smooth density function $P(x)$ as $p_i\simeq P(i)(\Delta x)^n$. 
A transition can be designated by its starting point and the direction $d=\pm1,\dots,\pm n$. With $w_{i}^{(d)}$ being the transition rate of the jump from $i$ in direction $d$, we have the master equation
\begin{align}
    \frac{dp_i}{dt}=\sum_{d=\pm1}^{\pm n}\left(w_{i+d}^{(-d)}p_{i+d}-w_i^{(d)}p_i\right), \label{eq:latticemaster}
\end{align}
where we make the time dependence implicit and $i+d$ is the point next to $i$ in the direction of $d$. Note that now each edge is determined by the combination of the starting point and the direction of movement.

As in previous work~\cite{vanvu2022thermodynamic,van2010three}, we assume the transition rates are expanded as the series
\begin{align}
    w_{i}^{(d)}=\frac{D_d(i)}{\Delta x^2}+\frac{f_d(i)}{\Delta x}
    +o(1).
\end{align}
Here, we assume $D_d(x)$ to be a symmetric function ($D_{-d}(x)=D_{d}(x)$) 
and $f_d(x)$ anti-symmetric ($f_{-d}(x)=-f_{d}(x)$). Then the master equation reads
\begin{align}
    \frac{dp_i}{dt}
    &=\frac{1}{\Delta x^2}\sum_{d=\pm 1}^{\pm n}\left(D_{-d}(i+d)p_{i+d}-D_d(i)p_i\right)
    \notag\\
    &\phantom{=}+\frac{1}{2\Delta x}\sum_{d=\pm 1}^{\pm n}\left(f_{-d}(i+d)p_{i+d}-f_d(i)p_i\right)
    +o(1)\notag\\
    &=\sum_{d=1}^n\frac{D_{d}(i+d)p_{i+d}+D_{d}(i-d)p_{i-d}-2D_d(i)p_i}{\Delta x^2}\notag\\
    &\phantom{=}-\sum_{d=1}^n \frac{f_{d}(i+d)p_{i+d}-f_{d}(i-d)p_{i-d}}{2\Delta x}+o(1).
\end{align}
By taking the limit $\Delta x \to 0$, we have the Fokker--Planck equation
\begin{align}
    \frac{\partial}{\partial t}P(x)
    =-\sum_{d=1}^n \frac{\partial}{\partial x_d}\left(f_d(x)P(x)\right)
    +\sum_{d=1}^n \frac{\partial^2}{\partial x_d^2}\left(D_d(x)P(x)\right).
\end{align}
Now, we can show that to the leading order, the current and the force on edge $(i,d)$ are given as
\begin{align}
    J_d(i)&\simeq \frac{f_d(i)p_i}{\Delta x}
    -\frac{\partial_d(D_d(i)p_i)}{\Delta x},\\
    F_d(i)&\simeq \frac{f_d(i)}{D_d(i)}\Delta x
    -\frac{\partial_d(D_d(i)p_i)}{D_d(i)p_i}\Delta x,
\end{align}
where $\partial_d(D_d(i)p_i)$ stands for $(D_d(i+d)p_{i+d}-D_d(i)p_i)/\Delta x$.
Thus we have
\begin{align}
    \ell_d(i)= \frac{D_d(i)p_i}{\Delta x^2}+o(\Delta x^{-2}),
\end{align}
which shows that fixing the functional form of the Onsager coefficients is similar to fixing the diffusion coefficient in a Langevin system.

\section{Detailed comparison with HS decomposition}
\label{app:HS}
In addition to the discussion given in Sec.~\ref{sec:decomp:hs}, we can further establish a connection between the HS decomposition and Onsager-projective decomposition. 
Let $P_\Psi$ be the orthogonal projection operator onto the vector $\mathbb{S}^\transpose \Psi$ with respect to the metric $L$, with $\Psi$ given by Eq.~\eqref{eq:HS-potential},
\begin{align}
    P_\Psi:=\frac{\mathbb{S}^\transpose \Psi\Psi^\transpose \mathbb{S}}{\|\mathbb{S}^\transpose \Psi\|_L^2}L. 
\end{align}
We define the pseudo-HS decomposition by
\begin{align}
    \sigma_\mathrm{pHS}^\mathrm{ex}&:=\|P_\Psi F\|_L^2,\\
    \sigma_\mathrm{pHS}^\mathrm{hk}&:=\|(I-P_\Psi) F\|_L^2. 
\end{align}
Because $P_\Psi$ is an orthogonal projection operator, the sum of the two terms is equal to the total EPR: 
\begin{align}
    \sigma=\sigma_\mathrm{pHS}^\mathrm{ex}+\sigma_\mathrm{pHS}^\mathrm{hk}. 
\end{align}
Since $PP_\Psi=P_\Psi$, we have the inequality
\begin{align}
    \sigma_\mathrm{pHS}^\mathrm{ex}\leq \sigma^\mathrm{ex}, 
\end{align}
which allows us to define a coupling term as in Refs.~\cite{dechant2022geometric1,dechant2022geometric2}
\begin{align}
    \sigma^\mathrm{cpl}:=\sigma^\mathrm{ex}-\sigma_\mathrm{pHS}^\mathrm{ex}\geq 0, 
\end{align}
and decompose the EPR into three positive contributions
\begin{align}
    \sigma = \sigma_\mathrm{pHS}^\mathrm{ex} + \sigma^\mathrm{cpl} + \sigma^\mathrm{hk} .
\end{align}

However, the pseudo-HS decomposition does not generally coincide with the genuine HS decomposition. 
A straightforward calculation reveals their connection
\begin{align}
    \sigma_\mathrm{pHS}^\mathrm{ex}=\frac{(\sigma_\mathrm{HS}^\mathrm{ex})^2}{\|\mathbb{S}^\transpose \Psi\|_L^2}
\end{align}
If $\sigma_\mathrm{HS}^\mathrm{ex}=\|\mathbb{S}^\transpose \Psi\|_L^2$ holds, the two give the same value, but it is not the case in general.

Moreover, the difference between $\sigma_\mathrm{HS}^\mathrm{ex}$ and $\|\mathbb{S}^\transpose \Psi\|_L^2$ is of the same order as each individual term even when the system is near a steady state; in other words, not small.
Let $\sup_{\alpha}|x_\alpha-x_\alpha^\mathrm{st}|=\epsilon$ be very small, with $x^\mathrm{st}$ a steady state and assume $d_t x=\mathbb{S}J$ is of the same order. 
Then, $\sigma^\mathrm{ex}_\mathrm{HS}$ and $\|\mathbb{S}^\transpose \Psi\|_L^2$ are of order of $\epsilon^2$ because $\Psi$ and $\mathbb{S}J$ are both of order $\epsilon$.
On the other hand, because $\mathbb{S}L^\mathrm{st}F^\mathrm{st}=\mathbb{S}J^\mathrm{st}=0$, the difference between $\sigma_\mathrm{HS}^\mathrm{ex}$ and $\|\mathbb{S}^\transpose \Psi\|_L^2$ becomes
\begin{align}
    \sigma_\mathrm{HS}^\mathrm{ex}-\|\mathbb{S}^\transpose \Psi\|_L^2
    &=-\Psi^\transpose \mathbb{S}(J+L\mathbb{S}^\transpose \Psi)\notag\\
    &=-\Psi^\transpose \mathbb{S}L(F+\mathbb{S}^\transpose \Psi)\notag\\
    &=-\Psi^\transpose \mathbb{S}LF^\mathrm{st}\notag\\
    &=-\Psi^\transpose \mathbb{S}(L-L^\mathrm{st})F^\mathrm{st},
\end{align}
where the third line comes from equation~\eqref{eq:ffpsi}. 
Here, $\Psi$ and $L-L^\mathrm{st}$ are of the order of $\epsilon$, while $F^\mathrm{st}$ is generally of the order of $1$, so that we see the difference is of the order of $\epsilon^2$, just as $\sigma_\mathrm{HS}^\mathrm{ex}$ and $\|\mathbb{S}^\transpose \Psi\|_L^2$. 
The equality only holds when we have the detailed balance condition, that is for conservative forces. In this case, we find $\sigma = \sigma^\text{ex}_\text{HS} = \sigma^\text{ex}_\text{pHS}$. 

\section{Derivations of finite-time TURs} \label{app:TUR}
\subsection{Housekeeping TUR}
To prove the housekeeping TUR~\eqref{eq:hktur}, let us consider a parametrization of transition rates by introducing the interpolated force  $F_e^\theta=F_e^*+\theta(F_e-F_e^*)$ as 
\begin{gather}
    k_e^\theta p_{\iota(e)}
    =\ell_e g(F_e^\theta), \label{eq:tilted1}\\
    \text{with}\quad g(x) = \frac{x}{1-e^{-x}},
\end{gather}
where $\theta\in\mathbb{R}$ is an interpolation parameter.
We also define $J_e^\theta:=k_e^{\theta}p_{\iota(e)}-k_{-e}^{\theta}p_{\iota'(e)}$.
One can easily show that $J_e^\theta =\ell_e  F_e^\theta$ and $J_e^\theta =J_e^*+\theta(J_e-J_e^*)$ hold. 
Because $BJ^\theta=BJ$ for any $\theta$ and any time, the solution of the original ($\theta=1$) master equation $\{p(t)\}_{t\in[0,\tau]}$ solves $d_tp(t)=BJ^\theta(t)$. 

In general, we denote the path probability for the master equation by $\mathbb{P}$ and that for the master equation with modified rate constants $k^\theta$, such as those defined in Eq.~\eqref{eq:tilted1}, by $\mathbb{P}^\theta$. 
The KL divergence between $\mathbb{P}^\theta$ and $\mathbb{P}^{\theta'}$ is given by
\begin{align}
    D(\mathbb{P}^\theta\|\mathbb{P}^{\theta'})&= \int d \Gamma\; \mathbb{P}^\theta (\Gamma) \ln \frac{\mathbb{P}^\theta(\Gamma)}{\mathbb{P}^{\theta'}(\Gamma)} \notag\\
    &=\int_0^\tau dt\sum_{\pm e}\pqty{\ln\frac{k_e^\theta}{k_e^{\theta'}}+\frac{k_e^{\theta'}}{k_e^{\theta}}-1}k_e^\theta p_{\iota(e)}^\theta, \label{eq:klpath}
\end{align}
where $\Gamma$ indicates a path, $\sum_{\pm e}$ denotes the summation over both directions of edges, and $p^\theta$ is the solution of the master equation given by $k^\theta$. We always assume the initial distribution does not depend on parameter $\theta$.
The KL divergence can also be expressed as
\begin{align}
    D(\mathbb{P}^\theta\|\mathbb{P}^{\theta'})&= \int_0^\tau dt\; D(k^{\theta}p^\theta|| k^{\theta'}p^\theta),
\end{align}
where $D(k^{\theta}p^\theta|| k^{\theta'}p^\theta)$ is the generalized KL divergence between the two ``distributions'' $\{k_e^\theta p_{\iota(e)}^\theta\}_{e=\pm1,\pm 2,\dots}$ and $\{k_e^{\theta'} p_{\iota(e)}^\theta\}_{e=\pm1,\pm 2,\dots}$, that is, 
\begin{align}
    &D(k^{\theta}p^\theta|| k^{\theta'}p^\theta) \notag\\
    &=\sum_{\pm e}\left[ k_e^\theta p_{\iota(e)}^\theta \ln\frac{k_e^\theta p_{\iota(e)}^\theta}{k_e^{\theta'}p_{\iota(e)}^\theta}-k_e^\theta p_{\iota(e)}^\theta + k_e^{\theta'} p_{\iota(e)}^\theta \right].
\end{align}
When we consider $\theta'=\theta+d\theta$ with a very small $d\theta$, the KL divergence becomes
\begin{align}
&D(\mathbb{P}^\theta\|\mathbb{P}^{\theta+d\theta})
=\int_0^\tau dt\; D(k^{\theta}p^\theta|| k^{\theta+d\theta}p^\theta) \nonumber \\
&=\frac{d\theta^2}{2}\int_0^\tau dt \sum_{\pm e} k_e^\theta p_{\iota(e)}^\theta  \left( \partial_{\theta} \ln k_e^\theta \right)^2 +o(d\theta^2).
\label{eq:kltaylor}
\end{align}

In the parametrization~\eqref{eq:tilted1}, the relation
\begin{align}
    \partial_\theta\ln k_e^\theta
    =(F_e-F_e^*) \eval{\frac{d}{dx}\ln g(x)}_{x=F_e^\theta}
\end{align}
holds. 
Thus, the KL divergence becomes
\begin{align}
&D(\mathbb{P}^\theta\|\mathbb{P}^{\theta+d\theta}) \notag\\
&=\frac{d\theta^2}{2}\int_0^\tau dt \sum_{\pm e} \ell_e (F_e-F_e^*)^2 h(F_e^\theta)  +o(d\theta^2),
\end{align}
with
\begin{align}
\quad h(x)&=g(x)\pqty{\frac{d}{dx}\ln g(x)}^2 \nonumber \\
    &=\frac{x}{1-e^{-x}}\pqty{\frac{e^x-x-1}{x(e^x-1)}}^2.
\end{align}
One can easily show that $h(x)+h(-x)\leq 1/2$. Therefore, we have the following inequality regardless of the value of $\theta$
\begin{align}
    D(\mathbb{P}^\theta\|\mathbb{P}^{\theta+d\theta})
    \leq \frac{d\theta^2}{4}\int_0^\tau dt\;\sigma^\mathrm{hk}. \label{eq:klhkep}
\end{align}
The fluctuation-response inequality~(12) of Ref.~\cite{dechant2020fluctuation} tells us that for any current observable $\mathcal{J}_w$ defined in Sec.~\ref{sec:tur:finitemarkov},
\begin{align}
    \frac{(\langle \mathcal{J}_{w}\rangle_\tau^{\theta+d\theta}-\langle \mathcal{J}_{w}\rangle_\tau^{\theta})^2}{2\mathrm{Var}^{\theta+d\theta}(\mathcal{J}_{w})}\leq 
    D(\mathbb{P}^\theta\|\mathbb{P}^{\theta+d\theta}), \label{eq:fri}
\end{align}
where $\langle\cdot\rangle_\tau^\theta$ and $\mathrm{Var}^\theta(\cdot)$ are the average and variance regarding $\mathbb{P}^\theta$. 
Because
\begin{align}
    \langle \mathcal{J}_{w}\rangle_\tau^{\theta}
    =\langle\mathcal{J}_{w}^\mathrm{ex}\rangle_\tau+\theta\langle\mathcal{J}_{w}^\mathrm{hk}\rangle_\tau, 
\end{align}
setting $\theta =1$ in Eq.~\eqref{eq:fri} and combining with Eq.~\eqref{eq:klhkep}, we obtain the desired inequality in Eq.~\eqref{eq:hktur}.

\subsection{Excess TUR}
Let us prove the relation~\eqref{eq:finiteturex}. We separate the time dependence as $t=\tau s$, so that $s\in[0,1]$, and write $\bar{f}(s;\tau)=f(\tau s)$ for any function of time $f$. The assumption on the functional form of the transition rate $k$ leads to that $\bar{k}$ only depends on $s$. The master equation reads
\begin{gather}
    \partial_s \bar{p}(s;\tau)
    =\tau B\bar{J}(s;\tau), \label{masteroriginal1}\\
    \bar{J}_e(s;\tau)=\bar{k}_{e}(s)\bar{p}_{\iota(e)}(s;\tau)-\bar{k}_{-e}(s)\bar{p}_{\iota'(e)}(s;\tau). \label{masteroriginal2}
\end{gather}
It is a natural assumption that the initial distribution $\bar{p}(0;\tau)$ does not depend on $\tau$. 

We first consider the tilting of the transition rates with $\tau$ fixed
\begin{align}
    \bar{k}_e^{\theta}(s;\tau)
    =\bar{k}_e(s)\exp(\frac{\bar{J}_e^*(s;\tau)}{\bar{\chi}_e(s;\tau)} \theta), \label{eq:tilting}
\end{align}
where $\theta$ is the parameter of the tilted transition rate. 
From the general formula~\eqref{eq:kltaylor}, 
the KL divergence between $\mathbb{P}^{0}$ and $\mathbb{P}^{d\theta}$ given by the tilting~\eqref{eq:tilting} can be calculated as
\begin{align}
    D(\mathbb{P}^0 \|\mathbb{P}^{d\theta})
    =\frac{\tau}{2}d\theta^2\int_0^1 ds
    \sum_e \frac{(\bar{J}_e^*(s;\tau))^2}{\bar{\chi}_e(s;\tau)}+o(d\theta^2)
\end{align}
because $\partial_\theta\ln \bar{k}_e^\theta(s)=\bar{J}^*_e(s;\tau)/\bar{\chi}_e(s;\tau)$. 
From the log mean inequality, we also have
\begin{align}
    \frac{\tau}{2}\int_0^1 ds
    \sum_e \frac{(\bar{J}_e^*(s;\tau))^2}{\bar{\chi}_e(s;\tau)}
    &\leq \frac{\tau}{4}\int_0^1 ds
    \sum_e \frac{(\bar{J}_e^*(s;\tau))^2}{\bar{\ell}_e(s;\tau)}\notag\\
    &=\frac{1}{4}\int_0^\tau dt
    \sum_e \frac{(J_e^*(t))^2}{\ell_e(t)}\notag\\
    &=\frac{1}{4}\int_0^\tau dt\;
    \sigma^\mathrm{ex}.
\end{align}
Thus, the fluctuation-response inequality provides
\begin{align}
    \frac{(\left. \partial_{\theta} \langle \mathcal{J}_{w}\rangle_\tau^{\theta} \right|_{\theta =0})^2}{\mathrm{Var}(\mathcal{J}_{w})}\leq 
    \frac{1}{2}\int_0^\tau dt\;
    \sigma^\mathrm{ex}. \label{eq:frifinitetime}
\end{align}

Next, we identify the quantity $\left. \partial_{\theta} \langle \mathcal{J}_{w}\rangle_\tau^{\theta} \right|_{\theta =0}$. Let $\{\bar{p}^{\theta}(s;\tau)\}_{s\in[0,1]}$ be the solution of the master equation given by $\bar{k}^{\theta}_e(s;\tau)$,
\begin{gather}
    \partial_s \bar{p}^{\theta}(s;\tau)
    =\tau B\bar{J}^{\theta}(s;\tau),\\
    \bar{J}_e^{\theta}(s;\tau)=\bar{k}_{e}^{\theta}(s;\tau)\bar{p}_{\iota(e)}^{\theta}(s;\tau)-\bar{k}_{-e}^{\theta}(s;\tau)\bar{p}_{\iota'(e)}^{\theta}(s;\tau),
\end{gather}
where $\theta =0$ reproduces the original dynamics Eqs.~(\ref{masteroriginal1}) and (\ref{masteroriginal2}).
If we use the notation
\begin{align}
    \left. \partial_{\theta} \bar{p}_i^{\theta}(s;\tau) \right|_{\theta =0}= r_i(s;\tau),
\end{align}
a straightforward calculation shows
\begin{align}
    &\left. \partial_{\theta} \bar{J}_e^{\theta}(s;\tau) \right|_{\theta =0} \nonumber \\
    &=
    \bar{J}_e^*(s;\tau)
    +[\bar{k}_e(s)r_{\iota(e)}(s;\tau)
    -\bar{k}_{-e}(s)r_{\iota'(e)}(s;\tau)]. \label{eq:perturbedcurrent}
\end{align}
Because $\tau B\bar{J}^*=\tau B\bar{L}\bar{F}^*=\tau B\bar{J}=\partial_s\bar{p}(s;\tau)$, 
we find that $r$ satisfies
\begin{align}
    &\partial_s r_i(s;\tau) -\partial_s \bar{p}_i (s;\tau) \notag \\
    &=\tau\sum_e B_{ie}[\bar{k}_e(s)r_{\iota(e)}(s;\tau)-\bar{k}_{-e}(s)r_{\iota'(e)}(s;\tau)]. \label{eq:eqforr}
\end{align}
We also consider another infinitesimal perturbation by changing $\tau$ to $\theta \tau$ where $\theta =1$ also reproduces the original dynamics Eqs.~(\ref{masteroriginal1}) and (\ref{masteroriginal2}).
The master equation reads
\begin{gather}
    \partial_s \bar{p}(s;\theta \tau)
    = \theta\tau B\bar{J}(s;\theta \tau).
\end{gather}
By calculating
\begin{align}
    \left. \partial_{\theta} \partial_s \bar{p}(s;\theta \tau) \right|_{\theta =1}
    &= \left. \partial_{\theta} [\theta\tau B\bar{J}(s;\theta \tau)] \right|_{\theta =1},
\end{align}
we find 
\begin{align}
 &\partial_s [\tau\partial_{\tau}  \bar{p}_i(s; \tau)] -\partial_s \bar{p}_i(s;\tau)
 =\tau\sum_e B_{ie}\tau\partial_\tau \bar{J}_e(s;\tau) \nonumber\\
 &= \tau \sum_e B_{ie}\left[\bar{k}_e(s)[\tau\partial_{\tau}  \bar{p}_{\iota(e)}(s; \tau)]-\bar{k}_{-e}(s)[\tau\partial_{\tau}  \bar{p}_{\iota'(e)}(s; \tau)]\right].
\end{align}
It is the same equation as Eq.~\eqref{eq:eqforr}.
Because $\tau \partial_{\tau}\bar{p}(s; \tau)$ and $r(s;\tau)$ satisfies the same first order differential equation and they share the initial condition $\tau \partial_{\tau}\bar{p}(0; \tau)=r(0;\tau) = 0$, we conclude that $r(s;\tau)=\tau\partial_\tau \bar{p}(s;\tau)$ and 
\begin{align}
\bar{k}_e(s)r_{\iota(e)}(s;\tau)-\bar{k}_{-e}(s)r_{\iota'(e)}(s;\tau) 
=\tau\partial_\tau \bar{J}_e(s;\tau).
\end{align}
Thus, Eq.~(\ref{eq:perturbedcurrent}) reads
\begin{align}
\left. \partial_{\theta} \bar{J}_e^{\theta}(s;\tau) \right|_{\theta =0} &=\bar{J}_e^*(s;\tau)+\tau\partial_\tau \bar{J}_e(s;\tau),
\end{align}
and the current observable changes by tilting as
\begin{align}
    &\left. \partial_{\theta} \langle \mathcal{J}_{w}\rangle_\tau^{\theta} \right|_{\theta =0}
    =\tau\int_0^1ds\sum_e w_e(s) \left. \partial_{\theta} \bar{J}_e^{\theta}(s;\tau) \right|_{\theta =0} \notag\\
    &=  \langle\mathcal{J}_{w}^\mathrm{ex}\rangle_\tau
    +\tau\int_0^1ds\sum_e w_e(s)\tau\partial_\tau\bar{J}_e(s;\tau) \notag\\
    &= \langle\mathcal{J}_{w}^\mathrm{ex}\rangle_\tau + 
    \tau\partial_\tau\langle\mathcal{J}_{w}\rangle_\tau
    -
    \langle\mathcal{J}_{w} \rangle_\tau  \notag\\
    &=\tau\partial_\tau\langle\mathcal{J}_{w}\rangle_\tau  -
    \langle\mathcal{J}_{w}^{\rm hk} \rangle_\tau
    \label{eq:jdtheta}
\end{align}
Therefore, equation~\eqref{eq:frifinitetime} finally leads to the TUR
\begin{align}
    \frac{(\tau\partial_\tau\langle\mathcal{J}_{w}\rangle_\tau-
    \langle\mathcal{J}_{w}^\mathrm{hk}\rangle_\tau)^2}{\mathrm{Var}(\mathcal{J}_{w})}
    \leq \frac{1}{2}\int_0^\tau dt\; \sigma^\mathrm{ex}. 
\end{align}

\section{Derivation of Eq.~\eqref{eq:generalwasserstein}}
\label{app:wasserstein}
We prove that $\mathcal{W}$ provided in Eq.~\eqref{eq:generalwasserstein2} is also given by Eq.~\eqref{eq:generalwasserstein}. 
Let us introduce the functional 
\begin{widetext}
\begin{align}
    \mathcal{I}[x,f,\psi]
    =\int_0^\mathcal{T} dt\pqty{\frac{1}{2}\|f(t)\|_{L(x(t))}^2
    +\psi(t)^\transpose\bqty{d_t x(t)-\mathbb{S}L(x(t))f(t)}}.
\end{align}
With $\psi$ being a Lagrange multiplier,
equation~\eqref{eq:generalwasserstein2} is rewritten as
\begin{align}
    \frac{1}{2\mathcal{T}}\mathcal{W}(x^{(0)},x^{(1)})^2
    =\inf_{x,f}\sup_\psi \mathcal{I}[x,f,\psi]. 
\end{align}
We assume that there exists a triple $(x^*,f^*,\psi^*)$ that provides the optimal value of the right-hand side. Then, it satisfies 
\begin{align}
    \frac{\delta \mathcal{I}}{\delta f_e(t)}
    =\ell_e(x^*(t))f^*_e(t)-[\psi^*(t)^\transpose \mathbb{S}]_e\ell_e(x^*(t))=0,
\end{align}
so we have
\begin{align}
    f^*(t)=\mathbb{S}^\transpose\psi^*(t). 
\end{align}
Therefore, the optimal force is conservative, and it is sufficient to consider conservative forces when we calculate the Wasserstein distance. 

\section{Derivation of Eq.~\eqref{eq:wasscsineqopposite}}
\label{app:reparametrization}
Here we prove the inequality in Eq.~\eqref{eq:wasscsineqopposite} by tracing the proof of Theorem 5.4 in Ref.~\cite{dolbeault2009new}. 
Let $\{x^*,\psi^*\}$ be a minimizer of the Wasserstein distance~\eqref{eq:generalwasserstein}. 
We define a function with sufficiently small $\epsilon>0$ as
\begin{align}
    \tilde{s}_\epsilon(t):=\int_0^t dt'\;(\epsilon+\|\mathbb{S}^\transpose\psi^*(t')\|_{L(x^*(t'))}^2)^{1/2}.
\end{align}
Then, $d_t\tilde{s}_\epsilon(t)>\epsilon$, and its inverse $\tilde{t}_\epsilon=\tilde{s}_\epsilon^{-1}:[0,\tilde{s}_\epsilon(\mathcal{T})]\to[0,\mathcal{T}]$ is well defined; thus, $\tilde{s}_\epsilon$ works as a time coordinate. 
The functions satisfy
\begin{align}
    \frac{d\tilde{t}_\epsilon}{ds}(\tilde{s}_\epsilon(t))=\pqty{\frac{d\tilde{s}_\epsilon}{dt}(t)}^{-1}=\frac{1}{(\epsilon+\|\mathbb{S}^\transpose\psi^*(t)\|_{L(x^*(t))}^2)^{1/2}}.
\end{align}
If we change the time variable as $\tilde{x}(s):=x^*(\tilde{t}_\epsilon(s))$ and $\tilde{\psi}(s):=(d_s\tilde{t}_\epsilon(s))\psi^*(\tilde{t}_\epsilon(s))$, we find
\begin{align}
    d_s\tilde{x}(s)
    =d_s\tilde{t}_\epsilon(s)d_tx^*(\tilde{t}_\epsilon(s))
    =d_s\tilde{t}_\epsilon(s)\mathbb{S}L(\tilde{x}(s))\mathbb{S}^\transpose\psi^*(\tilde{t}_\epsilon(s))
    =\mathbb{S}L(\tilde{x}(s))\mathbb{S}^\transpose\tilde{\psi}(s),
\end{align}
so that $(\tilde{x},\tilde{\psi})$ falls in the feasible set of the minimization~\eqref{eq:generalwasserstein} with time duration $\tilde{s}_\epsilon(\mathcal{T})$. Therefore, we have 
\begin{align}
    \mathcal{W}(x^{(0)},x^{(1)})^2
    &\leq \tilde{s}_\epsilon(\mathcal{T})
    \int_0^{\tilde{s}_\epsilon(\mathcal{T})}ds\; \|\mathbb{S}^\transpose\tilde{\psi}(s)\|_{L(\tilde{x}(s))}^2\notag\\
    &=\tilde{s}_\epsilon(\mathcal{T})
    \int_0^\mathcal{T} dt\; \frac{d\tilde{s}_\epsilon}{dt}(t)
    \times \pqty{\frac{d\tilde{t}_\epsilon}{ds}(\tilde{s}_\epsilon(t))}^2
    \|\mathbb{S}^\transpose\psi^*(t)\|_{L(x^*(t))}^2\notag\\
    &=\tilde{s}_\epsilon(\mathcal{T})
    \int_0^\mathcal{T} dt\;
    \frac{\|\mathbb{S}^\transpose\psi^*(t)\|_{L(x^*(t))}^2}{\epsilon+\|\mathbb{S}^\transpose\psi^*(t)\|_{L(x^*(t))}^2}
    (\epsilon+\|\mathbb{S}^\transpose\psi^*(t)\|_{L(x^*(t))}^2)^{1/2}\notag\\
    &\leq \tilde{s}_\epsilon(\mathcal{T})^2.
\end{align}
By taking the limit $\epsilon\to 0$, we obtain the inequality 
\begin{align}
    \mathcal{W}(x^{(0)},x^{(1)})\leq 
    \tilde{s}_0(\mathcal{T})
    =\int_0^\mathcal{T} dt\;\|\mathbb{S}^\transpose\psi^*(t)\|_{L(x^*(t))}. 
\end{align}

\section{Derivation of Eq.~\eqref{eq:dual}}
\label{app:dual}
We give a proof of the duality formula~\eqref{eq:dual}. 
We first prepare some relations regarding the functional $\mathcal{I}$ defined in Appendix~\ref{app:wasserstein}. 
Here, we concentrate on probability distributions, so the argument of the functional is $(p,f,\psi)$. 
We again assume there exists an optimal solution $(p^*,f^*,\psi^*)$ that provides the Wasserstein distance as
\begin{align}
    \frac{1}{2\mathcal{T}}\mathcal{W}(p^{(0)},p^{(1)})^2
    =\mathcal{I}[p^*,f^*,\psi^*]
    =\frac{1}{2}\int_0^\mathcal{T} dt\;\|B^\transpose \psi^*(t)\|^2_{L(p^*(t))},
\end{align}
where we used the relation $f^*(t)=B^\transpose\psi^*(t)$, which was derived in Appendix~\ref{app:wasserstein}. 
Consider the functional derivative 
\begin{align}
    \frac{\delta\mathcal{I}}{\delta p_i(t)}
    &=\frac{1}{2}\sum_e\frac{\partial\ell_e}{\partial p_i}(p(t))|f_e(t)|^2
    -d_t\psi_i(t)-\frac{\partial}{\partial p_i}\psi^\transpose(t)BL(p(t))f(t). 
\end{align}
It vanishes when $(p,f,\psi)=(p^*,f^*,\psi^*)$, so we have 
\begin{align}
    d_t\psi_i^*(t)+\frac{1}{2}\frac{\partial}{\partial p_i}\|f^*(t)\|_{L(p^*(t))}^2=0. \label{eq:appccond}
\end{align}
Combining it with $f^*(t)=B^\transpose\psi^*(t)$, we find the equation
\begin{align}
    d_t\psi_i^*(t)+\frac{1}{2}\frac{\partial}{\partial p_i}\|B^\transpose \psi^*(t)\|_{L(p^*(t))}^2=0.
    \label{eq:appccond2}
\end{align}

We next show $\psi^*$ gives the distance by the duality formula. 
Because of the equality for the log mean 
\begin{align}
    a\frac{\partial}{\partial a}\mathbb{L}(a,b)+b\frac{\partial}{\partial b}\mathbb{L}(a,b)=\mathbb{L}(a,b)
\end{align}
shown in Ref.~\cite{erbar2012ricci}, we have the relation
\begin{align}
    \sum_i p_i\frac{\partial}{\partial p_i}\ell_e(p)
    =W_{e}p_{\iota(e)}
    \frac{\partial}{\partial u_1}\mathbb{L}(W_ep_{\iota(e)},W_{-e}p_{\iota'(e)})
    +W_{-e}p_{\iota'(e)}
    \frac{\partial}{\partial u_2}\mathbb{L}(W_ep_{\iota(e)},W_{-e}p_{\iota'(e)})
    =\ell_e(p),
\end{align}
where $u_1$ and $u_2$ denote the first and second argument of $\mathbb{L}$. 
Combining it with equation~\eqref{eq:appccond2}, we obtain the Hamilton--Jacobi equation
\begin{align}
    p^*(t)^\transpose d_t\psi^*(t)+\frac{1}{2}\|B^\transpose\psi^*(t)\|_{L(p^*(t))}^2=0.
\end{align}
Then, the Wasserstein distance becomes
\begin{align}
    \frac{1}{\mathcal{T}}\mathcal{W}(p^{(0)},p^{(1)})^2&=\int_0^\mathcal{T} dt\;\|B^\transpose \psi^*(t)\|_{L(p^*(t))}^2=\int_0^\mathcal{T} dt\; \psi^*(t)^\transpose d_tp^*(t)\notag\\
    &=\langle\psi^*(\mathcal{T})\rangle_{1} -\langle\psi^*(0)\rangle_{0}
    -\int_0^\mathcal{T} dt\; (d_t\psi^*(t))^\transpose p^*(t)\notag\\
    &=\langle\psi^*(\mathcal{T})\rangle_{1} -\langle\psi^*(0)\rangle_{0}
    +\frac{1}{2}\int_0^\mathcal{T} dt\;\|B^\transpose \psi^*(t)\|_{L(p^*(t))}^2,
\end{align}
where we used the continuity equation $d_tp^*=BL(p)B^\transpose\psi^*$ in the second equality of the first line, and did the integration by parts in the second line. 
Recall that $\langle\cdot\rangle_{i}$ indicates the expectation value under $p^{(i)}$. 
Therefore, we have the formula
\begin{align}
    \frac{1}{2\mathcal{T}}\mathcal{W}(p^{(0)},p^{(1)})^2
    =\langle\psi^*(\mathcal{T})\rangle_{1} -\langle\psi^*(0)\rangle_{0}, 
\end{align}

Finally, we prove that $\psi^*$ provides the maximum under the condition in Eq.~\eqref{eq:hjsubeq} (also shown in Eq.~\eqref{eq:appchjsubeq} below). 
To show that $\psi^*$ satisfies the condition, let $g(p):=p^\transpose d_t\psi^*(t)+(1/2)\|B^\transpose\psi^*(t)\|_{L(p)}^2$. Then, 
\begin{align}
    \frac{\partial g}{\partial p_i}(p^*)=d_t\psi_i^*+\frac{1}{2}\frac{\partial}{\partial p_i}\|B^\transpose\psi^*\|_{L(p^*)}^2,
\end{align}
and it is zero because $\psi^*$ satisfies Eq.~\eqref{eq:appccond2}. 
As we will show later, since $\ell_e(p)$ is a concave function, $g(p)$ is also concave, so $p^*$ gives the maximum value of $g(p)$~\cite{rockafellar2009variational}.
Therefore, we conclude $g(p)\leq 0$ and $\psi^*$ satisfies the condition. 
Let $\psi'$ also satisfy the condition. Then, 
\begin{align}
    \langle\psi'(\mathcal{T})\rangle_{1} -\langle\psi'(0)\rangle_{0}
    &=\int_0^\mathcal{T} dt\;d_t[(p^*)^\transpose\psi']\notag\\
    &=\int_0^\mathcal{T} dt\;\pqty{(\psi')^\transpose BL(p^*)B^\transpose\psi^*
    +(p^*)^\transpose d_t\psi'}\notag\\
    &\leq\int_0^\mathcal{T} dt\;\pqty{\langle B^\transpose\psi',B^\transpose\psi^*\rangle_{L(p^*)}-\frac{1}{2}\|B^\transpose\psi'\|_{L(p^*)}^2} \notag\\
    &=\int_0^\mathcal{T} dt\;\pqty{\frac{1}{2}\|B^\transpose\psi^*\|_{L(p^*)}^2
    -\frac{1}{2}\|B^\transpose(\psi^*-\psi')\|_{L(p^*)}^2}\notag\\
    &\leq 
    \frac{1}{2}\int_0^\mathcal{T} dt\;\|B^\transpose\psi^*\|_{L(p^*)}^2
    =
    \langle\psi^*(\mathcal{T})\rangle_{1} -\langle\psi^*(0)\rangle_{0},
\end{align}
where the third line follows from the condition~\eqref{eq:hjsubeq}, and the fourth line is derived from a general property of an inner product. 
Therefore, we obtain the expression
\begin{gather}
    \frac{1}{2\mathcal{T}}\mathcal{W}(p^{(0)},p^{(1)})^2
    =\sup_{\psi}(\langle\psi(\mathcal{T})\rangle_{1} -\langle\psi(0)\rangle_{0})\\
    \mathrm{s.t.}\quad
    q^\transpose d_t\psi(t)+\frac{1}{2}\|B^\transpose\psi(t)\|_{L(q)}^2\leq 0\quad
    \mathrm{for\;any\;probability\;distribution}\;q.\label{eq:appchjsubeq}
\end{gather}

The concavity of $\ell_e(p)$ follows from the concavity of the log mean because the map $p\mapsto (W_{e}p_{\iota(e)}, W_{-e}p_{\iota'(e)})$ is linear. 
Let us prove that the log mean is concave. 
The log mean has an integral form
\begin{align}
    \mathbb{L}(x,y)=\int_0^1 dt\;x^{t}y^{1-t}.
\end{align}
Then, the Hessian $H$ is given as
\begin{align}
    H_{xx}&=\frac{\partial^2}{\partial x^2}\mathbb{L}(x,y)
    =\int_0^1 dt\;t(t-1)x^{t-2}y^{1-t}\\
    H_{yy}&=\frac{\partial^2}{\partial y^2}\mathbb{L}(x,y)
    =\int_0^1 dt\;t(t-1)x^{t}y^{-1-t}\\
    H_{xy}=H_{yx}&=\frac{\partial^2}{\partial x\partial y}\mathbb{L}(x,y)
    =\int_0^1 dt\;t(1-t)x^{t-1}y^{-t}.
\end{align}
The determinant of the Hessian is positive because
\begin{align}
    \mathrm{det}\,H&=H_{xx}H_{yy}-H_{xy}^2\notag\\
    &=\int_0^1 dt\;t(t-1)x^{t-2}y^{1-t}
    \int_0^1 dt\;t(t-1)x^{t}y^{-1-t}
    -\pqty{\int_0^1 dt\;t(t-1)x^{t-1}y^{-t}}^2\notag\\
    &=\int_0^1 dt\;t(1-t)x^{t-2}y^{1-t}
    \int_0^1 dt\;t(1-t)x^{t}y^{-1-t}
    -\pqty{\int_0^1 dt\;\sqrt{t(1-t)x^{t-2}y^{-1-t}}\sqrt{t(1-t)x^{t}y^{1-t}}}^2\notag\\
    &\geq 0,
\end{align}
where we used the Cauchy--Schwarz inequality in the last line.
The trace is shown to be negative because $t(t-1)\leq 0$ and $x,y$ are positive by definition, so $H_{xx}$ and $H_{yy}$ are negative. 
Therefore, the Hessian $H$ is negative semidefinite and the log mean is concave.

\section{Derivation of Eq.~\eqref{eq:l1l2ineq}}
\label{app:l1l2ineq}
In this section, we provide the proof of the inequality in Eq.~\eqref{eq:l1l2ineq}. 
First, we characterize the $L^2$-Wasserstein distance by a functional slightly different from what we used in the previous sections. Next, we show relations between the $L^2$- and $L^1$-Wasserstein distance by using the derived expression. 

Let $\mathcal{F}$ be the functional
\begin{align}
    \mathcal{F}[q,\psi]
    =\int_0^\tau dt\;\pqty{
    -\|B^\transpose\psi(t)\|_{L(q(t))}^2
    +2\psi(t)^\transpose d_t q(t)}.
\end{align}
Completing the square shows that this functional is maximized with respect to $\psi$ when $d_tq(t)=BL(q(t))B^\transpose\psi(t)$, and then 
\begin{align}
    \mathcal{F}'[q]:=\sup_\psi \mathcal{F}[q,\psi]
    =\int_0^\tau dt\;\|B^\transpose\psi(t)\|_{L(q(t))}^2. 
\end{align}
If we choose a solution of the master equation $\hat{p}$ as $q$, this amounts to the Onsager excess EP, $\mathcal{F}'[\hat{p}]=\int_0^\tau dt\,\sigma^\mathrm{ex}$. 
The $L^2$-Wasserstein distance is given by
\begin{align}
    \mathcal{W}_2(\hat{p}(0),\hat{p}(\tau))^2
    =\tau\inf_q\sup_\psi \mathcal{F}[q,\psi]
\end{align}
with $q$ satisfying the initial and final conditions, $q(0)=\hat{p}(0)$ and $q(\tau)=\hat{p}(\tau)$.
We write the minimizer as $p^*$, so that $\mathcal{W}_2(\hat{p}(0),\hat{p}(\tau))^2=\tau\mathcal{F}'[p^*]$. 

Let $(q,\psi)$ satisfy the continuity equation $d_tq=BL(q)B^\transpose \psi$, so that $\mathcal{F}'[q]=\mathcal{F}[q,\psi]$. 
We now do not assume $q$ is a solution of the master equation. 
The Cauchy--Schwarz inequality leads to
\begin{align}
    \pqty{\int_0^\tau dt\; \varphi(t)^\transpose d_tq(t)}^2
    &=\pqty{\int_0^\tau dt\; \langle B^\transpose\varphi(t),B^\transpose\psi(t)\rangle_{L(q(t))}}^2\notag\\
    &\leq \int_0^\tau dt\; \|B^\transpose\varphi(t)\|_{L(q(t))}^2
    \int_0^\tau dt\; \|B^\transpose\psi(t)\|_{L(q(t))}^2
\end{align}
for an arbitrary $\varphi$. 
Hence, by choosing $\varphi(t)=\xi(t)$, the potential of the duality in Eq.~\eqref{eq:l1kantrovich} between $q(t)$ and $q(t+dt)$ with $dt$ an infinitesimal time interval, we have
\begin{align}
    \mathcal{F}'[q]\geq \frac{\pqty{\int_0^\tau dt\; \xi(t)^\transpose d_tq(t)}^2}{\int_0^\tau dt\; \|B^\transpose\xi(t)\|_{L(q(t))}^2}.
\end{align}
According to the duality~\eqref{eq:l1kantrovich}, if we assume $q(0)=\hat{p}(0)$ and $q(\tau)=\hat{p}(\tau)$, the numerator reads 
\begin{align}
    \int_0^\tau dt\; \xi(t)^\transpose d_tq(t)
    =\int_0^\tau dt\;
    \lim_{\Delta t\to 0} \frac{\mathcal{W}_1(q(t),q(t+\Delta t))}{\Delta t}
    \geq \mathcal{W}_1(\hat{p}(0),\hat{p}(\tau)),
\end{align}
where we used the triangle inequality in the last inequality. 
The denominator is also bounded as
\begin{align}
    \|B^\transpose\xi(t)\|_{L(q(t))}^2
    =\sum_e \ell_e(q(t))(\xi_{\iota'(e)}-\xi_{\iota(e)})^2
    \leq\sum_e\ell_e(q(t))
    \leq \frac{1}{2}\sum_e \chi_e(q(t)),
\end{align}
where we used $|\xi_i-\xi_j|\leq 1$. Recall that $\chi_e(q(t))=k_eq_{\iota(e)}(t)+k_{-e}q_{\iota'(e)}(t)$ is the edgewise dynamical activity and the log mean inequality $\ell_e(q(t))\leq (1/2)\chi_e(q(t))$. Therefore, we obtain
\begin{align}
    \mathcal{F}'[q]\geq
    \frac{\mathcal{W}_1(\hat{p}(0),\hat{p}(\tau))^2}{\tau\bar{\mathcal{A}}[q]/2}
\end{align}
for an arbitrary $q$ with the appropriate initial and final conditions. 
If we choose $q=p^*$, the minimizer of $\mathcal{F}'$ with the initial and final states $(p(0),p(\tau))$, we have the desired inequality 
\begin{align}
    \frac{\bar{\mathcal{A}}[p^*]}{2}
    \mathcal{W}_2(\hat{p}(0),\hat{p}(\tau))^2
    \geq \mathcal{W}_1(\hat{p}(0),\hat{p}(\tau))^2. 
\end{align}
We also find a lower bound for the Onsager excess EP as 
\begin{align}
    \int_0^\tau dt\;\sigma^\mathrm{ex}\geq 
    \frac{\mathcal{W}_1(\hat{p}(0),\hat{p}(\tau))^2}{\tau\bar{\mathcal{A}}[\hat{p}]/2}.
\end{align}
\end{widetext}

% \bibliographystyle{apsrev4-2}
% \bibliography{biblio}

\begin{thebibliography}{73}%
\makeatletter
\providecommand \@ifxundefined [1]{%
 \@ifx{#1\undefined}
}%
\providecommand \@ifnum [1]{%
 \ifnum #1\expandafter \@firstoftwo
 \else \expandafter \@secondoftwo
 \fi
}%
\providecommand \@ifx [1]{%
 \ifx #1\expandafter \@firstoftwo
 \else \expandafter \@secondoftwo
 \fi
}%
\providecommand \natexlab [1]{#1}%
\providecommand \enquote  [1]{``#1''}%
\providecommand \bibnamefont  [1]{#1}%
\providecommand \bibfnamefont [1]{#1}%
\providecommand \citenamefont [1]{#1}%
\providecommand \href@noop [0]{\@secondoftwo}%
\providecommand \href [0]{\begingroup \@sanitize@url \@href}%
\providecommand \@href[1]{\@@startlink{#1}\@@href}%
\providecommand \@@href[1]{\endgroup#1\@@endlink}%
\providecommand \@sanitize@url [0]{\catcode `\\12\catcode `\$12\catcode
  `\&12\catcode `\#12\catcode `\^12\catcode `\_12\catcode `\%12\relax}%
\providecommand \@@startlink[1]{}%
\providecommand \@@endlink[0]{}%
\providecommand \url  [0]{\begingroup\@sanitize@url \@url }%
\providecommand \@url [1]{\endgroup\@href {#1}{\urlprefix }}%
\providecommand \urlprefix  [0]{URL }%
\providecommand \Eprint [0]{\href }%
\providecommand \doibase [0]{https://doi.org/}%
\providecommand \selectlanguage [0]{\@gobble}%
\providecommand \bibinfo  [0]{\@secondoftwo}%
\providecommand \bibfield  [0]{\@secondoftwo}%
\providecommand \translation [1]{[#1]}%
\providecommand \BibitemOpen [0]{}%
\providecommand \bibitemStop [0]{}%
\providecommand \bibitemNoStop [0]{.\EOS\space}%
\providecommand \EOS [0]{\spacefactor3000\relax}%
\providecommand \BibitemShut  [1]{\csname bibitem#1\endcsname}%
\let\auto@bib@innerbib\@empty
%</preamble>
\bibitem [{\citenamefont {Oono}\ and\ \citenamefont
  {Paniconi}(1998)}]{oono1998steady}%
  \BibitemOpen
  \bibfield  {author} {\bibinfo {author} {\bibfnamefont {Y.}~\bibnamefont
  {Oono}}\ and\ \bibinfo {author} {\bibfnamefont {M.}~\bibnamefont
  {Paniconi}},\ }\bibfield  {title} {\bibinfo {title} {Steady state
  thermodynamics},\ }\href@noop {} {\bibfield  {journal} {\bibinfo  {journal}
  {Progress of Theoretical Physics Supplement}\ }\textbf {\bibinfo {volume}
  {130}},\ \bibinfo {pages} {29} (\bibinfo {year} {1998})}\BibitemShut
  {NoStop}%
\bibitem [{\citenamefont {Hatano}\ and\ \citenamefont
  {Sasa}(2001)}]{hatano2001steady}%
  \BibitemOpen
  \bibfield  {author} {\bibinfo {author} {\bibfnamefont {T.}~\bibnamefont
  {Hatano}}\ and\ \bibinfo {author} {\bibfnamefont {S.-i.}\ \bibnamefont
  {Sasa}},\ }\bibfield  {title} {\bibinfo {title} {Steady-state thermodynamics
  of langevin systems},\ }\href@noop {} {\bibfield  {journal} {\bibinfo
  {journal} {Phys. Rev. Lett.}\ }\textbf {\bibinfo {volume} {86}},\ \bibinfo
  {pages} {3463} (\bibinfo {year} {2001})}\BibitemShut {NoStop}%
\bibitem [{\citenamefont {Esposito}\ and\ \citenamefont {Van~den
  Broeck}(2010)}]{esposito2010three}%
  \BibitemOpen
  \bibfield  {author} {\bibinfo {author} {\bibfnamefont {M.}~\bibnamefont
  {Esposito}}\ and\ \bibinfo {author} {\bibfnamefont {C.}~\bibnamefont {Van~den
  Broeck}},\ }\bibfield  {title} {\bibinfo {title} {Three faces of the second
  law. {I}. master equation formulation},\ }\href@noop {} {\bibfield  {journal}
  {\bibinfo  {journal} {Phys. Rev. E}\ }\textbf {\bibinfo {volume} {82}},\
  \bibinfo {pages} {011143} (\bibinfo {year} {2010})}\BibitemShut {NoStop}%
\bibitem [{\citenamefont {Maes}\ and\ \citenamefont
  {Neto{\v{c}}n{\`y}}(2014)}]{maes2014nonequilibrium}%
  \BibitemOpen
  \bibfield  {author} {\bibinfo {author} {\bibfnamefont {C.}~\bibnamefont
  {Maes}}\ and\ \bibinfo {author} {\bibfnamefont {K.}~\bibnamefont
  {Neto{\v{c}}n{\`y}}},\ }\bibfield  {title} {\bibinfo {title} {A
  nonequilibrium extension of the {C}lausius heat theorem},\ }\href@noop {}
  {\bibfield  {journal} {\bibinfo  {journal} {J. Stat. Phys.}\ }\textbf
  {\bibinfo {volume} {154}},\ \bibinfo {pages} {188} (\bibinfo {year}
  {2014})}\BibitemShut {NoStop}%
\bibitem [{\citenamefont {Rao}\ and\ \citenamefont
  {Esposito}(2016)}]{rao2016nonequilibrium}%
  \BibitemOpen
  \bibfield  {author} {\bibinfo {author} {\bibfnamefont {R.}~\bibnamefont
  {Rao}}\ and\ \bibinfo {author} {\bibfnamefont {M.}~\bibnamefont {Esposito}},\
  }\bibfield  {title} {\bibinfo {title} {Nonequilibrium thermodynamics of
  chemical reaction networks: wisdom from stochastic thermodynamics},\
  }\href@noop {} {\bibfield  {journal} {\bibinfo  {journal} {Phys. Rev. X}\
  }\textbf {\bibinfo {volume} {6}},\ \bibinfo {pages} {041064} (\bibinfo {year}
  {2016})}\BibitemShut {NoStop}%
\bibitem [{\citenamefont {Ge}\ and\ \citenamefont
  {Qian}(2016)}]{ge2016nonequilibrium}%
  \BibitemOpen
  \bibfield  {author} {\bibinfo {author} {\bibfnamefont {H.}~\bibnamefont
  {Ge}}\ and\ \bibinfo {author} {\bibfnamefont {H.}~\bibnamefont {Qian}},\
  }\bibfield  {title} {\bibinfo {title} {Nonequilibrium thermodynamic formalism
  of nonlinear chemical reaction systems with waage--guldberg’s law of mass
  action},\ }\href@noop {} {\bibfield  {journal} {\bibinfo  {journal} {Chem.
  Phys.}\ }\textbf {\bibinfo {volume} {472}},\ \bibinfo {pages} {241} (\bibinfo
  {year} {2016})}\BibitemShut {NoStop}%
\bibitem [{Note1()}]{Note1}%
  \BibitemOpen
  \bibinfo {note} {More precisely, systems with complex balanced steady states
  are globally stable within each stoichiometric compatibility class~\cite
  {horn1972general,craciun2015toric,anderson2011proof}.}\BibitemShut {Stop}%
\bibitem [{\citenamefont {Feinberg}(2019)}]{feinberg2019foundations}%
  \BibitemOpen
  \bibfield  {author} {\bibinfo {author} {\bibfnamefont {M.}~\bibnamefont
  {Feinberg}},\ }\href@noop {} {\emph {\bibinfo {title} {Foundations of
  chemical reaction network theory}}}\ (\bibinfo  {publisher} {Springer},\
  \bibinfo {year} {2019})\BibitemShut {NoStop}%
\bibitem [{\citenamefont {Beard}\ and\ \citenamefont
  {Qian}(2008)}]{beard2008chemical}%
  \BibitemOpen
  \bibfield  {author} {\bibinfo {author} {\bibfnamefont {D.~A.}\ \bibnamefont
  {Beard}}\ and\ \bibinfo {author} {\bibfnamefont {H.}~\bibnamefont {Qian}},\
  }\href@noop {} {\emph {\bibinfo {title} {Chemical biophysics: quantitative
  analysis of cellular systems}}},\ Vol.\ \bibinfo {volume} {126}\ (\bibinfo
  {publisher} {Cambridge University Press Cambridge},\ \bibinfo {year}
  {2008})\BibitemShut {NoStop}%
\bibitem [{\citenamefont {Schnakenberg}(1976)}]{schnakenberg1976network}%
  \BibitemOpen
  \bibfield  {author} {\bibinfo {author} {\bibfnamefont {J.}~\bibnamefont
  {Schnakenberg}},\ }\bibfield  {title} {\bibinfo {title} {Network theory of
  microscopic and macroscopic behavior of master equation systems},\
  }\href@noop {} {\bibfield  {journal} {\bibinfo  {journal} {Rev. Mod. Phys.}\
  }\textbf {\bibinfo {volume} {48}},\ \bibinfo {pages} {571} (\bibinfo {year}
  {1976})}\BibitemShut {NoStop}%
\bibitem [{\citenamefont {Mielke}(2011)}]{mielke2011gradient}%
  \BibitemOpen
  \bibfield  {author} {\bibinfo {author} {\bibfnamefont {A.}~\bibnamefont
  {Mielke}},\ }\bibfield  {title} {\bibinfo {title} {A gradient structure for
  reaction--diffusion systems and for energy-drift-diffusion systems},\
  }\href@noop {} {\bibfield  {journal} {\bibinfo  {journal} {Nonlinearity}\
  }\textbf {\bibinfo {volume} {24}},\ \bibinfo {pages} {1329} (\bibinfo {year}
  {2011})}\BibitemShut {NoStop}%
\bibitem [{\citenamefont {Mielke}(2013)}]{mielke2013geodesic}%
  \BibitemOpen
  \bibfield  {author} {\bibinfo {author} {\bibfnamefont {A.}~\bibnamefont
  {Mielke}},\ }\bibfield  {title} {\bibinfo {title} {Geodesic convexity of the
  relative entropy in reversible {M}arkov chains},\ }\href@noop {} {\bibfield
  {journal} {\bibinfo  {journal} {Calc. Var. Partial Differ. Equ.}\ }\textbf
  {\bibinfo {volume} {48}},\ \bibinfo {pages} {1} (\bibinfo {year}
  {2013})}\BibitemShut {NoStop}%
\bibitem [{\citenamefont {Gingrich}\ \emph {et~al.}(2016)\citenamefont
  {Gingrich}, \citenamefont {Horowitz}, \citenamefont {Perunov},\ and\
  \citenamefont {England}}]{gingrich2016dissipation}%
  \BibitemOpen
  \bibfield  {author} {\bibinfo {author} {\bibfnamefont {T.~R.}\ \bibnamefont
  {Gingrich}}, \bibinfo {author} {\bibfnamefont {J.~M.}\ \bibnamefont
  {Horowitz}}, \bibinfo {author} {\bibfnamefont {N.}~\bibnamefont {Perunov}},\
  and\ \bibinfo {author} {\bibfnamefont {J.~L.}\ \bibnamefont {England}},\
  }\bibfield  {title} {\bibinfo {title} {Dissipation bounds all steady-state
  current fluctuations},\ }\href@noop {} {\bibfield  {journal} {\bibinfo
  {journal} {Phys. Rev. Lett.}\ }\textbf {\bibinfo {volume} {116}},\ \bibinfo
  {pages} {120601} (\bibinfo {year} {2016})}\BibitemShut {NoStop}%
\bibitem [{\citenamefont {Pietzonka}\ \emph {et~al.}(2016)\citenamefont
  {Pietzonka}, \citenamefont {Barato},\ and\ \citenamefont
  {Seifert}}]{pietzonka2016universal}%
  \BibitemOpen
  \bibfield  {author} {\bibinfo {author} {\bibfnamefont {P.}~\bibnamefont
  {Pietzonka}}, \bibinfo {author} {\bibfnamefont {A.~C.}\ \bibnamefont
  {Barato}},\ and\ \bibinfo {author} {\bibfnamefont {U.}~\bibnamefont
  {Seifert}},\ }\bibfield  {title} {\bibinfo {title} {Universal bound on the
  efficiency of molecular motors},\ }\href@noop {} {\bibfield  {journal}
  {\bibinfo  {journal} {J. Stat. Mech.}\ }\textbf {\bibinfo {volume} {2016}},\
  \bibinfo {pages} {124004} (\bibinfo {year} {2016})}\BibitemShut {NoStop}%
\bibitem [{\citenamefont {Horowitz}\ and\ \citenamefont
  {Gingrich}(2017)}]{horowitz2017proof}%
  \BibitemOpen
  \bibfield  {author} {\bibinfo {author} {\bibfnamefont {J.~M.}\ \bibnamefont
  {Horowitz}}\ and\ \bibinfo {author} {\bibfnamefont {T.~R.}\ \bibnamefont
  {Gingrich}},\ }\bibfield  {title} {\bibinfo {title} {Proof of the finite-time
  thermodynamic uncertainty relation for steady-state currents},\ }\href@noop
  {} {\bibfield  {journal} {\bibinfo  {journal} {Phys. Rev. E}\ }\textbf
  {\bibinfo {volume} {96}},\ \bibinfo {pages} {020103} (\bibinfo {year}
  {2017})}\BibitemShut {NoStop}%
\bibitem [{\citenamefont {Proesmans}\ and\ \citenamefont {Van~den
  Broeck}(2017)}]{proesmans2017discrete}%
  \BibitemOpen
  \bibfield  {author} {\bibinfo {author} {\bibfnamefont {K.}~\bibnamefont
  {Proesmans}}\ and\ \bibinfo {author} {\bibfnamefont {C.}~\bibnamefont
  {Van~den Broeck}},\ }\bibfield  {title} {\bibinfo {title} {Discrete-time
  thermodynamic uncertainty relation},\ }\href@noop {} {\bibfield  {journal}
  {\bibinfo  {journal} {Europhys. Lett.}\ }\textbf {\bibinfo {volume} {119}},\
  \bibinfo {pages} {20001} (\bibinfo {year} {2017})}\BibitemShut {NoStop}%
\bibitem [{\citenamefont {Pietzonka}\ \emph {et~al.}(2017)\citenamefont
  {Pietzonka}, \citenamefont {Ritort},\ and\ \citenamefont
  {Seifert}}]{pietzonka2017finite}%
  \BibitemOpen
  \bibfield  {author} {\bibinfo {author} {\bibfnamefont {P.}~\bibnamefont
  {Pietzonka}}, \bibinfo {author} {\bibfnamefont {F.}~\bibnamefont {Ritort}},\
  and\ \bibinfo {author} {\bibfnamefont {U.}~\bibnamefont {Seifert}},\
  }\bibfield  {title} {\bibinfo {title} {Finite-time generalization of the
  thermodynamic uncertainty relation},\ }\href@noop {} {\bibfield  {journal}
  {\bibinfo  {journal} {Phys. Rev. E}\ }\textbf {\bibinfo {volume} {96}},\
  \bibinfo {pages} {012101} (\bibinfo {year} {2017})}\BibitemShut {NoStop}%
\bibitem [{\citenamefont {Dechant}\ and\ \citenamefont
  {Sasa}(2018)}]{dechant2018current}%
  \BibitemOpen
  \bibfield  {author} {\bibinfo {author} {\bibfnamefont {A.}~\bibnamefont
  {Dechant}}\ and\ \bibinfo {author} {\bibfnamefont {S.-i.}\ \bibnamefont
  {Sasa}},\ }\bibfield  {title} {\bibinfo {title} {Current fluctuations and
  transport efficiency for general langevin systems},\ }\href@noop {}
  {\bibfield  {journal} {\bibinfo  {journal} {J. Stat. Mech.}\ }\textbf
  {\bibinfo {volume} {2018}},\ \bibinfo {pages} {063209} (\bibinfo {year}
  {2018})}\BibitemShut {NoStop}%
\bibitem [{\citenamefont {Dechant}(2018)}]{dechant2018multidimensional}%
  \BibitemOpen
  \bibfield  {author} {\bibinfo {author} {\bibfnamefont {A.}~\bibnamefont
  {Dechant}},\ }\bibfield  {title} {\bibinfo {title} {Multidimensional
  thermodynamic uncertainty relations},\ }\href@noop {} {\bibfield  {journal}
  {\bibinfo  {journal} {J. Phys. A}\ }\textbf {\bibinfo {volume} {52}},\
  \bibinfo {pages} {035001} (\bibinfo {year} {2018})}\BibitemShut {NoStop}%
\bibitem [{\citenamefont {Hasegawa}\ and\ \citenamefont
  {Van~Vu}(2019)}]{hasegawa2019fluctuation}%
  \BibitemOpen
  \bibfield  {author} {\bibinfo {author} {\bibfnamefont {Y.}~\bibnamefont
  {Hasegawa}}\ and\ \bibinfo {author} {\bibfnamefont {T.}~\bibnamefont
  {Van~Vu}},\ }\bibfield  {title} {\bibinfo {title} {Fluctuation theorem
  uncertainty relation},\ }\href@noop {} {\bibfield  {journal} {\bibinfo
  {journal} {Phys. Rev. Lett.}\ }\textbf {\bibinfo {volume} {123}},\ \bibinfo
  {pages} {110602} (\bibinfo {year} {2019})}\BibitemShut {NoStop}%
\bibitem [{\citenamefont {Falasco}\ \emph {et~al.}(2020)\citenamefont
  {Falasco}, \citenamefont {Esposito},\ and\ \citenamefont
  {Delvenne}}]{falasco2020unifying}%
  \BibitemOpen
  \bibfield  {author} {\bibinfo {author} {\bibfnamefont {G.}~\bibnamefont
  {Falasco}}, \bibinfo {author} {\bibfnamefont {M.}~\bibnamefont {Esposito}},\
  and\ \bibinfo {author} {\bibfnamefont {J.-C.}\ \bibnamefont {Delvenne}},\
  }\bibfield  {title} {\bibinfo {title} {Unifying thermodynamic uncertainty
  relations},\ }\href@noop {} {\bibfield  {journal} {\bibinfo  {journal} {New
  J. Phys.}\ }\textbf {\bibinfo {volume} {22}},\ \bibinfo {pages} {053046}
  (\bibinfo {year} {2020})}\BibitemShut {NoStop}%
\bibitem [{\citenamefont {Wolpert}(2020)}]{wolpert2020uncertainty}%
  \BibitemOpen
  \bibfield  {author} {\bibinfo {author} {\bibfnamefont {D.~H.}\ \bibnamefont
  {Wolpert}},\ }\bibfield  {title} {\bibinfo {title} {Uncertainty relations and
  fluctuation theorems for {B}ayes nets},\ }\href@noop {} {\bibfield  {journal}
  {\bibinfo  {journal} {Phys. Rev. Lett.}\ }\textbf {\bibinfo {volume} {125}},\
  \bibinfo {pages} {200602} (\bibinfo {year} {2020})}\BibitemShut {NoStop}%
\bibitem [{\citenamefont {Otsubo}\ \emph {et~al.}(2020)\citenamefont {Otsubo},
  \citenamefont {Ito}, \citenamefont {Dechant},\ and\ \citenamefont
  {Sagawa}}]{otsubo2020estimating}%
  \BibitemOpen
  \bibfield  {author} {\bibinfo {author} {\bibfnamefont {S.}~\bibnamefont
  {Otsubo}}, \bibinfo {author} {\bibfnamefont {S.}~\bibnamefont {Ito}},
  \bibinfo {author} {\bibfnamefont {A.}~\bibnamefont {Dechant}},\ and\ \bibinfo
  {author} {\bibfnamefont {T.}~\bibnamefont {Sagawa}},\ }\bibfield  {title}
  {\bibinfo {title} {Estimating entropy production by machine learning of
  short-time fluctuating currents},\ }\href@noop {} {\bibfield  {journal}
  {\bibinfo  {journal} {Phys. Rev. E}\ }\textbf {\bibinfo {volume} {101}},\
  \bibinfo {pages} {062106} (\bibinfo {year} {2020})}\BibitemShut {NoStop}%
\bibitem [{\citenamefont {Manikandan}\ \emph {et~al.}(2020)\citenamefont
  {Manikandan}, \citenamefont {Gupta},\ and\ \citenamefont
  {Krishnamurthy}}]{manikandan2020inferring}%
  \BibitemOpen
  \bibfield  {author} {\bibinfo {author} {\bibfnamefont {S.~K.}\ \bibnamefont
  {Manikandan}}, \bibinfo {author} {\bibfnamefont {D.}~\bibnamefont {Gupta}},\
  and\ \bibinfo {author} {\bibfnamefont {S.}~\bibnamefont {Krishnamurthy}},\
  }\bibfield  {title} {\bibinfo {title} {Inferring entropy production from
  short experiments},\ }\href@noop {} {\bibfield  {journal} {\bibinfo
  {journal} {Phys. Rev. Lett.}\ }\textbf {\bibinfo {volume} {124}},\ \bibinfo
  {pages} {120603} (\bibinfo {year} {2020})}\BibitemShut {NoStop}%
\bibitem [{\citenamefont {Liu}\ \emph {et~al.}(2020)\citenamefont {Liu},
  \citenamefont {Gong},\ and\ \citenamefont {Ueda}}]{liu2020thermodynamic}%
  \BibitemOpen
  \bibfield  {author} {\bibinfo {author} {\bibfnamefont {K.}~\bibnamefont
  {Liu}}, \bibinfo {author} {\bibfnamefont {Z.}~\bibnamefont {Gong}},\ and\
  \bibinfo {author} {\bibfnamefont {M.}~\bibnamefont {Ueda}},\ }\bibfield
  {title} {\bibinfo {title} {Thermodynamic uncertainty relation for arbitrary
  initial states},\ }\href@noop {} {\bibfield  {journal} {\bibinfo  {journal}
  {Phys. Rev. Lett.}\ }\textbf {\bibinfo {volume} {125}},\ \bibinfo {pages}
  {140602} (\bibinfo {year} {2020})}\BibitemShut {NoStop}%
\bibitem [{\citenamefont {Dechant}\ and\ \citenamefont
  {Sasa}(2020)}]{dechant2020fluctuation}%
  \BibitemOpen
  \bibfield  {author} {\bibinfo {author} {\bibfnamefont {A.}~\bibnamefont
  {Dechant}}\ and\ \bibinfo {author} {\bibfnamefont {S.-i.}\ \bibnamefont
  {Sasa}},\ }\bibfield  {title} {\bibinfo {title} {Fluctuation--response
  inequality out of equilibrium},\ }\href@noop {} {\bibfield  {journal}
  {\bibinfo  {journal} {Proceedings of the National Academy of Sciences}\
  }\textbf {\bibinfo {volume} {117}},\ \bibinfo {pages} {6430} (\bibinfo {year}
  {2020})}\BibitemShut {NoStop}%
\bibitem [{\citenamefont {Dechant}\ and\ \citenamefont
  {Sasa}(2021)}]{dechant2021continuous}%
  \BibitemOpen
  \bibfield  {author} {\bibinfo {author} {\bibfnamefont {A.}~\bibnamefont
  {Dechant}}\ and\ \bibinfo {author} {\bibfnamefont {S.-i.}\ \bibnamefont
  {Sasa}},\ }\bibfield  {title} {\bibinfo {title} {Continuous time reversal and
  equality in the thermodynamic uncertainty relation},\ }\href@noop {}
  {\bibfield  {journal} {\bibinfo  {journal} {Phys. Rev. Research}\ }\textbf
  {\bibinfo {volume} {3}},\ \bibinfo {pages} {L042012} (\bibinfo {year}
  {2021})}\BibitemShut {NoStop}%
\bibitem [{\citenamefont {Dechant}\ \emph
  {et~al.}(2022{\natexlab{a}})\citenamefont {Dechant}, \citenamefont {Sasa},\
  and\ \citenamefont {Ito}}]{dechant2022geometric1}%
  \BibitemOpen
  \bibfield  {author} {\bibinfo {author} {\bibfnamefont {A.}~\bibnamefont
  {Dechant}}, \bibinfo {author} {\bibfnamefont {S.-i.}\ \bibnamefont {Sasa}},\
  and\ \bibinfo {author} {\bibfnamefont {S.}~\bibnamefont {Ito}},\ }\bibfield
  {title} {\bibinfo {title} {Geometric decomposition of entropy production in
  out-of-equilibrium systems},\ }\href@noop {} {\bibfield  {journal} {\bibinfo
  {journal} {Phys. Rev. Research}\ }\textbf {\bibinfo {volume} {4}},\ \bibinfo
  {pages} {L012034} (\bibinfo {year} {2022}{\natexlab{a}})}\BibitemShut
  {NoStop}%
\bibitem [{\citenamefont {Dechant}\ \emph
  {et~al.}(2022{\natexlab{b}})\citenamefont {Dechant}, \citenamefont {Sasa},\
  and\ \citenamefont {Ito}}]{dechant2022geometric2}%
  \BibitemOpen
  \bibfield  {author} {\bibinfo {author} {\bibfnamefont {A.}~\bibnamefont
  {Dechant}}, \bibinfo {author} {\bibfnamefont {S.-i.}\ \bibnamefont {Sasa}},\
  and\ \bibinfo {author} {\bibfnamefont {S.}~\bibnamefont {Ito}},\ }\bibfield
  {title} {\bibinfo {title} {Geometric decomposition of entropy production into
  excess, housekeeping, and coupling parts},\ }\href@noop {} {\bibfield
  {journal} {\bibinfo  {journal} {Phys. Rev. E}\ }\textbf {\bibinfo {volume}
  {106}},\ \bibinfo {pages} {024125} (\bibinfo {year}
  {2022}{\natexlab{b}})}\BibitemShut {NoStop}%
\bibitem [{\citenamefont {Yoshimura}\ and\ \citenamefont
  {Ito}(2021)}]{yoshimura2021thermodynamic}%
  \BibitemOpen
  \bibfield  {author} {\bibinfo {author} {\bibfnamefont {K.}~\bibnamefont
  {Yoshimura}}\ and\ \bibinfo {author} {\bibfnamefont {S.}~\bibnamefont
  {Ito}},\ }\bibfield  {title} {\bibinfo {title} {Thermodynamic uncertainty
  relation and thermodynamic speed limit in deterministic chemical reaction
  networks},\ }\href@noop {} {\bibfield  {journal} {\bibinfo  {journal} {Phys.
  Rev. Lett.}\ }\textbf {\bibinfo {volume} {127}},\ \bibinfo {pages} {160601}
  (\bibinfo {year} {2021})}\BibitemShut {NoStop}%
\bibitem [{\citenamefont {Van~Vu}\ and\ \citenamefont
  {Hasegawa}(2021)}]{van2021geometrical}%
  \BibitemOpen
  \bibfield  {author} {\bibinfo {author} {\bibfnamefont {T.}~\bibnamefont
  {Van~Vu}}\ and\ \bibinfo {author} {\bibfnamefont {Y.}~\bibnamefont
  {Hasegawa}},\ }\bibfield  {title} {\bibinfo {title} {Geometrical bounds of
  the irreversibility in {M}arkovian systems},\ }\href@noop {} {\bibfield
  {journal} {\bibinfo  {journal} {Phys. Rev. Lett.}\ }\textbf {\bibinfo
  {volume} {126}},\ \bibinfo {pages} {010601} (\bibinfo {year}
  {2021})}\BibitemShut {NoStop}%
\bibitem [{\citenamefont {Villani}(2009)}]{villani2009optimal}%
  \BibitemOpen
  \bibfield  {author} {\bibinfo {author} {\bibfnamefont {C.}~\bibnamefont
  {Villani}},\ }\href@noop {} {\emph {\bibinfo {title} {Optimal transport: old
  and new}}},\ Vol.\ \bibinfo {volume} {338}\ (\bibinfo  {publisher}
  {Springer},\ \bibinfo {year} {2009})\BibitemShut {NoStop}%
\bibitem [{\citenamefont {Aurell}\ \emph {et~al.}(2011)\citenamefont {Aurell},
  \citenamefont {Mej{\'\i}a-Monasterio},\ and\ \citenamefont
  {Muratore-Ginanneschi}}]{aurell2011optimal}%
  \BibitemOpen
  \bibfield  {author} {\bibinfo {author} {\bibfnamefont {E.}~\bibnamefont
  {Aurell}}, \bibinfo {author} {\bibfnamefont {C.}~\bibnamefont
  {Mej{\'\i}a-Monasterio}},\ and\ \bibinfo {author} {\bibfnamefont
  {P.}~\bibnamefont {Muratore-Ginanneschi}},\ }\bibfield  {title} {\bibinfo
  {title} {Optimal protocols and optimal transport in stochastic
  thermodynamics},\ }\href@noop {} {\bibfield  {journal} {\bibinfo  {journal}
  {Phys. Rev. Lett.}\ }\textbf {\bibinfo {volume} {106}},\ \bibinfo {pages}
  {250601} (\bibinfo {year} {2011})}\BibitemShut {NoStop}%
\bibitem [{\citenamefont {Aurell}\ \emph {et~al.}(2012)\citenamefont {Aurell},
  \citenamefont {Gaw\c{e}dzki}, \citenamefont {Mej\'{\i}a-Monasterio},
  \citenamefont {Mohayaee},\ and\ \citenamefont
  {Muratore-Ginanneschi}}]{aurell2012refined}%
  \BibitemOpen
  \bibfield  {author} {\bibinfo {author} {\bibfnamefont {E.}~\bibnamefont
  {Aurell}}, \bibinfo {author} {\bibfnamefont {K.}~\bibnamefont
  {Gaw\c{e}dzki}}, \bibinfo {author} {\bibfnamefont {C.}~\bibnamefont
  {Mej\'{\i}a-Monasterio}}, \bibinfo {author} {\bibfnamefont {R.}~\bibnamefont
  {Mohayaee}},\ and\ \bibinfo {author} {\bibfnamefont {P.}~\bibnamefont
  {Muratore-Ginanneschi}},\ }\bibfield  {title} {\bibinfo {title} {Refined
  second law of thermodynamics for fast random processes},\ }\href@noop {}
  {\bibfield  {journal} {\bibinfo  {journal} {J. Stat. Phys.}\ }\textbf
  {\bibinfo {volume} {147}} (\bibinfo {year} {2012})}\BibitemShut {NoStop}%
\bibitem [{\citenamefont {Bo}\ \emph {et~al.}(2013)\citenamefont {Bo},
  \citenamefont {Aurell}, \citenamefont {Eichhorn},\ and\ \citenamefont
  {Celani}}]{bo2013optimal}%
  \BibitemOpen
  \bibfield  {author} {\bibinfo {author} {\bibfnamefont {S.}~\bibnamefont
  {Bo}}, \bibinfo {author} {\bibfnamefont {E.}~\bibnamefont {Aurell}}, \bibinfo
  {author} {\bibfnamefont {R.}~\bibnamefont {Eichhorn}},\ and\ \bibinfo
  {author} {\bibfnamefont {A.}~\bibnamefont {Celani}},\ }\bibfield  {title}
  {\bibinfo {title} {Optimal stochastic transport in inhomogeneous thermal
  environments},\ }\href@noop {} {\bibfield  {journal} {\bibinfo  {journal}
  {Europhys. Lett.}\ }\textbf {\bibinfo {volume} {103}},\ \bibinfo {pages}
  {10010} (\bibinfo {year} {2013})}\BibitemShut {NoStop}%
\bibitem [{\citenamefont {Dechant}\ and\ \citenamefont
  {Sakurai}(2019)}]{dechant2019thermodynamic}%
  \BibitemOpen
  \bibfield  {author} {\bibinfo {author} {\bibfnamefont {A.}~\bibnamefont
  {Dechant}}\ and\ \bibinfo {author} {\bibfnamefont {Y.}~\bibnamefont
  {Sakurai}},\ }\bibfield  {title} {\bibinfo {title} {{Thermodynamic
  interpretation of {W}asserstein distance}}} (\bibinfo {year}
  {2019})\BibitemShut {NoStop}%
\bibitem [{\citenamefont {Chen}\ \emph {et~al.}(2019)\citenamefont {Chen},
  \citenamefont {Georgiou},\ and\ \citenamefont
  {Tannenbaum}}]{chen2019stochastic}%
  \BibitemOpen
  \bibfield  {author} {\bibinfo {author} {\bibfnamefont {Y.}~\bibnamefont
  {Chen}}, \bibinfo {author} {\bibfnamefont {T.~T.}\ \bibnamefont {Georgiou}},\
  and\ \bibinfo {author} {\bibfnamefont {A.}~\bibnamefont {Tannenbaum}},\
  }\bibfield  {title} {\bibinfo {title} {Stochastic control and nonequilibrium
  thermodynamics: Fundamental limits},\ }\href@noop {} {\bibfield  {journal}
  {\bibinfo  {journal} {IEEE Trans. Automat. Contr.}\ }\textbf {\bibinfo
  {volume} {65}},\ \bibinfo {pages} {2979} (\bibinfo {year}
  {2019})}\BibitemShut {NoStop}%
\bibitem [{\citenamefont {Fu}\ \emph {et~al.}(2021)\citenamefont {Fu},
  \citenamefont {Taghvaei}, \citenamefont {Chen},\ and\ \citenamefont
  {Georgiou}}]{fu2021maximal}%
  \BibitemOpen
  \bibfield  {author} {\bibinfo {author} {\bibfnamefont {R.}~\bibnamefont
  {Fu}}, \bibinfo {author} {\bibfnamefont {A.}~\bibnamefont {Taghvaei}},
  \bibinfo {author} {\bibfnamefont {Y.}~\bibnamefont {Chen}},\ and\ \bibinfo
  {author} {\bibfnamefont {T.~T.}\ \bibnamefont {Georgiou}},\ }\bibfield
  {title} {\bibinfo {title} {Maximal power output of a stochastic thermodynamic
  engine},\ }\href@noop {} {\bibfield  {journal} {\bibinfo  {journal}
  {Automatica}\ }\textbf {\bibinfo {volume} {123}},\ \bibinfo {pages} {109366}
  (\bibinfo {year} {2021})}\BibitemShut {NoStop}%
\bibitem [{\citenamefont {Nakazato}\ and\ \citenamefont
  {Ito}(2021)}]{nakazato2021geometrical}%
  \BibitemOpen
  \bibfield  {author} {\bibinfo {author} {\bibfnamefont {M.}~\bibnamefont
  {Nakazato}}\ and\ \bibinfo {author} {\bibfnamefont {S.}~\bibnamefont {Ito}},\
  }\bibfield  {title} {\bibinfo {title} {Geometrical aspects of entropy
  production in stochastic thermodynamics based on {W}asserstein distance},\
  }\href@noop {} {\bibfield  {journal} {\bibinfo  {journal} {Phys. Rev.
  Research}\ }\textbf {\bibinfo {volume} {3}},\ \bibinfo {pages} {043093}
  (\bibinfo {year} {2021})}\BibitemShut {NoStop}%
\bibitem [{\citenamefont {Miangolarra}\ \emph {et~al.}(2022)\citenamefont
  {Miangolarra}, \citenamefont {Taghvaei}, \citenamefont {Chen},\ and\
  \citenamefont {Georgiou}}]{miangolarra2022geometry}%
  \BibitemOpen
  \bibfield  {author} {\bibinfo {author} {\bibfnamefont {O.~M.}\ \bibnamefont
  {Miangolarra}}, \bibinfo {author} {\bibfnamefont {A.}~\bibnamefont
  {Taghvaei}}, \bibinfo {author} {\bibfnamefont {Y.}~\bibnamefont {Chen}},\
  and\ \bibinfo {author} {\bibfnamefont {T.~T.}\ \bibnamefont {Georgiou}},\
  }\bibfield  {title} {\bibinfo {title} {Geometry of finite-time thermodynamic
  cycles with anisotropic thermal fluctuations},\ }\href@noop {} {\bibfield
  {journal} {\bibinfo  {journal} {arXiv preprint arXiv:2203.12483}\ } (\bibinfo
  {year} {2022})}\BibitemShut {NoStop}%
\bibitem [{\citenamefont {Van~Vu}\ and\ \citenamefont
  {Hasegawa}(2020)}]{van2020unified}%
  \BibitemOpen
  \bibfield  {author} {\bibinfo {author} {\bibfnamefont {T.}~\bibnamefont
  {Van~Vu}}\ and\ \bibinfo {author} {\bibfnamefont {Y.}~\bibnamefont
  {Hasegawa}},\ }\bibfield  {title} {\bibinfo {title} {Unified approach to
  classical speed limit and thermodynamic uncertainty relation},\ }\href@noop
  {} {\bibfield  {journal} {\bibinfo  {journal} {Phys. Rev. E}\ }\textbf
  {\bibinfo {volume} {102}},\ \bibinfo {pages} {062132} (\bibinfo {year}
  {2020})}\BibitemShut {NoStop}%
\bibitem [{\citenamefont {Maas}(2011)}]{maas2011gradient}%
  \BibitemOpen
  \bibfield  {author} {\bibinfo {author} {\bibfnamefont {J.}~\bibnamefont
  {Maas}},\ }\bibfield  {title} {\bibinfo {title} {Gradient flows of the
  entropy for finite {M}arkov chains},\ }\href@noop {} {\bibfield  {journal}
  {\bibinfo  {journal} {J. Funct. Anal.}\ }\textbf {\bibinfo {volume} {261}},\
  \bibinfo {pages} {2250} (\bibinfo {year} {2011})}\BibitemShut {NoStop}%
\bibitem [{\citenamefont {Erbar}\ and\ \citenamefont
  {Maas}(2012)}]{erbar2012ricci}%
  \BibitemOpen
  \bibfield  {author} {\bibinfo {author} {\bibfnamefont {M.}~\bibnamefont
  {Erbar}}\ and\ \bibinfo {author} {\bibfnamefont {J.}~\bibnamefont {Maas}},\
  }\bibfield  {title} {\bibinfo {title} {Ricci curvature of finite {M}arkov
  chains via convexity of the entropy},\ }\href@noop {} {\bibfield  {journal}
  {\bibinfo  {journal} {Arch. Ration. Mech. and Anal.}\ }\textbf {\bibinfo
  {volume} {206}},\ \bibinfo {pages} {997} (\bibinfo {year}
  {2012})}\BibitemShut {NoStop}%
\bibitem [{\citenamefont {Erbar}\ \emph {et~al.}(2019)\citenamefont {Erbar},
  \citenamefont {Maas},\ and\ \citenamefont {Wirth}}]{erbar2019geometry}%
  \BibitemOpen
  \bibfield  {author} {\bibinfo {author} {\bibfnamefont {M.}~\bibnamefont
  {Erbar}}, \bibinfo {author} {\bibfnamefont {J.}~\bibnamefont {Maas}},\ and\
  \bibinfo {author} {\bibfnamefont {M.}~\bibnamefont {Wirth}},\ }\bibfield
  {title} {\bibinfo {title} {On the geometry of geodesics in discrete optimal
  transport},\ }\href@noop {} {\bibfield  {journal} {\bibinfo  {journal} {Calc.
  Var. Partial Differ. Equ.}\ }\textbf {\bibinfo {volume} {58}},\ \bibinfo
  {pages} {1} (\bibinfo {year} {2019})}\BibitemShut {NoStop}%
\bibitem [{\citenamefont {Avanzini}\ \emph {et~al.}(2021)\citenamefont
  {Avanzini}, \citenamefont {Penocchio}, \citenamefont {Falasco},\ and\
  \citenamefont {Esposito}}]{avanzini2021nonequilibrium}%
  \BibitemOpen
  \bibfield  {author} {\bibinfo {author} {\bibfnamefont {F.}~\bibnamefont
  {Avanzini}}, \bibinfo {author} {\bibfnamefont {E.}~\bibnamefont {Penocchio}},
  \bibinfo {author} {\bibfnamefont {G.}~\bibnamefont {Falasco}},\ and\ \bibinfo
  {author} {\bibfnamefont {M.}~\bibnamefont {Esposito}},\ }\bibfield  {title}
  {\bibinfo {title} {Nonequilibrium thermodynamics of non-ideal chemical
  reaction networks},\ }\href@noop {} {\bibfield  {journal} {\bibinfo
  {journal} {J. Chem. Phys.}\ }\textbf {\bibinfo {volume} {154}},\ \bibinfo
  {pages} {094114} (\bibinfo {year} {2021})}\BibitemShut {NoStop}%
\bibitem [{\citenamefont {Polettini}\ and\ \citenamefont
  {Esposito}(2014)}]{polettini2014irreversible}%
  \BibitemOpen
  \bibfield  {author} {\bibinfo {author} {\bibfnamefont {M.}~\bibnamefont
  {Polettini}}\ and\ \bibinfo {author} {\bibfnamefont {M.}~\bibnamefont
  {Esposito}},\ }\bibfield  {title} {\bibinfo {title} {Irreversible
  thermodynamics of open chemical networks. i. emergent cycles and broken
  conservation laws},\ }\href@noop {} {\bibfield  {journal} {\bibinfo
  {journal} {J. Chem. Phys.}\ }\textbf {\bibinfo {volume} {141}},\ \bibinfo
  {pages} {07B610\_1} (\bibinfo {year} {2014})}\BibitemShut {NoStop}%
\bibitem [{\citenamefont {Andrieux}\ and\ \citenamefont
  {Gaspard}(2007)}]{andrieux2007fluctuation}%
  \BibitemOpen
  \bibfield  {author} {\bibinfo {author} {\bibfnamefont {D.}~\bibnamefont
  {Andrieux}}\ and\ \bibinfo {author} {\bibfnamefont {P.}~\bibnamefont
  {Gaspard}},\ }\bibfield  {title} {\bibinfo {title} {Fluctuation theorem for
  currents and schnakenberg network theory},\ }\href@noop {} {\bibfield
  {journal} {\bibinfo  {journal} {J. Stat. Phys.}\ }\textbf {\bibinfo {volume}
  {127}},\ \bibinfo {pages} {107} (\bibinfo {year} {2007})}\BibitemShut
  {NoStop}%
\bibitem [{\citenamefont {Horowitz}\ and\ \citenamefont
  {Esposito}(2014)}]{horowitz2014thermodynamics}%
  \BibitemOpen
  \bibfield  {author} {\bibinfo {author} {\bibfnamefont {J.~M.}\ \bibnamefont
  {Horowitz}}\ and\ \bibinfo {author} {\bibfnamefont {M.}~\bibnamefont
  {Esposito}},\ }\bibfield  {title} {\bibinfo {title} {Thermodynamics with
  continuous information flow},\ }\href@noop {} {\bibfield  {journal} {\bibinfo
   {journal} {Phys. Rev. X}\ }\textbf {\bibinfo {volume} {4}},\ \bibinfo
  {pages} {031015} (\bibinfo {year} {2014})}\BibitemShut {NoStop}%
\bibitem [{\citenamefont {van~der Meer}\ \emph {et~al.}(2022)\citenamefont
  {van~der Meer}, \citenamefont {Ertel},\ and\ \citenamefont
  {Seifert}}]{van2022thermodynamic}%
  \BibitemOpen
  \bibfield  {author} {\bibinfo {author} {\bibfnamefont {J.}~\bibnamefont
  {van~der Meer}}, \bibinfo {author} {\bibfnamefont {B.}~\bibnamefont
  {Ertel}},\ and\ \bibinfo {author} {\bibfnamefont {U.}~\bibnamefont
  {Seifert}},\ }\bibfield  {title} {\bibinfo {title} {Thermodynamic inference
  in partially accessible {M}arkov networks: A unifying perspective from
  transition-based waiting time distributions},\ }\href@noop {} {\bibfield
  {journal} {\bibinfo  {journal} {arXiv preprint arXiv:2203.12020}\ } (\bibinfo
  {year} {2022})}\BibitemShut {NoStop}%
\bibitem [{Note2()}]{Note2}%
  \BibitemOpen
  \bibinfo {note} {We note that in a withdrawn preprint~\cite
  {ge2011thermodynamics}, Ge and coauthors also found this characterization of
  the EPR~\protect \eqref {eq:epassqnorm} as the squared norm of the
  thermodynamic force with the Onsager coefficient being the metric. They tried
  to express the HS decomposition as an orthogonal decomposition, but
  discovered an error in their proof and withdrew their preprint. In Sec.~\ref
  {sec:decomp:hs}, we discuss that, in general, the HS decomposition is not
  given geometrically.}\BibitemShut {Stop}%
\bibitem [{\citenamefont {Godsil}\ and\ \citenamefont
  {Royle}(2001)}]{godsil2001algebraic}%
  \BibitemOpen
  \bibfield  {author} {\bibinfo {author} {\bibfnamefont {C.}~\bibnamefont
  {Godsil}}\ and\ \bibinfo {author} {\bibfnamefont {G.~F.}\ \bibnamefont
  {Royle}},\ }\href@noop {} {\emph {\bibinfo {title} {Algebraic graph
  theory}}},\ Vol.\ \bibinfo {volume} {207}\ (\bibinfo  {publisher} {Springer
  Science \& Business Media},\ \bibinfo {year} {2001})\BibitemShut {NoStop}%
\bibitem [{Note3()}]{Note3}%
  \BibitemOpen
  \bibinfo {note} {Let $\{p^*(t)\}_{t\in [0,1]}$ and $\{\psi ^*(t)\}_{t\in
  [0,1]}$ give the Wasserstein distance $\protect \mathcal
  {W}(p^{(0)},p^{(1)})$. Then for any $\protect \mathcal {T}>0$,
  $p'(t'):=p^*(t'/\protect \mathcal {T})$ and $\psi '(t'):=\psi ^*(t'/\protect
  \mathcal {T})/\protect \mathcal {T}$ satisfy
  $d_{t'}p'(t')=BL(p'(t'))B^\protect \mathsf {T}\psi '(t')$ and provides
  $\protect \mathcal {W}(p^{(0)},p^{(1)})/\protect \mathcal {T}$ by integrating
  $\DOTSI \intop \ilimits@ _0^\protect \mathcal {T}dt'\|B^\protect \mathsf
  {T}\psi '(t')\|_{L(p'(t'))}^2$. This discussion is valid for more general
  settings like Eqs.~\protect \eqref {eq:generalwasserstein0} and~\protect
  \eqref {eq:generalwasserstein2}}\BibitemShut {NoStop}%
\bibitem [{\citenamefont {Puntanen}\ \emph {et~al.}(2011)\citenamefont
  {Puntanen}, \citenamefont {Styan},\ and\ \citenamefont
  {Isotalo}}]{puntanen2011matrix}%
  \BibitemOpen
  \bibfield  {author} {\bibinfo {author} {\bibfnamefont {S.}~\bibnamefont
  {Puntanen}}, \bibinfo {author} {\bibfnamefont {G.~P.}\ \bibnamefont
  {Styan}},\ and\ \bibinfo {author} {\bibfnamefont {J.}~\bibnamefont
  {Isotalo}},\ }\href@noop {} {\emph {\bibinfo {title} {Matrix tricks for
  linear statistical models: our personal top twenty}}}\ (\bibinfo  {publisher}
  {Springer},\ \bibinfo {year} {2011})\BibitemShut {NoStop}%
\bibitem [{Note4()}]{Note4}%
  \BibitemOpen
  \bibinfo {note} {It is shown under the assumption that the dynamics obeys the
  law of mass action by seeing the equivalence between the following two
  conditions. One is that the force is conservative, which is equivalent to
  $\sigma ^\protect \mathrm {hk}=0$. The other is the so-called generalized
  Wegscheider condition $\ln (k^+/k^-)\in \protect \mathrm {im}\protect
  \,\protect \mathbb {S}^\protect \mathsf {T}$, which is a necessary and
  sufficient condition for the steady state to be detailed balanced~\cite
  {schuster1989generalization}.}\BibitemShut {Stop}%
\bibitem [{\citenamefont {Dechant}(2022)}]{dechant2022minimum}%
  \BibitemOpen
  \bibfield  {author} {\bibinfo {author} {\bibfnamefont {A.}~\bibnamefont
  {Dechant}},\ }\bibfield  {title} {\bibinfo {title} {Minimum entropy
  production, detailed balance and {W}asserstein distance for continuous-time
  {M}arkov processes},\ }\href@noop {} {\bibfield  {journal} {\bibinfo
  {journal} {J. Phys. A}\ } (\bibinfo {year} {2022})}\BibitemShut {NoStop}%
\bibitem [{\citenamefont {Remlein}\ and\ \citenamefont
  {Seifert}(2021)}]{remlein2021optimality}%
  \BibitemOpen
  \bibfield  {author} {\bibinfo {author} {\bibfnamefont {B.}~\bibnamefont
  {Remlein}}\ and\ \bibinfo {author} {\bibfnamefont {U.}~\bibnamefont
  {Seifert}},\ }\bibfield  {title} {\bibinfo {title} {Optimality of
  nonconservative driving for finite-time processes with discrete states},\
  }\href@noop {} {\bibfield  {journal} {\bibinfo  {journal} {Phys. Rev. E}\
  }\textbf {\bibinfo {volume} {103}},\ \bibinfo {pages} {L050105} (\bibinfo
  {year} {2021})}\BibitemShut {NoStop}%
\bibitem [{\citenamefont {Ilker}\ \emph {et~al.}(2022)\citenamefont {Ilker},
  \citenamefont {G{\"u}ng{\"o}r}, \citenamefont {Kuznets-Speck}, \citenamefont
  {Chiel}, \citenamefont {Deffner},\ and\ \citenamefont
  {Hinczewski}}]{ilker2022shortcuts}%
  \BibitemOpen
  \bibfield  {author} {\bibinfo {author} {\bibfnamefont {E.}~\bibnamefont
  {Ilker}}, \bibinfo {author} {\bibfnamefont {{\"O}.}~\bibnamefont
  {G{\"u}ng{\"o}r}}, \bibinfo {author} {\bibfnamefont {B.}~\bibnamefont
  {Kuznets-Speck}}, \bibinfo {author} {\bibfnamefont {J.}~\bibnamefont
  {Chiel}}, \bibinfo {author} {\bibfnamefont {S.}~\bibnamefont {Deffner}},\
  and\ \bibinfo {author} {\bibfnamefont {M.}~\bibnamefont {Hinczewski}},\
  }\bibfield  {title} {\bibinfo {title} {Shortcuts in stochastic systems and
  control of biophysical processes},\ }\href@noop {} {\bibfield  {journal}
  {\bibinfo  {journal} {Physical Review X}\ }\textbf {\bibinfo {volume} {12}},\
  \bibinfo {pages} {021048} (\bibinfo {year} {2022})}\BibitemShut {NoStop}%
\bibitem [{\citenamefont {Van~Vu}\ and\ \citenamefont
  {Saito}(2022)}]{vanvu2022thermodynamic}%
  \BibitemOpen
  \bibfield  {author} {\bibinfo {author} {\bibfnamefont {T.}~\bibnamefont
  {Van~Vu}}\ and\ \bibinfo {author} {\bibfnamefont {K.}~\bibnamefont {Saito}},\
  }\bibfield  {title} {\bibinfo {title} {Thermodynamic unification of optimal
  transport: Thermodynamic uncertainty relation, minimum dissipation, and
  thermodynamic speed limits},\ }\href@noop {} {\bibfield  {journal} {\bibinfo
  {journal} {arXiv preprint arXiv:2206.02684}\ } (\bibinfo {year}
  {2022})}\BibitemShut {NoStop}%
\bibitem [{Note5()}]{Note5}%
  \BibitemOpen
  \bibinfo {note} {This relation was proved for one-dimensional systems in
  \cite {vanvu2022thermodynamic}. In Appendix~\ref {app:onsdiff}, we
  generalized it to multidimensional cases.}\BibitemShut {Stop}%
\bibitem [{\citenamefont {Maes}\ and\ \citenamefont
  {Neto{\v{c}}n{\`y}}(2008)}]{maes2008canonical}%
  \BibitemOpen
  \bibfield  {author} {\bibinfo {author} {\bibfnamefont {C.}~\bibnamefont
  {Maes}}\ and\ \bibinfo {author} {\bibfnamefont {K.}~\bibnamefont
  {Neto{\v{c}}n{\`y}}},\ }\bibfield  {title} {\bibinfo {title} {Canonical
  structure of dynamical fluctuations in mesoscopic nonequilibrium steady
  states},\ }\href@noop {} {\bibfield  {journal} {\bibinfo  {journal} {EPL
  (Europhysics Letters)}\ }\textbf {\bibinfo {volume} {82}},\ \bibinfo {pages}
  {30003} (\bibinfo {year} {2008})}\BibitemShut {NoStop}%
\bibitem [{\citenamefont {Koyuk}\ and\ \citenamefont
  {Seifert}(2020)}]{koyuk2020thermodynamic}%
  \BibitemOpen
  \bibfield  {author} {\bibinfo {author} {\bibfnamefont {T.}~\bibnamefont
  {Koyuk}}\ and\ \bibinfo {author} {\bibfnamefont {U.}~\bibnamefont
  {Seifert}},\ }\bibfield  {title} {\bibinfo {title} {Thermodynamic uncertainty
  relation for time-dependent driving},\ }\href@noop {} {\bibfield  {journal}
  {\bibinfo  {journal} {Phys. Rev. Lett.}\ }\textbf {\bibinfo {volume} {125}},\
  \bibinfo {pages} {260604} (\bibinfo {year} {2020})}\BibitemShut {NoStop}%
\bibitem [{\citenamefont {Dolbeault}\ \emph {et~al.}(2009)\citenamefont
  {Dolbeault}, \citenamefont {Nazaret},\ and\ \citenamefont
  {Savar{\'e}}}]{dolbeault2009new}%
  \BibitemOpen
  \bibfield  {author} {\bibinfo {author} {\bibfnamefont {J.}~\bibnamefont
  {Dolbeault}}, \bibinfo {author} {\bibfnamefont {B.}~\bibnamefont {Nazaret}},\
  and\ \bibinfo {author} {\bibfnamefont {G.}~\bibnamefont {Savar{\'e}}},\
  }\bibfield  {title} {\bibinfo {title} {A new class of transport distances
  between measures},\ }\href@noop {} {\bibfield  {journal} {\bibinfo  {journal}
  {Calc. Var. Partial Differ. Equ.}\ }\textbf {\bibinfo {volume} {34}},\
  \bibinfo {pages} {193} (\bibinfo {year} {2009})}\BibitemShut {NoStop}%
\bibitem [{\citenamefont {Shiraishi}(2021)}]{shiraishi2021optimal}%
  \BibitemOpen
  \bibfield  {author} {\bibinfo {author} {\bibfnamefont {N.}~\bibnamefont
  {Shiraishi}},\ }\bibfield  {title} {\bibinfo {title} {Optimal thermodynamic
  uncertainty relation in {M}arkov jump processes},\ }\href@noop {} {\bibfield
  {journal} {\bibinfo  {journal} {J. Stat. Phys.}\ }\textbf {\bibinfo {volume}
  {185}},\ \bibinfo {pages} {1} (\bibinfo {year} {2021})}\BibitemShut {NoStop}%
\bibitem [{\citenamefont {Chow}\ \emph {et~al.}(2012)\citenamefont {Chow},
  \citenamefont {Huang}, \citenamefont {Li},\ and\ \citenamefont
  {Zhou}}]{chow2012fokker}%
  \BibitemOpen
  \bibfield  {author} {\bibinfo {author} {\bibfnamefont {S.-N.}\ \bibnamefont
  {Chow}}, \bibinfo {author} {\bibfnamefont {W.}~\bibnamefont {Huang}},
  \bibinfo {author} {\bibfnamefont {Y.}~\bibnamefont {Li}},\ and\ \bibinfo
  {author} {\bibfnamefont {H.}~\bibnamefont {Zhou}},\ }\bibfield  {title}
  {\bibinfo {title} {Fokker--planck equations for a free energy functional or
  {M}arkov process on a graph},\ }\href@noop {} {\bibfield  {journal} {\bibinfo
   {journal} {Arch. Ration. Mech. and Anal.}\ }\textbf {\bibinfo {volume}
  {203}},\ \bibinfo {pages} {969} (\bibinfo {year} {2012})}\BibitemShut
  {NoStop}%
\bibitem [{\citenamefont {Muratore-Ginanneschi}\ \emph
  {et~al.}(2013)\citenamefont {Muratore-Ginanneschi}, \citenamefont
  {Mej{\'\i}a-Monasterio},\ and\ \citenamefont {Peliti}}]{muratore2013heat}%
  \BibitemOpen
  \bibfield  {author} {\bibinfo {author} {\bibfnamefont {P.}~\bibnamefont
  {Muratore-Ginanneschi}}, \bibinfo {author} {\bibfnamefont {C.}~\bibnamefont
  {Mej{\'\i}a-Monasterio}},\ and\ \bibinfo {author} {\bibfnamefont
  {L.}~\bibnamefont {Peliti}},\ }\bibfield  {title} {\bibinfo {title} {Heat
  release by controlled continuous-time {M}arkov jump processes},\ }\href@noop
  {} {\bibfield  {journal} {\bibinfo  {journal} {J. Stat. Phys.}\ }\textbf
  {\bibinfo {volume} {150}},\ \bibinfo {pages} {181} (\bibinfo {year}
  {2013})}\BibitemShut {NoStop}%
\bibitem [{\citenamefont {Kolchinsky}\ \emph {et~al.}(2022)\citenamefont
  {Kolchinsky}, \citenamefont {Dechant}, \citenamefont {Yoshimura},\ and\
  \citenamefont {Ito}}]{Kolchinsky2022information}%
  \BibitemOpen
  \bibfield  {author} {\bibinfo {author} {\bibfnamefont {A.}~\bibnamefont
  {Kolchinsky}}, \bibinfo {author} {\bibfnamefont {A.}~\bibnamefont {Dechant}},
  \bibinfo {author} {\bibfnamefont {K.}~\bibnamefont {Yoshimura}},\ and\
  \bibinfo {author} {\bibfnamefont {S.}~\bibnamefont {Ito}},\ }\bibfield
  {title} {\bibinfo {title} {Information geometry of excess and housekeeping
  entropy production},\ }\href@noop {} {\bibfield  {journal} {\bibinfo
  {journal} {arXiv preprint arXiv:2206.14599}\ } (\bibinfo {year}
  {2022})}\BibitemShut {NoStop}%
\bibitem [{\citenamefont {Van~den Broeck}\ and\ \citenamefont
  {Esposito}(2010)}]{van2010three}%
  \BibitemOpen
  \bibfield  {author} {\bibinfo {author} {\bibfnamefont {C.}~\bibnamefont
  {Van~den Broeck}}\ and\ \bibinfo {author} {\bibfnamefont {M.}~\bibnamefont
  {Esposito}},\ }\bibfield  {title} {\bibinfo {title} {Three faces of the
  second law. {II}. fokker-planck formulation},\ }\href@noop {} {\bibfield
  {journal} {\bibinfo  {journal} {Physical Review E}\ }\textbf {\bibinfo
  {volume} {82}},\ \bibinfo {pages} {011144} (\bibinfo {year}
  {2010})}\BibitemShut {NoStop}%
\bibitem [{\citenamefont {Rockafellar}\ and\ \citenamefont
  {Wets}(2009)}]{rockafellar2009variational}%
  \BibitemOpen
  \bibfield  {author} {\bibinfo {author} {\bibfnamefont {R.~T.}\ \bibnamefont
  {Rockafellar}}\ and\ \bibinfo {author} {\bibfnamefont {R.~J.-B.}\
  \bibnamefont {Wets}},\ }\href@noop {} {\emph {\bibinfo {title} {Variational
  analysis}}},\ Vol.\ \bibinfo {volume} {317}\ (\bibinfo  {publisher} {Springer
  Science \& Business Media},\ \bibinfo {year} {2009})\BibitemShut {NoStop}%
\bibitem [{\citenamefont {Horn}\ and\ \citenamefont
  {Jackson}(1972)}]{horn1972general}%
  \BibitemOpen
  \bibfield  {author} {\bibinfo {author} {\bibfnamefont {F.}~\bibnamefont
  {Horn}}\ and\ \bibinfo {author} {\bibfnamefont {R.}~\bibnamefont {Jackson}},\
  }\bibfield  {title} {\bibinfo {title} {General mass action kinetics},\
  }\href@noop {} {\bibfield  {journal} {\bibinfo  {journal} {Arch. Ration.
  Mech. and Anal.}\ }\textbf {\bibinfo {volume} {47}},\ \bibinfo {pages} {81}
  (\bibinfo {year} {1972})}\BibitemShut {NoStop}%
\bibitem [{\citenamefont {Craciun}(2015)}]{craciun2015toric}%
  \BibitemOpen
  \bibfield  {author} {\bibinfo {author} {\bibfnamefont {G.}~\bibnamefont
  {Craciun}},\ }\bibfield  {title} {\bibinfo {title} {Toric differential
  inclusions and a proof of the global attractor conjecture},\ }\href@noop {}
  {\bibfield  {journal} {\bibinfo  {journal} {arXiv preprint arXiv:1501.02860}\
  } (\bibinfo {year} {2015})}\BibitemShut {NoStop}%
\bibitem [{\citenamefont {Anderson}(2011)}]{anderson2011proof}%
  \BibitemOpen
  \bibfield  {author} {\bibinfo {author} {\bibfnamefont {D.~F.}\ \bibnamefont
  {Anderson}},\ }\bibfield  {title} {\bibinfo {title} {A proof of the global
  attractor conjecture in the single linkage class case},\ }\href@noop {}
  {\bibfield  {journal} {\bibinfo  {journal} {SIAM Journal on Applied
  Mathematics}\ }\textbf {\bibinfo {volume} {71}},\ \bibinfo {pages} {1487}
  (\bibinfo {year} {2011})}\BibitemShut {NoStop}%
\bibitem [{\citenamefont {Ge}\ \emph {et~al.}(2011)\citenamefont {Ge},
  \citenamefont {Kim},\ and\ \citenamefont {Qian}}]{ge2011thermodynamics}%
  \BibitemOpen
  \bibfield  {author} {\bibinfo {author} {\bibfnamefont {H.}~\bibnamefont
  {Ge}}, \bibinfo {author} {\bibfnamefont {W.~H.}\ \bibnamefont {Kim}},\ and\
  \bibinfo {author} {\bibfnamefont {H.}~\bibnamefont {Qian}},\ }\bibfield
  {title} {\bibinfo {title} {Thermodynamics and geometry of reversible and
  irreversible {M}arkov processes},\ }\href@noop {} {\bibfield  {journal}
  {\bibinfo  {journal} {arXiv preprint arXiv:1108.4055}\ } (\bibinfo {year}
  {2011})}\BibitemShut {NoStop}%
\bibitem [{\citenamefont {Schuster}\ and\ \citenamefont
  {Schuster}(1989)}]{schuster1989generalization}%
  \BibitemOpen
  \bibfield  {author} {\bibinfo {author} {\bibfnamefont {S.}~\bibnamefont
  {Schuster}}\ and\ \bibinfo {author} {\bibfnamefont {R.}~\bibnamefont
  {Schuster}},\ }\bibfield  {title} {\bibinfo {title} {A generalization of
  {W}egscheider's condition. implications for properties of steady states and
  for quasi-steady-state approximation},\ }\href@noop {} {\bibfield  {journal}
  {\bibinfo  {journal} {Journal of Mathematical Chemistry}\ }\textbf {\bibinfo
  {volume} {3}},\ \bibinfo {pages} {25} (\bibinfo {year} {1989})}\BibitemShut
  {NoStop}%
\end{thebibliography}

\end{document}